\documentclass{emulateapj}
\usepackage{natbib}

\shorttitle{Radio variability of young stars in Orion}
\shortauthors{Rivilla et al.}

\begin{document}

\title{Short- and long-term radio variability of young stars in the Orion Nebula Cluster and Molecular Cloud}

\author{V. M. Rivilla\altaffilmark{1,2},  C.J. Chandler\altaffilmark{3}, J. Sanz-Forcada\altaffilmark{4}, I. Jim\'{e}nez-Serra\altaffilmark{5,6}, J. Forbrich\altaffilmark{7} and J. Mart\'{\i}n-Pintado\altaffilmark{1}}

\altaffiltext{1}{Osservatorio Astrofisico di Arcetri, Largo Enrico Fermi 5, I-50125, Firenze, Italia;
rivilla@arcetri.astro.it}

\altaffiltext{2}{Centro de Astrobiolog\'{\i}a (CSIC/INTA),
Ctra. de Torrej\'on a Ajalvir km 4,
E-28850 Torrej\'on de Ardoz, Madrid, Spain}

\altaffiltext{3}{National Radio Astronomy Observatory, P.O. Box O,
Socorro, NM 87801, USA}

\altaffiltext{4}{Centro de Astrobiolog\'{\i}a (CSIC/INTA), ESAC Campus, PO Box 78, 28691 Villanueva de la Ca\~nada, Madrid, Spain}

\altaffiltext{5}{European Southern Observatory, Karl-Schwarzschild-Str. 2, 85748, Garching, Germany}

\altaffiltext{6}{Department of Physics and Astronomy, 
University College London,
132 Hampstead Road,
London NW1 2PS, UK}

\altaffiltext{7}{Department of Astrophysics, University of Vienna, T\"urkenschanzstr. 17, 1180 Vienna, Austria }

\begin{abstract}

We have used the Karl G. Jansky Very Large Array (VLA) to carry out a multi-epoch radio continuum monitoring of the Orion Nebula Cluster (ONC) and the background
Orion Molecular Cloud (OMC) (3 epochs at Q-band and 11 epochs at Ka-band).
Our new observations reveal the presence of 19 radio sources, mainly concentrated in the Trapezium Cluster and the Orion Hot Core (OHC) regions. 
With the exception of the Becklin-Neugebauer (BN) object and the source C (which we identify here as dust emission associated with a proplyd) the sources all show radio variability between the different epochs. We have found tentative evidence of variability in the emission from the massive object related with source I. Our observations also confirm radio flux density variations of a factor $>$2 on timescales of hours to days in 5 sources. One of these flaring sources, OHC-E, has been detected
for the first time.  
We conclude that the radio emission can be attributed to two different components: i) highly-variable (flaring) non-thermal radio gyrosynchrotron emission produced by electrons accelerated in the magnetospheres of pre-main sequence low-mass stars; ii) thermal emission due to free-free radiation from ionized gas and/or heated dust around embedded massive objects and proplyds.
Combining our sample with other radio monitoring at 8.3 GHz and the X-ray catalog provided by Chandra, we have studied the properties of the entire sample of radio/X-ray stars in the ONC/OMC region (51 sources).    
We have found several hints of a relation between the X-ray activity and the mechanisms responsible for (at least some fraction of) the radio emission.
We have estimated a radio flaring rate of $\sim$0.14 flares day$^{-1}$ in the dense stellar cluster embedded in the OHC region.  This suggests that radio flares are more common events during the first stages of stellar evolution than previously thought.  
The advent of improved sensitivity with the new VLA and ALMA will dramatically increase the number of stars in young clusters detected at radio wavelengths, which will help us to improve our understanding of the origin and nature of the radio emission.

\end{abstract}


\begin{table*}
\caption{Multi-epoch VLA radio continuum observations.}
\label{tableobservations}
\tabcolsep 2.pt
\centering
\hspace*{-6mm}
\begin{tabular}{c c c c c c c c c c c c c c c c}
\hline
Config. & Project & Band & Freq. & BW & Epoch & JD & \multicolumn{2}{c}{Pointing center} & Obs.  & beam & RMS & Primary & Gain calibrator \\
        & ID & name & (GHz) & (MHz) &     & & RA$_{\rm J2000}$ & DEC$_{\rm
J2000}$ & length & ($\arcsec\times\arcsec$) &  (mJy) & flux density & flux density (Jy) \\
  & & & & & & & 5$h$ 35$m$ & -5$^{\rm o}$ 22$\arcmin$ & (min) &  &  & calibrator & (J0541-0541) \\
\hline
B    & AJ356  & Q  & 45.6 & 25  &  2009 Mar 9  & 2454900    & 14.60$s$ &  30.0$\arcsec$ & 120 & 0.23$\times$0.15   & 0.45 & J0137+3309  & 0.63 \\
B    & AJ356  & Q  & 45.6 & 25  &  2009 Mar 19 & 2454910    & 14.60$s$ &  30.0$\arcsec$ & 240 & 0.22$\times$0.15   & 0.37 & J0137+3309  & 0.63 \\
D    & AR712  & Q  & 43.3 & 100 &  2009 Dec 22 & 2455188    & 14.50$s$ &  31.0$\arcsec$ & 60  & 1.9$\times$1.4     & 0.54 & J0137+3309  & 0.46 \\
C    & 10B-175 & Ka & 33.6 & 256 &  2010 Oct 24 & 2455494    & 14.50$s$ &  30.0$\arcsec$ & 30  & 1.3$\times$0.65    & 0.36 & J0542+4951  & 0.68 \\
C    & 10B-175 & Ka & 33.6 & 256 &  2010 Nov 23 & 2455523.76 & 14.50$s$ &  30.0$\arcsec$ & 30  & 0.92$\times$0.60   & 0.40 & J0542+4951  & 0.68 \\
C    & 10B-175 & Ka & 33.6 & 256 &  2010 Nov 23 & 2455523.89 & 14.50$s$ &  30.0$\arcsec$ & 30  & 0.75$\times$0.56   & 0.44 & J0137+3309  & 0.69 \\
C    & 10B-175 & Ka & 33.6 & 256 &  2011 Jan 8  & 2455569    & 14.50$s$ &  30.0$\arcsec$ & 30  & 0.86$\times$0.56   & 0.39 & J0137+3309  & 0.62 \\
CnB-B& 10B-175 & Ka & 33.6 & 256 &  2011 Feb 8  & 2455601    & 14.50$s$ &  30.0$\arcsec$ & 30  & 0.30$\times$0.24   & 0.32 & J0137+3309  & 0.58 \\
B    & 10B-175 & Ka & 33.6 & 256 &  2011 Mar 28 & 2455649    & 14.50$s$ &  30.0$\arcsec$ & 30  & 0.28$\times$0.22   & 0.22 & J0542+4951  & 0.50 \\
BnA  & 10B-175 & Ka & 33.6 & 256 &  2011 May 27 & 2455709    & 14.50$s$ &  30.0$\arcsec$ & 30  & 0.62$\times$0.077  & 0.22 & J0542+4951  & 0.54 \\
BnA-A& 10B-175 & Ka & 37.5 & 128 &  2011 Jun 4  & 2455717    & 14.50$s$ &  30.0$\arcsec$ & 120 & 0.096$\times$0.073 & 0.13 & J0542+4951  & 0.50 \\         
BnA-A& 10B-175 & Ka & 30.5 & 128 &  2011 Jun 4  & 2455717    & 14.50$s$ &  30.0$\arcsec$ & 120 & 0.13$\times$0.083  & 0.10 & J0542+4951  & 0.52 \\         
A    & 10B-175 & Ka & 33.6 & 256 &  2011 Jun 11 & 2455724    & 14.50$s$ &  30.0$\arcsec$ & 30  & 0.14$\times$0.062  & 0.16 & J0137+3309  & 0.54 \\
A    & 10B-175 & Ka & 33.6 & 256 &  2011 Jul 09 & 2455752    & 14.50$s$ &  30.0$\arcsec$ & 30  & 0.091$\times$0.060 & 0.14 & J0542+4951  & 0.48 \\

\hline
\end{tabular}
\end{table*}

\begin{figure*}
\centering
\includegraphics[angle=0,width=8.9cm]{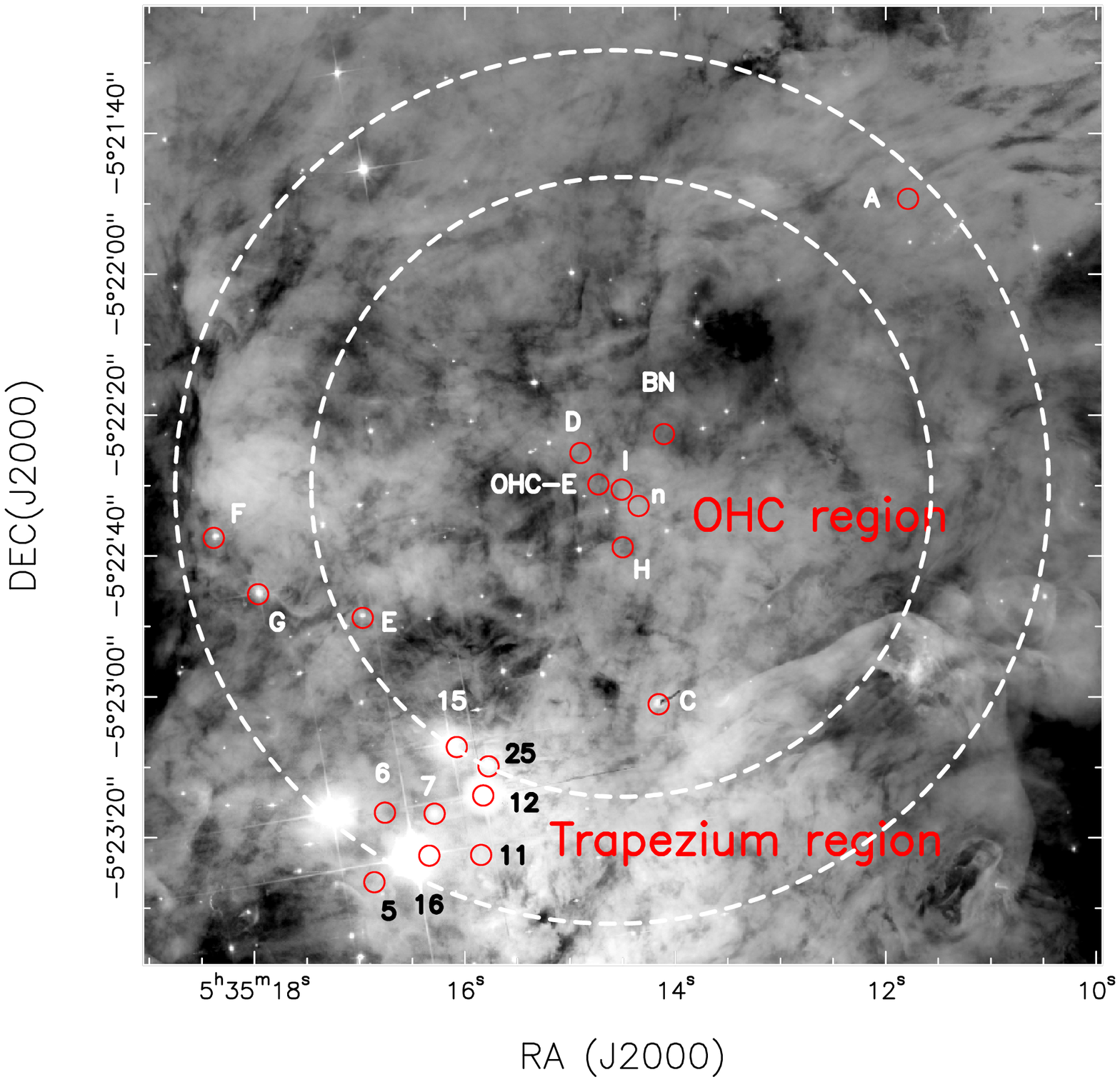}
\hspace*{0.3cm}
\includegraphics[angle=0,width=7.8cm]{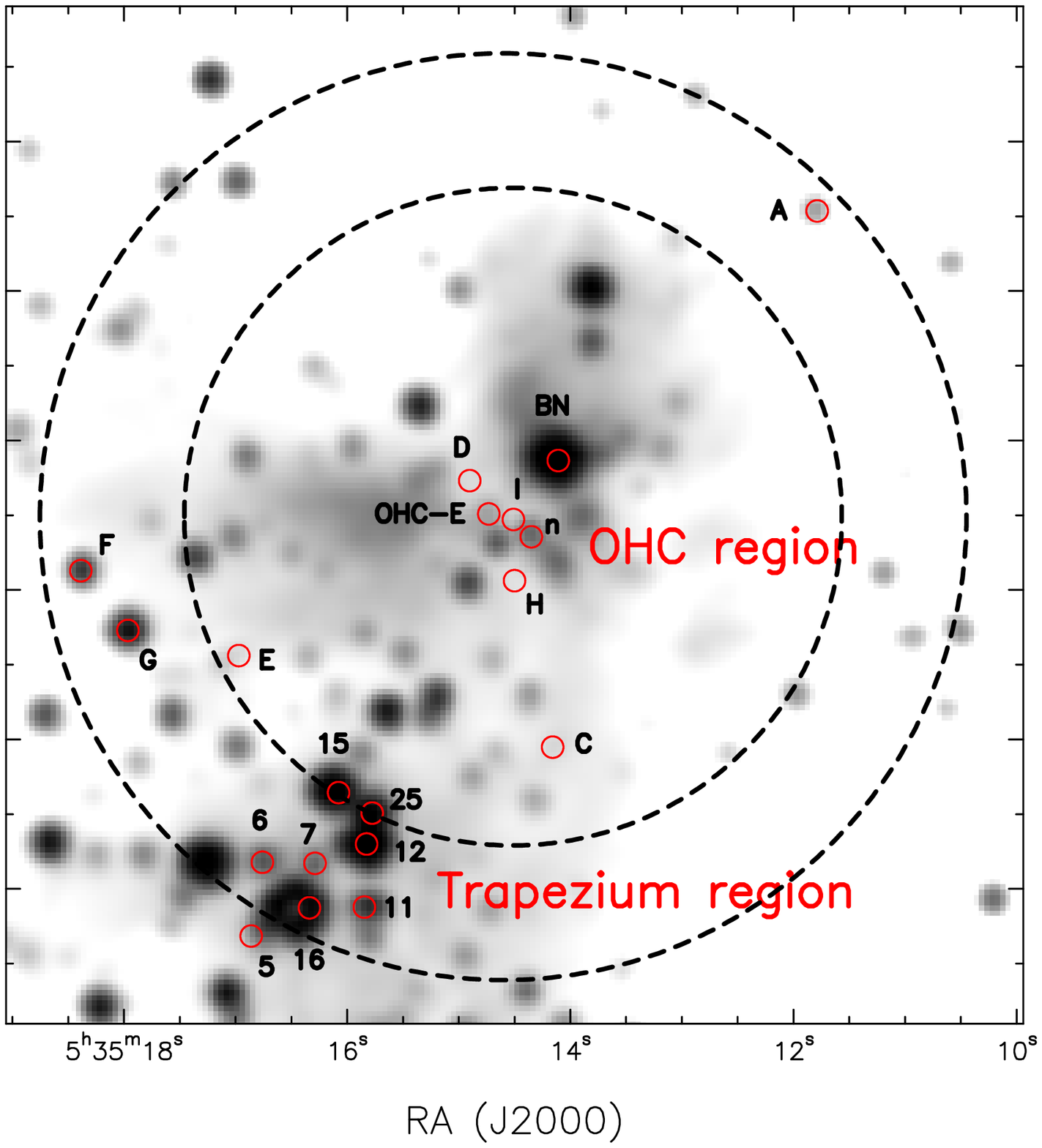}
\caption{Position of the 19 radio sources detected in our monitoring program, overplotted on the R-band ACS/WFC HST image (left panel) and on the K-band 2 Micron All Sky Survey (2MASS) image (right
panel). The dotted circles indicate the regions of the field where the primary beam responses are $>$0.15 at Q-band (43.3 GHz) and Ka-band (33.6 GHz).}
\label{fig-our-sources}
\end{figure*}

\section{Introduction}
\label{intro}

High energy processes during the first evolutionary stages of star
formation are responsible for both radio and X-ray emission
(\citealt{feigelson99}). Low-mass pre-main sequence (PMS) stars are
well known strong X-ray emitters. Their enhanced magnetic activity
with respect to more evolved stars produces violent reconnection
events in the corona of the stars, where the plasma heated to high
temperatures strongly emits variable X-ray emission. Massive stars
also emit X-ray radiation, usually related to wind shocks.

Our understanding of the X-ray emission from young stars has
dramatically increased in the recent years due to {\it Chandra} and
{\it XMM-Newton} (\citealt{getman08}, \citealt{arzner07}). X-ray
observations have revealed thousands of PMS stars in tens of stellar
clusters, resulting in good constraints on their X-ray properties
such as plasma temperatures, levels of variability, luminosities
and X-ray flare rate (see, e.g., \citealt{wolk05}).

In contrast, the physics associated with the radio events (nature
and origin of the emission, variability, timescales, flaring rate)
are still poorly constrained. \citet{drake89} proposed that radio
flares might be produced by the same coronal activity that is responsible for bright X-ray emission (see review by \citealt{gudel02}).
It would be expected then that electrons spiraling in the magnetic
field of the corona produce non-thermal and highly variable
gyrosynchrotron radiation. Moreover, ionized material in the vicinity
of stars, in circumstellar disks or envelopes or at the base of
bipolar outflows, also produce thermal free-free ({\it bremsstrahlung})
radiation.

Long-term radio variability on timescales of months to
years has been observed in star-forming regions
(\citealt{felli93,zapata04a,forbrich07,choi08}). However, it is still not clear
whether these variations are caused by long-term mechanisms, or
they are indeed produced by a sequence of events occurring
on shorter timescales. Systematic observations looking for short-term
variability are required to answer this question.

Recently, \citet{liu14} detected radio variability on hour timescales in the young stellar cluster R Coronae Australis.
The most powerful radio flares\footnote{\hspace{1mm} In this work we will use the term {\em flare} to refer to flux
density variations of a factor of $>$2 on timescales from hours
to days.} have been serendipitously reported so far toward the Orion
Nebula Cluster (ONC) and the background Orion Molecular Cloud (OMC).
\citet{bower03} reported a strong radio flare at 86 GHz arising
from a PMS star in the ONC, and \citet{forbrich08} presented an
even stronger radio flare at 22 GHz, originating from a young star
deeply embedded in the OMC previously detected through its X-ray
emission.  The low number of observed events could indicate that
radio flares are a rare phenomenon (\citealt{andre96}), but at the
same time prevents a proper statistical analysis of short-term
variability phenomena.

Since the typical timescales of radio variability are poorly known,
we have carried out a monitoring program comprising various cadences
ranging from 3 hours to several months. The new capabilities of the
Karl G. Jansky VLA now allows the scheduling of multiple, short
snapshots with good sensitivity in a reasonable amount of observing
time. The only two examples of powerful radio flares are
located in Orion, and this region also harbors a rich cluster
of low-mass stars (\citealt{rivilla13a}); hence, this region is an excellent
target for the detection of many sources in a single pointing.  

We have carried out a multi-epoch radio continuum monitoring of the ONC/OMC region using the Karl G. Jansky Very Large Array (VLA). This is the first radio monitoring at high centimeter frequencies in Orion, with 3 epochs at Q-band and 11 epochs at Ka-band.
Our data allow us to study for the first time both the short
(hours to days) and long (months) timescale variability of the radio
sources in Orion.

The paper is laid out as follows. In Section \ref{observations} we
present the details of the observations. In Section \ref{results}
we show the results of the monitoring at Ka and Q bands.  In Section \ref{full-sample}
we compare our results with a previous monitoring at lower frequency (8.3 GHz). We also compile the full sample of radio/X-ray sources in the ONC/OMC region, and then compare the radio and X-ray properties with the aim of better understanding the link between radio and X-ray emission. In Section
\ref{discussionE} we discuss in more detail the new radio flaring
source detected by our monitoring, OHC-E.  In Section \ref{source-12}
we analyze in particular the radio variability observed in the binary
system $\theta^{1}$ {\it Ori A}.  In Section \ref{flaring-rate} we estimate the rate of radio
flaring activity of young stars in the Orion Hot Core (OHC). Finally, in Section \ref{conclusions} we summarize
the conclusions of our work.


\begin{figure*}
\centering
\includegraphics[angle=0,width=17cm]{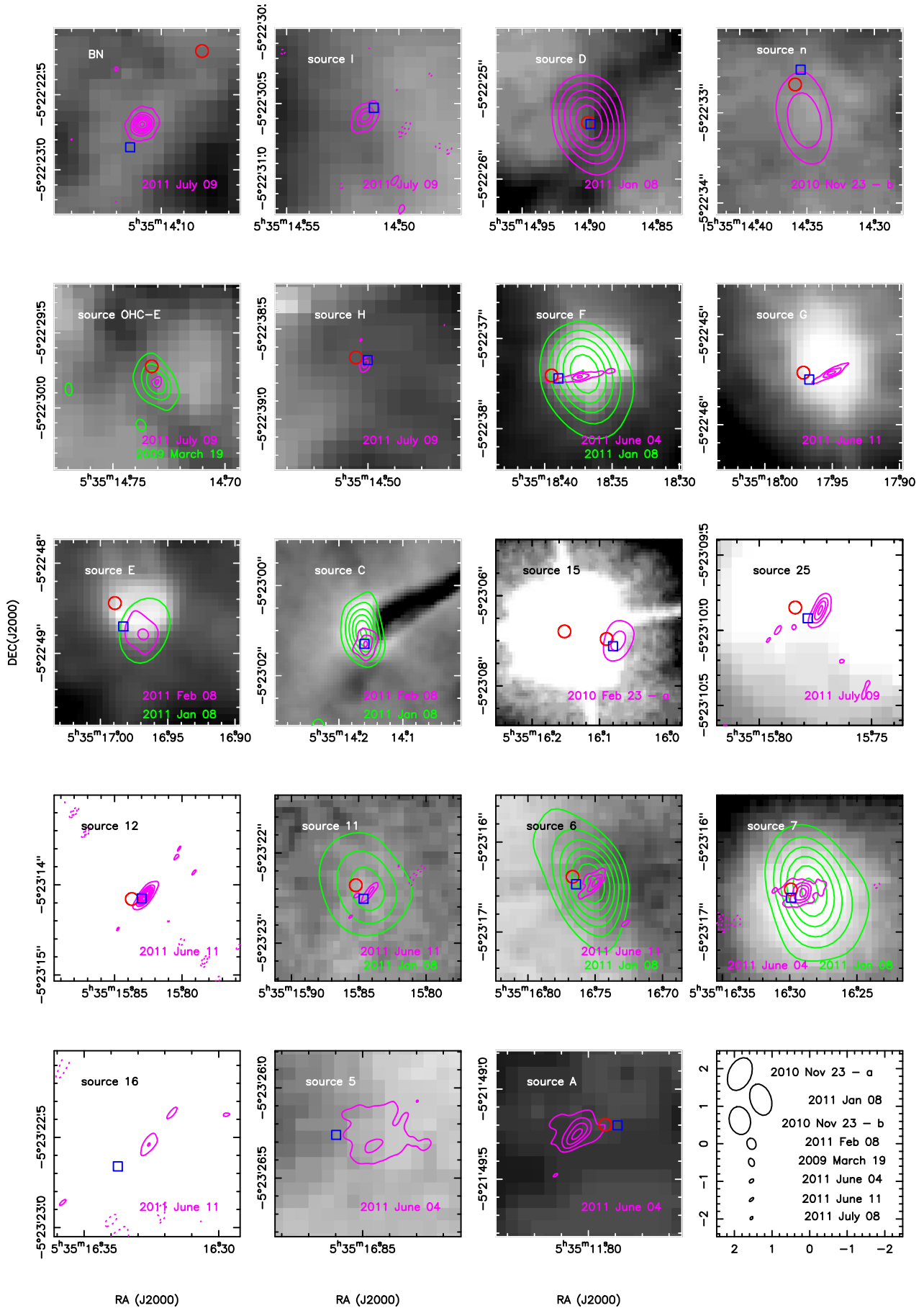}       
\caption{Images of the 19 radio sources detected in our radio
monitoring.  The name of the source is indicated in the top left
corner of each panel. In the lower right corners we indicate the
epoch of the observation (we have selected the epochs with the best
spatial resolution and good detections, and in some cases we show
the detections in two different epochs with different colors). The
first contour level is 3$\sigma$ and the step between successive
contours is 2$\sigma$, with the exception of BN, I, 25 and 12 (which
have steps of 10$\sigma$), and the 2009 March 19 epoch for OHC-E
(steps of 5$\sigma$). The open blue squares indicate the position
of known radio sources detected by \citet{zapata04a} at 8.3 GHz,
and the red circles denote the position of X-ray stars detected by
Chandra (COUP Project). The background greyscale image is the ACS/WFC
HST image at R-band, from the Hubble Legacy Archive. In the lower
right panel we show the synthesized beams of the different images
used.}
\label{fig-2}
\end{figure*}

\begin{figure}
\hskip8mm
\includegraphics[angle=0,width=7cm]{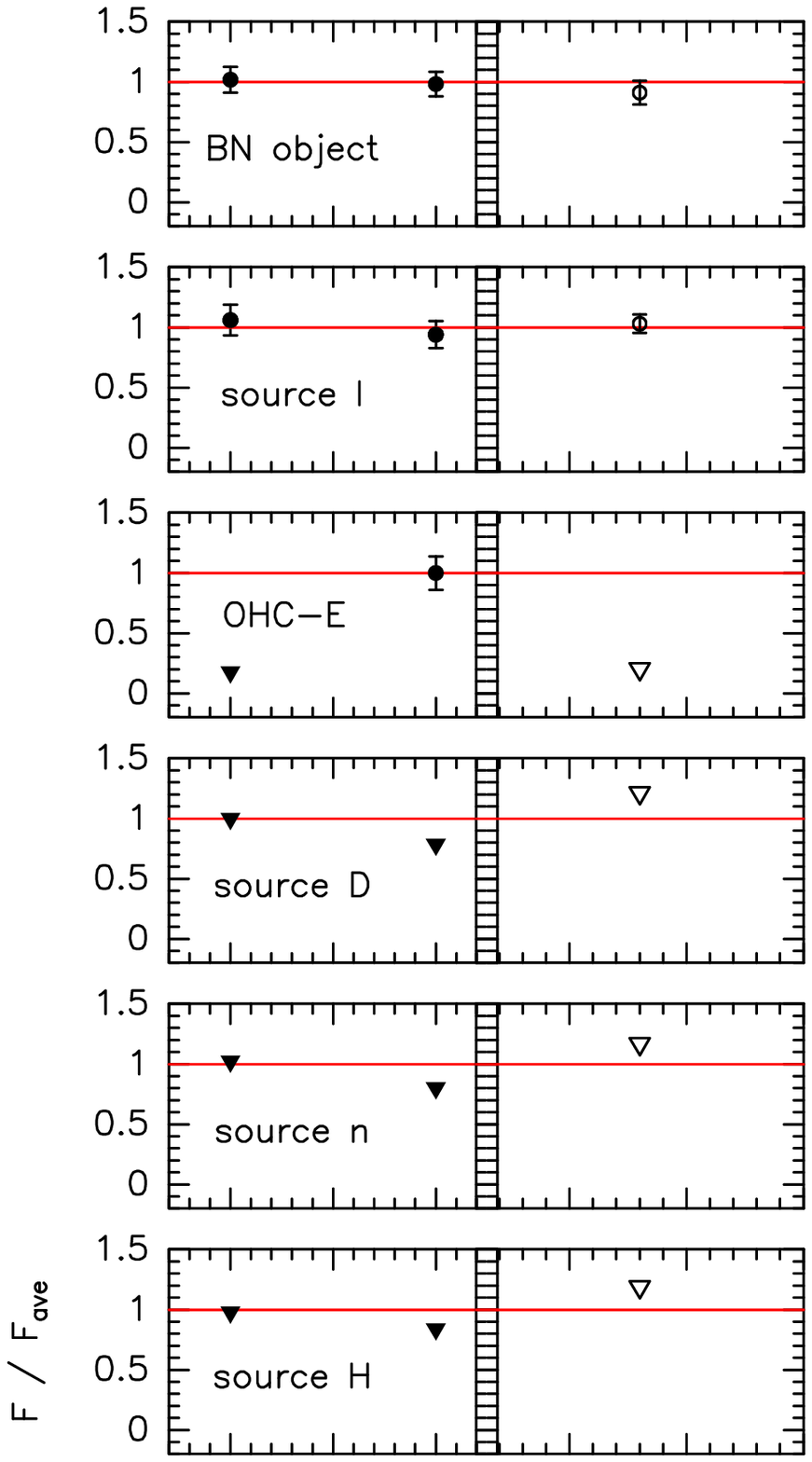}
\vskip2.5mm
\hskip15.5mm
\includegraphics[angle=0,width=6.6cm]{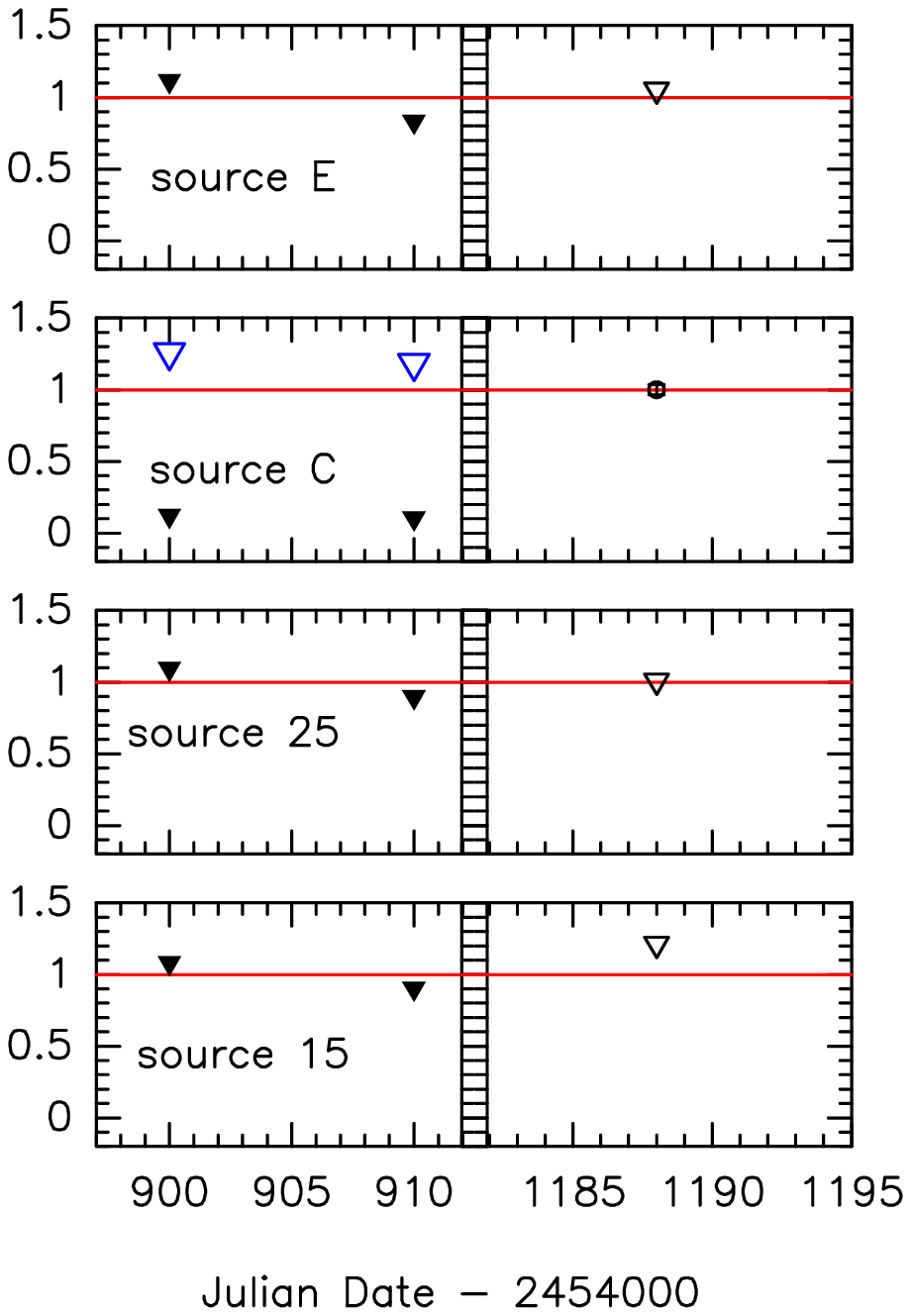}
\caption{Q-band light curves for those sources that fall within the
Q-band primary beam. In the case of non-detections, 3$\sigma$ upper
limits are indicated with triangles. The flux densities have been
normalized by the average value ($F_{\rm av}$) calculated with the
positive detections (or the average value of the upper limits if
the source is not detected). The error bars indicate the flux density
uncertainties. The red horizontal line indicates the value at which
the flux density is equal to the average. The first two observations
were carried out at 45.6 GHz (black filled symbols), while the last
one was carried out at 43.3 GHz (black open symbols). For source
C, the blue open triangles indicate upper limits for extended
emission derived from smoothed images (see Section
\ref{variability-months}).}
\label{fig-Q-variability}
\end{figure}

\begin{figure*}
\centering
\includegraphics[angle=0,width=18cm]{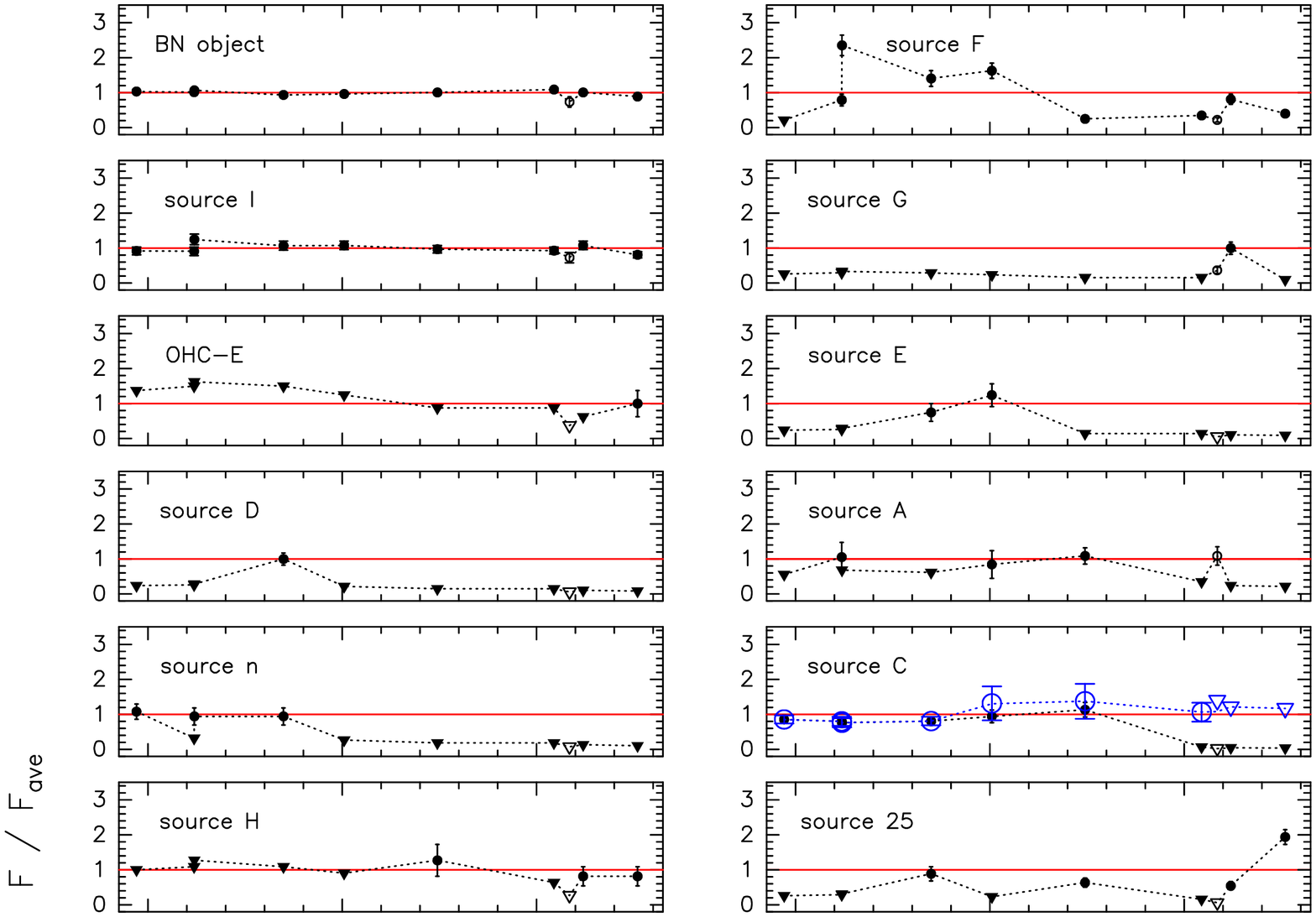}
\vskip3mm
\hskip11.4mm
\includegraphics[angle=0,width=16.75cm]{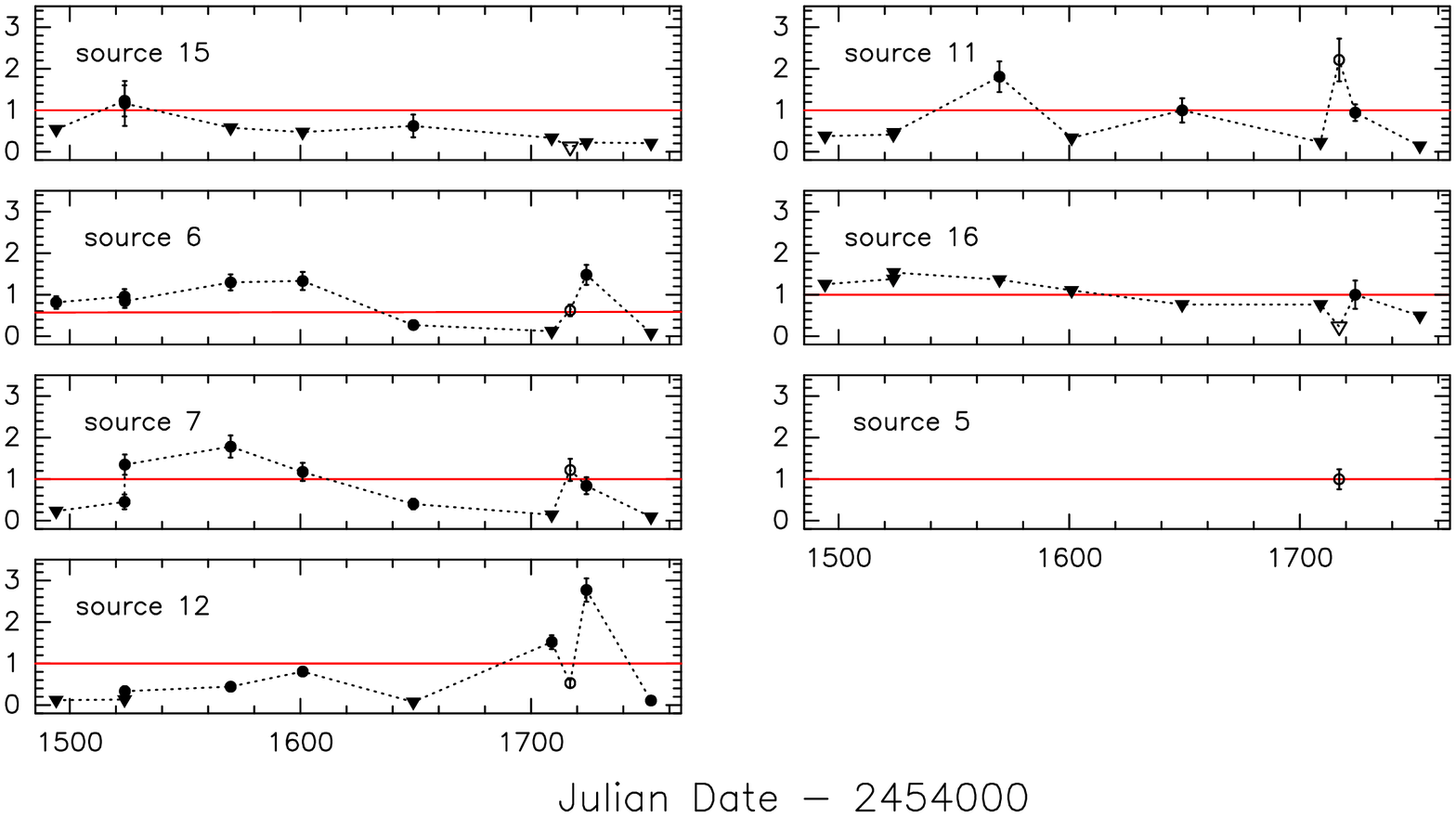}
\caption{Ka-band light curves of all the 19 sources detected
throughout the Ka-band monitoring. In the case of non-detections,
3$\sigma$ upper limits are indicated with triangles. The flux
densities have been normalized by the average value ($F_{\rm av}$)
calculated with the positive detections (or the average value of
the upper limits if the source is not detected). The error bars
indicate the flux density uncertainties. The red horizontal line
indicates the value at which the flux density is equal to the average
flux density. The dotted line merely joins the flux densities at
different epochs, and is not indicative of the evolution of the
flux density between epochs. The frequency of the observations
plotted is 33.6 GHz (black symbols), with the exception of the 2011
June 4 observation, which corresponds to a frequency of 30.5 GHz
(indicated with open symbols). The second and third epoch on 2010
November 23 are only separated by $\sim$3 hours. We discuss this
short-term variability in detail in Section \ref{short-variability}.
For source C, the blue open symbols indicate values for extended
emission derived from smoothed images (see Section
\ref{variability-months}).}
\label{fig-Ka-variability}
\end{figure*}

\begin{figure}
\centering
\includegraphics[angle=0,width=8.0cm]{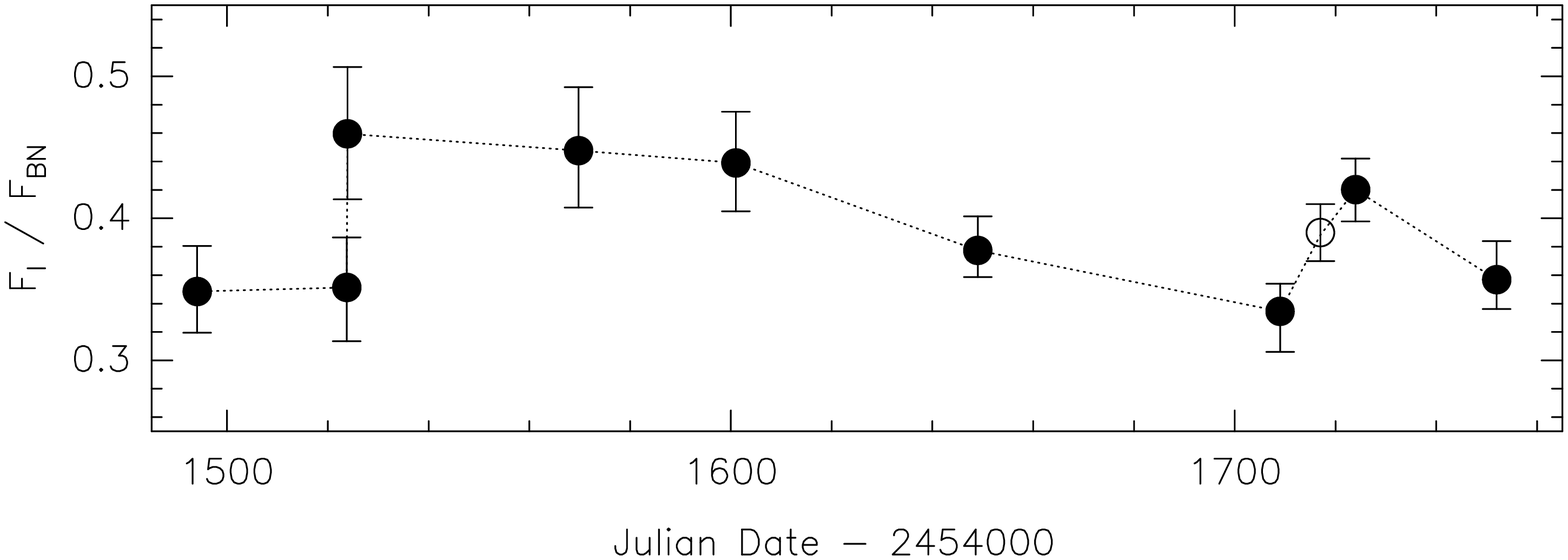}
\caption{Ratio between the flux densities of sources I and BN versus
time at 33.6 GHz (filled circles) and at 30.5 GHz (open circle).}
\label{fig-ratio-BN-I}
\end{figure}

\begin{figure*}
\centering
\includegraphics[angle=0,width=18.5cm]{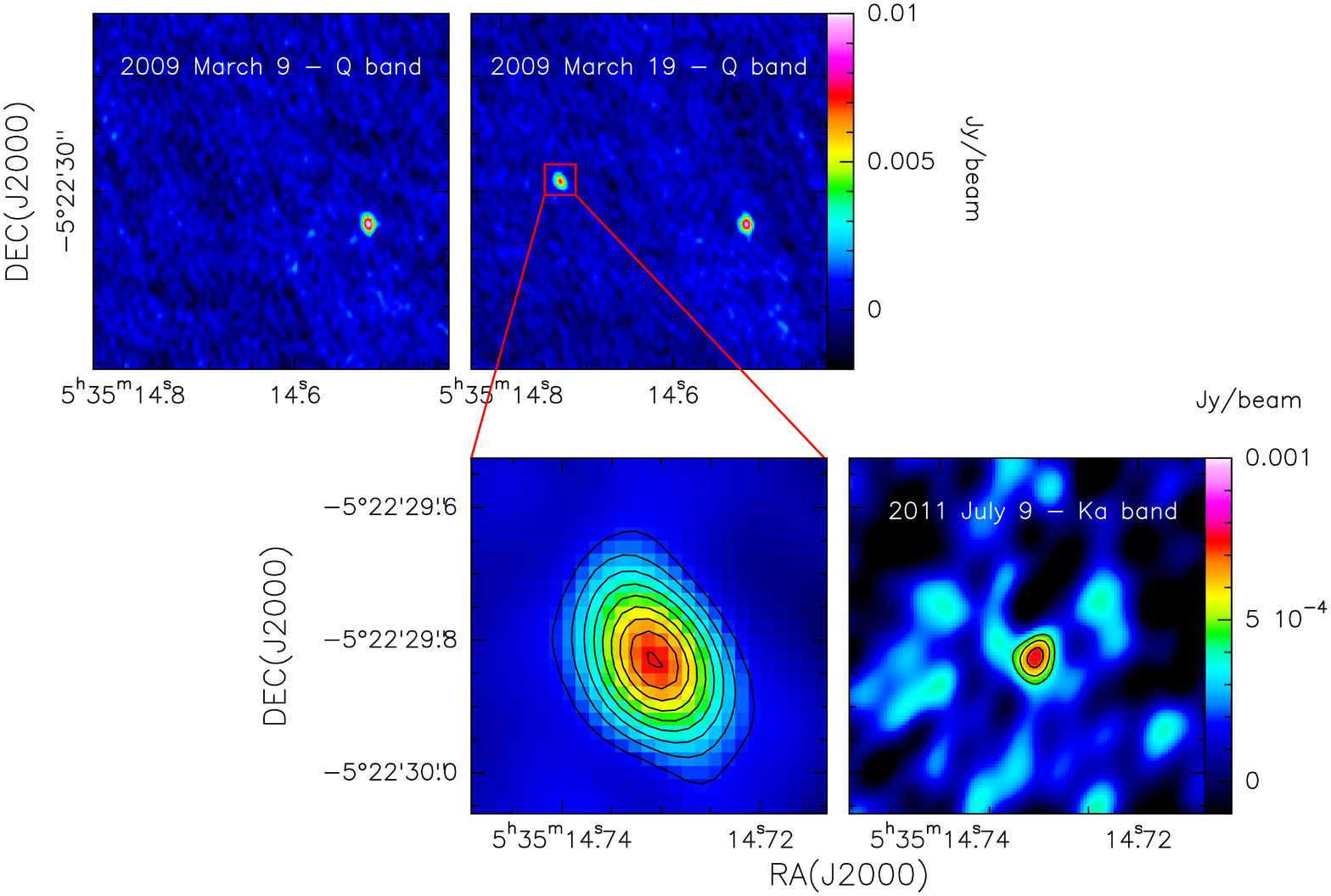} \caption{{\em
Upper panels:} VLA observations of the OHC region at 45.6 GHz from
2009 March 9 (left) and 2009 March 19 (right). The source I is
detected in both images, while the new radio flaring source OHC-E
appeared 3.3$\arcsec$ toward the northeast on 2009 March 19. {\em
Lower left panel:} Zoom-in view of the detection of the source OHC-E
from the 2009 March 19 observation.  The flux density scale is the
same as in upper panels. The contours indicate emission from 3$\sigma$
to 19$\sigma$ in steps of 2$\sigma$.  {\em Lower right panel:}
Second detection of OHC-E in the 2011 July 9 observation. The
contours indicate emission at 3$\sigma$, 4$\sigma$ and 5$\sigma$.}
\label{fig-OHC-E}
\end{figure*}

\section{Observations and data reduction}
\label{observations}

The radio observations of the ONC/OMC region were made with the VLA
in the A, B, C and D configurations at Q-band (3 epochs) and Ka-band
(11 epochs) from 2009 to 2011.  Table \ref{tableobservations}
summarizes the observational details for each of the epochs: array
configuration, project ID, band name, frequency, observing bandwidth,
date, pointing center, observation length (including all calibrations),
synthesized beam, RMS noise at the center of the primary beam,
primary flux density calibrator used, and the flux density of the
complex gain calibrator.  
The separation between the epochs is different: the shortest one is only 3 hours, others are separated by several days, and some other by several months. This will allow to trace radio variability at different timescales.
The first two observations at Q-band (45.6 GHz)
were made in spectral line mode using the old VLA correlator, single
polarization. 
The third Q-band observation (45.6 GHz, AR712) used the standard
dual polarization continuum set-up with the old correlator (100 MHz
total bandwidth).  For the data taken under observing code 10B-175
the new WIDAR correlator was used, with two 128 MHz sub-bands centered on 33.56 GHz placed
contiguously in frequency for all observations except those of 2011
Jun 4, for which the sub-bands were separated by 7 GHz as noted in
Table \ref{tableobservations}.  The observations at Q-band were
reduced using the Astronomical Image Processing System (AIPS)
package. The observations at Ka-band were reduced using the VLA
Calibration
Pipeline\footnote{https://science.nrao.edu/facilities/vla/data-processing/pipeline},
which uses the Common Astronomy Software Applications (CASA)
package\footnote{http://casa.nrao.edu}.

Phase self-calibration was performed where possible.  However, the
tropospheric phase stability of the observations on 2011 Jun 4 was
poor, and the data were of insufficient signal-to-noise ratio to
enable phase self-calibration on short enough timescales to correct
for the resulting decorrelation.  While the typical uncertainty in
the absolute flux density scale is 10\% at these frequencies, the
uncertainty in the absolute flux density scale for the 2011 Jun 4
data is increased to 20\%.  Images were made using CASA, applying
an inner uv-cutoff ($>$50 k$\lambda$) to filter the extended emission from the foreground
HII region ionized by the Trapezium cluster, and natural weighting
to maximize sensitivity.  The images are corrected for the response
of the primary beam before being used for photometry.
The $FITS$ files of the reduced images are available in the electronic version of the paper.

\section{Results: radio stellar population detected at high centimeter frequencies}
\label{results}

The large fields of view (FoVs) covered by the images made here,
especially in the VLA's B and A configurations, requires a rigorous criterion
for the detection of sources.  Assuming a gaussian noise, and approximating the number of potential sources as the ratio between the area of the FoV and the solid angle of the beam ($\pi\theta_{\rm minor}\theta_{\rm major}/4Ln2$), we would expect $<$ 0.3 sources above 5 times the local RMS noise (5$\sigma$) in the worst case. Therefore, we only report detections with flux densities $>$ 5$\sigma$.

We detected a total of 19 sources (Table \ref{table-our-sources}),
18 of which have been previously detected by the radio monitorings
at lower frequencies (\citealt{felli93} at 5 and 15 GHz and
\citealt{zapata04a} at 8.3 GHz). Our observations have revealed the
presence of a new radio source, hereafter OHC-E, detected in the
OHC region in two different epochs. 
In Fig.\ \ref{fig-our-sources}
we show the positions of all the detected sources, overplotted on
the R-band Advanced Camera for Surveys/Wide Field Channel (ACS/WFC)
Hubble Space Telescope (HST) image and on the infrared K-band image
from the 2 Micron All Sky Survey (2MASS). The sources are mainly
concentrated in the OHC region and the Trapezium Cluster, which harbors the two highest stellar densities within our FoV (\citealt{rivilla13a}).

To measure the flux densities of the sources, we use the AIPS task
JMFIT\footnote{JMFIT fits a 2 dimensional gaussian models to the sources by least-squares.}.  We also add an absolute uncertainty of 10$\%$ in quadrature
(20$\%$ for the 2011 Jun 4 observations due to the poorer phase
stability of those data; see Section \ref{observations}).  Table
\ref{table-fluxes} summarizes the results. Only the sources BN and
source I are detected in all epochs, showing nearly constant flux
densities, and verifying the reproduceability of the flux density
scale.  Source C is consistent with a constant flux density, although
is not detected in all epochs.
The other sources exhibit clear flux density variation between
epochs.  In the case of non-detections, we quote 3$\sigma$ upper
limits.

\begin{table}
\caption{Positions of the radio sources detected in our Ka-band and
Q-band monitoring.}
\label{table-our-sources}
\tabcolsep 6.pt
\centering
\begin{tabular}{c c c}     
\hline\hline
Source & $RA_{J2000}$ & $DEC_{J2000}$ \\
& 5 35 & -5  \\
\hline\hline
BN     & 14.11 & 22 22.69  \\
I      & 14.51 & 22 30.58  \\
OHC-E  & 14.73 & 22 29.83  \\
D      & 14.90 & 22 25.38  \\
n$^{a}$& 14.35 & 22 32.89  \\
H      & 14.50 & 22 38.76  \\
A      & 11.80 & 21 49.29  \\
C      & 14.16 & 23 01.04  \\
F      & 18.37 & 22 37.43  \\
G      & 17.95 & 22 45.42  \\
E      & 16.96 & 22 48.78  \\
15     & 16.07 & 23 07.12  \\
6      & 16.75 & 23 16.44  \\
7      & 16.28 & 23 16.58  \\
25     & 15.77 & 23 09.86  \\
12     & 15.82 & 23 14.00  \\
11     & 15.84 & 23 22.40  \\
16     & 16.33 & 23 22.54  \\
5      & 16.85 & 23 26.31  \\
\hline
\end{tabular}
\begin{list}{}{}
\item[$^{\mathrm{a}}$]{This source was called $L$ by \citet{garay87},
but its more common name is n (\citealt{menten95}).}
\end{list}
\end{table}

\subsection{Long-term variability: month timescales}
\label{variability-months}

In this section we study the behavior of the radio emission throughout the full monitoring. In Figs. \ref{fig-Q-variability} and
\ref{fig-Ka-variability} we show the measured integrated flux
densities of all sources during the different epochs of our monitoring
for the Q-band and Ka-band, respectively.  We note that in the
figures the observation on JD 2455188 (2009 Dec 22) at Q-band
corresponds to a slightly different frequency (43.3 GHz) from the
first two epochs (45.6 GHz).  Also, the flux density from the
observation on JD 2455717 (2011 June 4) at Ka-band (shown in Fig.\
\ref{fig-Ka-variability}) corresponds to 30.5 GHz, and not to 33.6
GHz as for the others.

With the aim of quantify the radio variability we study two parameters: i) the standard deviation $\Delta F$
of the flux densities from the average flux $F_{\rm av}$, which measure the absolute variability; and ii) $\beta$, which is defined as $\beta=\Delta
F/F_{\rm av}$, following \citet{felli93}, which measures the relative variability.
 Given that some sources remain undetected in some epochs, and hence have flux densities below the sensitivity limit, we consider in the calculation of $\Delta F$ (and hence of $\beta$) the positive detections as well as the lowest upper limit. In Table \ref{table-big} we show the values of $F_{\rm av}$, $\Delta F$ and $\beta$ for the sources detected.

The sources BN and I are associated with massive stars
(\citealt{reid07}, \citealt{goddi11}), and are expected to emit
mainly thermal (and constant) emission arising from ionized gas
surrounding the central object. Our monitoring at Q-band and Ka-band
(Figs. \ref{fig-Q-variability} and \ref{fig-Ka-variability}) shows
indeed that the flux density of BN is nearly constant, with a very
low month-timescale radio variability parameter at 33.6 GHz of
$\beta$=0.06.

In the case of source I the radio variability is higher, $\beta$=0.13.
\citet{zapata04a} also reported some variability towards this
source, with flux density variations of a factor $\sim$2. Furthermore,
\citet{plambeck13}, using observations separated by 15 yr also found evidence of a gradual flux density increase from source I with respect the more steady flux
density of BN (see their Fig. 5).  From our Ka-band monitoring,
we have studied the month-timescale evolution of the ratio between
the flux densities of sources I and BN\@. Fig.\ \ref{fig-ratio-BN-I}
shows that this ratio exhibits variations larger than the statistical
uncertainties (we do not include the uncertainty in the absolute
flux density scale in calculating the error in these ratios),
suggesting that real variability is present.  
 The origin of this confirmed long-term variability toward source I could be due to ionization of infalling accretion flows onto the massive star (\citealt{galvan-madrid11,depree14}). 

Source C is only detected in the epochs for which the VLA was in its most compact configurations. This suggests that the emission from source C is extended, and that the higher resolution observations have filtered it out. To derive a proper sensitivity level for extended emission we have smoothed the B-configuration images at Q-band to the resolution of the D-configuration image
(1.9$\arcsec\times$1.4$\arcsec$). For the Ka-band monitoring, we have
smoothed the higher resolution images (from 2011 Feb 8 to 2011 Jul
09) to a C-configuration resolution of 0.8$\arcsec\times$0.8$\arcsec$.
We have inspected the smoothed images, and measured the flux density
(or 3$\sigma$ upper limit) at the location of source C. The
resulting light curves (Figs. \ref{fig-Q-variability} and
\ref{fig-Ka-variability}) are in agreement with nearly constant emission during the monitoring.
This radio source is associated with one of the proplyds\footnote{{\it Proplyds} are objects for which circumstellar material is being ionized by the ultraviolet radiation from massive stars.} revealed by the HST (see Fig.\ \ref{fig-2}). Using the flux densities detected at 33.6
GHz and 43.3 GHz, we obtain a spectral index ($F\sim\nu^{\alpha}$)
of $\alpha\sim$3, consistent with optically-thin dust emission from
the proplyd. 

The other sources detected show clear variations at Ka-band between
epochs ($\beta>$0.29). {Many of them are detected only in some epochs, remaining below the sensitivity limit at other epochs}. There is no clear trend or pattern in the
variability, which appears to be stochastic for many of the sources. 
This highly variable emission suggests non-thermal processes. Indeed, \citet{menten07} detected 4 of these sources (A, F, G and 12) with Very Long Baseline Array (VLBA) observations, confirming their compactness and hence the non-thermal nature of the emission.

Using epochs separated by $\sim$1 month, it is not possible to determine whether flux density variation is smooth during this period, or whether it happens in shorter timescales. Observations with shorter separations are needed. We address this issue in Section \ref{short-variability}.

\begin{table*}
\caption{Multi-epoch radio continuum flux densities$^{a}$. The
values in parenthesis correspond to 3$\sigma$ upper limits.}
\label{table-fluxes}
\hspace*{-0.75cm}
\tabcolsep 3.pt
\centering
\begin{scriptsize}
\begin{tabular}{c| c c c| c c c c c c c c c c c}
        & \multicolumn{13}{c}{Epoch: JD - 2454000}  \\
Source  & 900 & 910 &   1188      & 1494  & 1523.76 & 1523.89 & 1569 & 1601 & 1649 & 1709 & \multicolumn{2}{c}{1717} & 1724 & 1752 \\ \cline{2-15}
        & \multicolumn{3}{c|}{Q-band (GHz)} & \multicolumn{11}{c}{Ka-band (GHz)}  \\ 
        & 45.6  & 45.6 & 43.3 & 33.6  & 33.6 & 33.6  & 33.6 & 33.6 & 33.6 & 33.6 & 37.5 & 30.5 & 33.6  & 33.6 \\ \hline
BN  & 31.2$\pm$3.3 & 30.1$\pm$3.1   & 27.9$\pm$3.0 & 26.4$\pm$2.7 & 25.9$\pm$2.7 & 27.2$\pm$2.8 & 23.9$\pm$2.5 & 24.6$\pm$2.5 & 25.7$\pm$2.6 & 27.8$\pm$2.8 & 20.5$\pm$4.1 & 18.8$\pm$3.8 & 25.7$\pm$2.6 & 22.7$\pm$2.3 \\
I &  14.1$\pm$1.7  &  12.5$\pm$1.5  &  13.7$\pm$1.7   & 9.2$\pm$1.1 &   9.1$\pm$1.2  &  12.5$\pm$1.5   &   10.7$\pm$1.3  &   10.8$\pm$1.2   &  9.7$\pm$1.1   &   9.3$\pm$1.0  &   9.1$\pm$1.9  &  7.3$\pm$1.5   &  10.8$\pm$1.2   &    8.1$\pm$0.9 \\
OHC-E & (1.4) & 7.9$\pm$1.1 & (1.6) & (1.1) & (1.3) & (1.2)  & (1.3)  &  (1.0) &  (0.7) & (0.7)  & (0.4)  &  (0.3) & (0.5)  &  0.8$\pm$0.3 \\
D & (1.4) & (1.1)  & (1.7) & (1.1) & (1.2)  & (1.3)  & 4.6$\pm$0.8 &  (1.0) &  (0.7) &  (0.7) &  (0.4) & (0.3)  &  (0.5) &   (0.4) \\
n & (1.4) & (1.1) & (1.6) & 4.0$\pm$0.8 & (1.2) & 3.5$\pm$0.9 & 3.5$\pm$0.9 & (1.0)  & (0.7) & (0.7) & (0.4) & (0.3) & (0.5) & (0.4) \\
H & (1.4) & (1.2) & (1.7) & (1.1) & (1.2) & (1.4) & (1.2) & (1.0) & 1.4$\pm$ 0.5 & (0.7) &  (0.4) & (0.3) & 0.9$\pm$0.3 & 0.9$\pm$0.3 \\
A & - & - & - & (5.1) & 9.6$\pm$3.8 & (6.2) & (5.6) & 7.7$\pm$3.6 & 9.9$\pm$2.1 & (3.2) & - &  9.9$\pm$2.4 & (2.2) & (2.0) \\
C & (35) & (33) & 27.8$\pm$3.7 & 12.7$\pm$1.7 & 11.9$\pm$1.9 &
11.3$\pm$1.9 & 12.0$\pm$1.8 & 19.5$\pm$7.2 & 20.4$\pm$7.4 & 15.8$\pm$4.0 & (20.8) & (16.0) & (18.0) & (17.4) \\
F & - & - & - & (5.4) & 19.5$\pm$4.1 & 57.9$\pm$7.2 & 34.6$\pm$5.5 & 40.1$\pm$5.4 &  6.2$\pm$2.6 & 8.6$\pm$2.5 & - & 5.5$\pm$1.5 & 20.0$\pm$3.6 & 9.9$\pm$2.6 \\
G  & -  & - & - & (4.1) & (4.6) & (5.2) & (4.5) & (3.7) & (2.5) & (2.5) & (2.2) & 5.7$\pm$1.4 & 15.5$\pm$2.7 & (1.6) \\
E & (7.1) & (5.3) & (6.7) & (2.3) & (2.5) & (2.8) & 7.1$\pm$2.4 & 11.8$\pm$3.1 & (1.4) & (1.4) & (1.0)  & (0.5) & (1.0) & (0.9) \\
15 & (8.5) & (7.1) & (8.0) & (2.6) & 5.9$\pm$1.8 & 5.6$\pm$2.6  & (2.8) & (2.3) & 3.0$\pm$1.3 & (1.6) & (1.2) & (0.6) & (1.1) & (1.0) \\
6 & - & - & - & 20.9$\pm$3.9 & 24.5$\pm$4.6 & 21.7$\pm$4.1 & 33.2$\pm$4.9 & 34.1$\pm$5.6 & 7.0$\pm$2.4 & (3.2) & - & 16.0$\pm$ 3.6 & 37.9$\pm$6.2 & (2.0) \\
7 & - & - & - & (4.1) & 8.0$\pm$3.1 & 23.8$\pm$4.3 & 31.4$\pm$4.7 & 20.7$\pm$3.8 & 7.1$\pm$2.2 & (2.6) & 11.9$\pm$3.5 & 21.5$\pm$4.7 & 14.8$\pm$ 3.6 & (1.6) \\
25 & (8.9) & (7.3) & (8.2) & (2.6) & (2.9) & (3.1) & 8.8$\pm$2.0 & (2.3) & 6.3$\pm$1.3 & (1.6) & (1.2) & (0.6) & 5.4$\pm$1.2 & 19.2$\pm$2.1 \\
12 & - & - & - & (3.1) & (3.5) & 8.5$\pm$2.9 & 11.2$\pm$2.6 & 20.4$\pm$2.7 & (1.9) & 38.3$\pm$4.2 &  18.4$\pm$4.3 & 13.4$\pm$2.7 & 69.9 $\pm$7.1 & 2.8$\pm$0.9 \\
11 & - & - & - & (4.8) & (5.3) & (5.9) & 22.8$\pm$4.7 & (4.2) & 12.6$\pm$3.7 & (2.9) & - & 27.9$\pm$ 6.5 & 11.9$\pm$2.5 & (1.8) \\
16 &  - & - & - & (5.9) & (6.5) & (7.2) & (6.4) & (5.2) & (3.6) & (3.6) & - & (1.1) &  4.7$\pm$1.6 & (2.3) \\
5$^{b}$ & - & - & - & - & - & - & - & - & - & - & - & 30.9$\pm$7.5 & - & - \\
\hline
\end{tabular}
\begin{list}{}{}
\item[$^a$]{Integrated flux densities are presented in mJy.}
\item[$^b$]{Source 5 is outside the FoV for all but the 2011 Jun 4 at 30.5 GHz, whose FoV is larger.}
\end{list}
\end{scriptsize}
\end{table*}

\subsection{Short-term variability}
\label{short-variability}

\subsubsection{Day timescales}
\label{variability-days}

The first two Q-band observations are only separated by 10 days
(2009 March 9 and 19). The main result obtained by comparing these
observations is the discovery of a new radio source (hereafter OHC-E). Fig.\ \ref{fig-OHC-E} shows the
comparison of the two images. The source is detected at 3.3$\arcsec$
northeast of source I. The flux density variation between the two
epochs is $>$5.6 (using the 3$\sigma$ upper limit as the flux density
in the 2009 March 9 image).  Source OHC-E shows a constant flux
density during the 4 hours of the 2009 March 19 observation,
indicating that we might have detected a fraction of a powerful flare event.
This is consistent with the duration reported on other radio flares observed in Orion (source A and ORBS), which have timescales of hours to days.

\subsubsection{Hour timescales}
\label{variability-hours}

\begin{figure}
\centering
\includegraphics[angle=0,width=8.8cm]{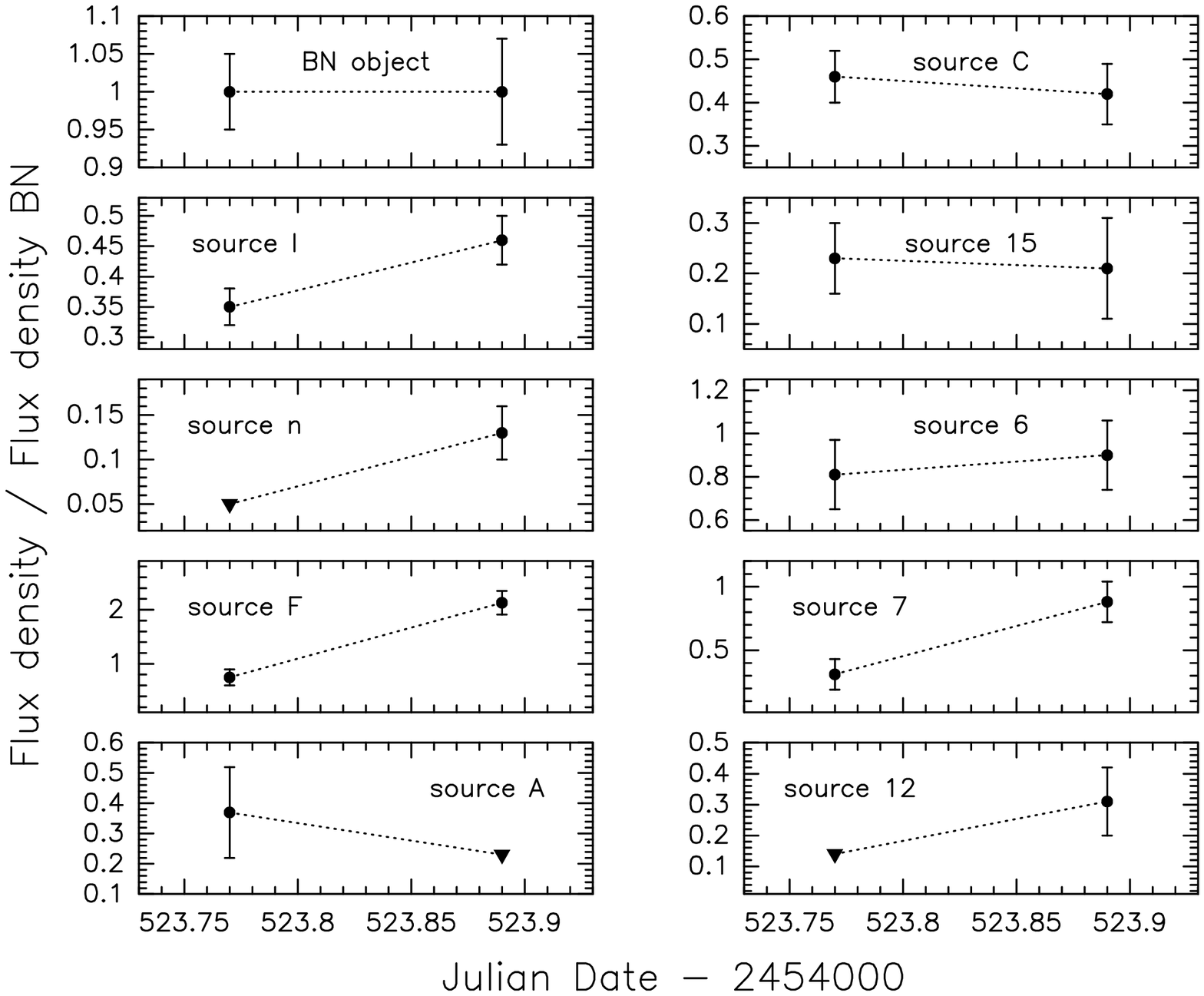}           
\caption{Flux density variation (normalized by the flux density of
BN) on hour timescales for sources detected in the two runs observed
on 2010 November 23.}
\label{fig-hours}
\end{figure}

Two of the Ka-band observations are separated by only $\sim$3 hours (2010 Nov 23), enabling the study of flux density variations on even shorter timescales. Fig. \ref{fig-hours} shows the flux density variation of the 10 sources detected in these two epochs, normalized by the flux density of BN.  In Table \ref{table-hours} we show the ratio between the flux densities of the two epochs ($F_{\rm
523.89}/F_{\rm 523.76}$, where the subscript is the JD date minus 2454000, and the variability parameter defined between these two epochs is $\beta$. The sources BN, C, 15, 6 are nearly constant within the flux density uncertainties, with variations within a factor 0.95$-$1.17 and $\beta<$ 0.11. The latter 2 sources, however, exhibit clear variations at other epochs (see Fig. \ref{fig-Ka-variability}).

The sources n, 7, F and 12 show clear radio flares (see Fig. \ref{fig-hours}), with flux density increases by factors $\geq$2.4 and variability parameters $\beta\geq$0.59 (Table \ref{table-hours}). The source A, a well known radio flaring source (\citealt{bower03,zapata04a,gomez08}), suffers a significant decrease of its flux density, being undetectable in the second image.

There is a tentative detection of variability in source I between the two epochs.  The flux densities measured are not within the flux density uncertainty limits (see Fig.\ \ref{fig-hours} and Table
\ref{table-fluxes}), the variability parameter is $\beta$=0.22, and the variation factor is 1.37 (while
in the case of the BN source is only 1.05).  We conclude that this
short-term variation of source I is not due to calibration
uncertainties, and might be real. 
As already commented in Sect. \ref{variability-months}, ionized gas of infalling accretion flows onto this massive star (\citealt{galvan-madrid11,depree14}) might be responsible of the long-term variability observed in source I (Fig. \ref{fig-ratio-BN-I}). However, other mechanism(s) would be needed to explain the variability observed in hours-timescales, like stochastic shocks in the radiative wind of the massive star (\citealt{stelzer05}), or the presence of
an unseen low-mass companion emitting non-thermal gyrosynchrotron radiation.
Indeed, \citet{goddi11} present arguments for source I harboring a
binary system based on the kinematic history of the region. The
interaction of a companion with the wind from a massive primary
could explain the variability observed here.

\begin{figure}
\centering
\includegraphics[angle=0,width=6.0cm]{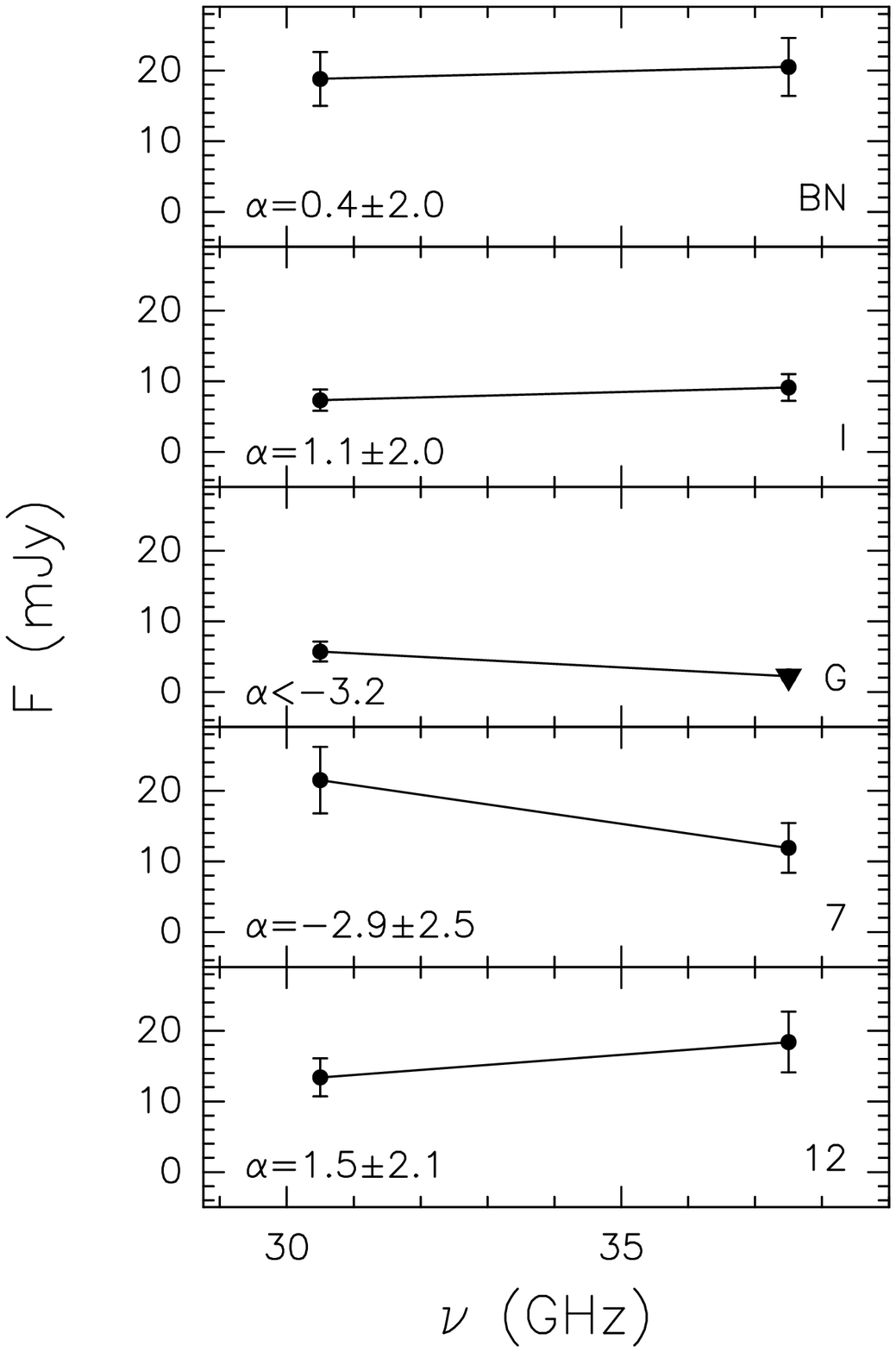}
\caption{Flux densities at 30.5 and 37.5 GHz from the 2011 June 4 observations.}
\label{fig-spectral-indeces}
\end{figure}

\subsection{Spectral indeces between 30.5 and 37.5 GHz}

One of the methods commonly used to unveil the nature of the radio emission is by studying the emission as a function of frequency, $F \propto \nu^{\alpha}$, where $\alpha$ is the spectral index. 

In the 2011 June 4 observations, we have observed simultaneously at two different frequencies, 30.5 and 37.5 GHz. The calibration uncertainty in this epoch is higher than in the other runs (see Section \ref{observations}), which implies large uncertainties in the derivation of $\alpha$. However, it is worth mentioning that source 7 and source G exhibit clear flux decreases with $\alpha<$-0.4, suggesting that the dominant emission mechanism seems to be non-thermal.

\begin{table*}
\caption{ List of radio sources in the ONC/OMC region considered our analysis}
\label{table-big}
\tabcolsep 1.0pt
\centering
\vspace{-0.1cm}
\hspace{-0.7cm}
\begin{tabular}{c c c c c c c c| c c c c c c c c| c c c| c| c}     
\hline\hline
  \multicolumn{8}{c|}{Radio sources} & \multicolumn{8}{c|}{X-ray sources} &  \multicolumn{3}{c|}{Optical$^b$} & IR$^c$ & Mem.$^d$ \\ \cline{1-8} \cline{9-16}
  \multicolumn{2}{c|}{ID} & \multicolumn{3}{c|}{33.6 GHz} & \multicolumn{3}{c|}{8.3 GHz} &  COUP  & \multicolumn{7}{c|}{Properties} & & &  & \\   \cline{1-2} \cline{3-5} \cline{6-8} \cline{10-16}
Source & Z04$^a$ & $F_{\rm av}$ & $\Delta F$ & $\beta$ & $F_{\rm av}$ & $\Delta F$ & $\beta$ & ID  & Med$E$ & $HR_{1}$& log$N_{\rm H}$ & log$P_{\rm KS}$ & BB & X-ray  & log$L_{X}$ & OW94 & H97 & K05 & HC00 & \\
  &  & (mJy) & (mJy) & & (mJy) & (mJy)   &  &   & (keV) & & (cm$^{-2}$) &  & Num & flare?$^e$ & (erg s$^{-1}$) & & & \tiny({propl.)} &  & \\
\hline\hline
 I & 19 & 10.0 & 1.3  & 0.13 & 0.64  & 0.23  & 0.36  &-$^f$    & -    & -    & -               & -     & -     & - & -  &  - & - & - & - & OMC \\
  C$^{g}$     & 14 & 14.8&3.7   & 0.25  & 4.97 & 1.61  & 0.32    & -     &  -   & -    &    -     &   -   & -     & - &   -  & -  & y & y & y & ONC \\
  16     & 53 &  4.7&1.7     & 0.36 &  3.14 & 0.43 & 0.14  & - & -  &  -   &  -     &   -   &    -  & - &   -   & y &  - & - & - &  ONC \\
 5     & 61 & - & -                & -  & 15.88 & 0.26  & 0.02 & -     & -    &  -   &  -   &   -   &    -  & - &   -   & y & y & y & y & ONC \\
\hline
 BN & 12 & 25.5 & 1.6  & 0.06 & 3.61 & 0.25  & 0.07 & 599b$^h$& -    & -    & -               & -     & -     & - & -  & -  & - & - & y & OMC \\
OHC-E & -  &  0.8 & 0.2     & 0.27 & - &  -&  -    & 655  & 3.58 & 0.89 & 22.85$\pm$0.01 & -4.00 & 9     & y & 31.37 & - &  - & - & y & OMC \\
 D & 21 &  4.6&3.0     & 0.65 & 1.04 & 0.68 &  0.65  & 662   & 4.50 & 0.98 & 23.22$\pm$0.03  & -4.00 & 13    & y & 31.29 & - &  - & - & - & OMC \\
 n & 17 &  3.7&1.7   & 0.45 & 1.12 & 0.37 &  0.33  & 621   & 3.57 & 0.84 & 22.74$\pm$0.01  & -4.00 & 13    & y & 30.97 & - &  - & - & y & OMC \\
 H & 18 &  1.1&0.3   & 0.28 &  0.66 & 0.50 &  0.75 & 639   & 3.99 & 0.96 & 23.06$\pm$0.04  & -4.00 & 8     & y & 30.90 & - & - & - & - & OMC \\
 A & 6  &  9.1&3.7   & 0.40 & 16.84  & 26.7 &  1.58 & 450   & 3.32 & 0.64 & 22.34$\pm$0.03  & -4.00 & 12    & y & 32.27 & - & - & - & y & ONC \\
 F     & 76 & 24.6&18.1  & 0.74 & 15.04 & 5.10 &  0.34  & 965   & 1.31 &-0.63 & 21.25$\pm$0.10  & -4.00 & 2     & n & 31.89 & -&  y & - & y & ONC \\
 G     & 73 & 15.5&9.8  & 0.63 & 2.48 & 2.09 &  0.84 & 932   & 1.29 &-0.64 & 21.17$\pm$0.09  & -4.00 & 12    & y & 32.18 &-  & y & - & y & ONC \\
 E     & 63 &  9.5&5.5   & 0.58 & 2.07 & 0.47 &  0.23 & 844   & 1.63 &-0.50 & 22.01$\pm$0.14  & -0.96 & 1     & y & 29.22 & - & y & y & y & ONC \\
15     & 49 &  4.8&2.3   & 0.48 & 4.35 & 0.83 &  0.19 & 766   & 1.58 &-0.32 & 21.54$\pm$0.03  & -4.00 & 20    & y & 30.71 & - & y & - & y & ONC \\
 6     & 59 & 25.6&12.8  & 0.50 & 22.23 & 0.33 &  0.01 & 826   & 2.31 & 0.16 & 22.06$\pm$0.04  & -4.00 & 11  & y & 30.35 & - & y & y & y & ONC \\
 7     & 52 & 17.6&10.6   & 0.60 & 10.12 & 0.87  &  0.09 & 787   & 2.43 & 0.29 & 22.40$\pm$0.04  & -4.00 & 5     & y & 30.22 & y & y & y & y & ONC \\
25     & 38 &  9.9&6.6   & 0.67 & 4.96 & 0.61 &  0.12 & 732   & 1.33 &-0.60 & 20.99$\pm$1.99  & -4.00 & 23    & y & 32.35 & - & y & - & y & ONC \\
12     & 41 & 25.2&24.6  & 0.98 & 12.09  & 7.89 & 0.65  & 745   & 1.33 &-0.59 & 20.79$\pm$0.08  & -4.00 & 20    & y & 32.33 & y & y & - & y & ONC \\
11     & 42 & 12.6&8.5   & 0.68 & 10.82  & 0.17 & 0.02  & 746   & 2.23 & 0.19 & 22.10$\pm$0.55  & -3.40 &  2    & n & 28.79 & y & y & y & y & ONC \\
\hline
ORBS   & -  & -   & - & -  & - &   -  & -    & 647  & 5.20 & 1.00 &  23.51$\pm$0.03 & -4.00 &  8    & y & 30.93 & - &  - & - & - & OMC \\
\hline
 -  & 1  & -  & - & -  & 2.84  &  0.12  & 0.04    & 229   & 4.05 & 0.98 & 22.96$\pm$0.19  & -0.05 &  1    & n & 29.28 & - &  - & - & - & EG  \\
 -  & 2  & -  & - & -  & 0.81  &  0.81 & 1.00 & 342   & 1.98 &-0.01 & 22.21$\pm$ 0.01 & -4.00 & 31    & y & 31.32 & - & y & - & y & ONC \\
 -  & 7  & -  & -   & -  & 0.78 &   0.42   & 0.54 & 510   & 4.75 & 1.00 & 23.54$\pm$ 0.06 & -2.59 &  3    & y & 31.26 & - & - & - & - & OMC \\
 -  & 9  & -  & -   & -  & 0.48 &  0.27   & 0.56 & 530   & 5.23 & 0.98 & 23.50$\pm$0.03  & -4.00 &  2    &(y)& 30.75 & - & - & - & - & OMC \\
 -  & 16 & -  & -  & -  &  0.33 &   0.18   & 0.56    & 625   & 4.81 & 0.98 & 23.42$\pm$0.05  & -4.00 &  4    & y & 30.95 &- &  - & - & - & OMC \\
 -  & 31 & -  & - & -  & 0.20 &  0.08   & 0.42    & 699   & 1.51 &-0.50 & 21.61$\pm$0.11  & -4.00 &  4    & y & 29.54 & - & y & y & y & ONC \\
 -  & 33 & -  & -  &-   & 2.90 & 1.79  & 0.62    & 708   & 1.54 &-0.36 & 21.46$\pm$0.05  & -4.00 & 10    & y & 30.19 & - &  y & - & y & ONC \\
 -  & 34 & -  & -  & - & 5.66 & 0.59   & 0.10    & 717   & 1.55 &-0.54 & 20.98$\pm$0.70  & -0.44 &  1    & (y) & 28.53 & y  & y & y & y & ONC \\
 -  & 37 & -  & - & -  &  1.44 & 0.04    & 0.03    & 733   & 2.49 & 0.19 & 22.54$\pm$0.23  & -2.33 &  2    & n & 29.40 & y &  - & y & y & ONC \\
 -  & 43 & -  & -  & -  & 4.05  & 0.43    & 0.11    & 747   & 3.47 & 0.33 & 22.25$\pm$0.35  & -1.06 &  1    & y & 28.88 & y & y & y & y & ONC \\
 -  & 44 & -  & -  & -  & 1.33 &  0.48   & 0.36    & 757   & 1.89 &-0.17 & 22.45$\pm$0.21  & -0.64 &  1    & (y) & 29.01 & y & - & y & y & ONC \\
 -  & 45 & -  & -  & - & 6.03 &   0.58    & 0.10    & 758   & 1.66 &-0.29 & 21.73$\pm$0.02  & -4.00 & 29    & y & 31.05 & y & y & y & y & ONC \\
-   & 46 & -  & -  & -  & 1.70 & 0.84    & 0.49    & 768   & 1.63 &-0.33 & 21.52$\pm$0.04  & -4.00 &  4    & y & 30.27 & y & y & y & y & ONC \\
-   & 54 & -  & - & -  &  1.02   &   0.61   & 0.60    & 800   & 1.93 &-0.14 & 22.36$\pm$0.21  & -0.85 &  2    & y & 28.86 & - & y & y & y & ONC \\
-   & 56 & -  & -  & -  & 0.65 &  0.45    & 0.69    & 807   & 1.37 &-0.60 & 21.43$\pm$0.10  & -4.00 &  7    & y & 29.73 & - & y & y & y & ONC \\
-   & 58 & -  & - & - & 1.58  &   0.16    & 0.10    & 820   & 3.64 & 0.95 & 22.86$\pm$0.13  & -4.00 &  3    & y & 30.07 & y & - & y & y & ONC \\
-   & 60 & -  & - & -  &  3.12   & 0.41    & 0.13    & 827   & 1.63 &-0.41 & 22.24$\pm$0.03  & -4.00 &  2    & n & 30.46 & y &  y & y & y & ONC \\
-   & 64 & -  & - &-  &  7.65 &  0.45    & 0.06    & 847   & 1.35 &-0.64 & 21.62$\pm$0.35  & -2.00 &  1    & (y) & 29.22 & y & y & y & y & ONC \\
-   & 65 & -  & - & -  & 3.82  &  0.18    & 0.05    & 855   & 1.96 &-0.01 & 21.69$\pm$0.09  & -4.00 &  8    & y & 31.79 & y & y & y & y & ONC \\
-   & 69 & -  & -  & -  &  2.03 &  0.38   & 0.19    & 876   & 3.22 & 0.70 & 22.69$\pm$0.06  & -2.60 &  3    & y & 29.81 & - & y & y & y & ONC \\
-   & 71 & -  & -  & -  &   4.05 &   0.39   & 0.10    & 900   & 4.04 & 0.46 & 23.54$\pm$0.46  & -4.00 &  2    & y & 30.44 & - & y & y & y & ONC \\
-   & 75 & -   & -  & -  &  0.43 & 0.14    & 0.33 & 955   & 1.70 &-0.31 & 21.46$\pm$0.19  & -0.36 &  2    & y & 28.78 & - & y & y & y & ONC \\
-   & 77 & -  & -  &-   &  2.81 & 1.71   & 0.61    & 1130  & 1.23 &-0.71 & 21.28$\pm$0.14  & -4.00 &  7    & (y) & 31.67 & - & y & - & y & ONC \\
\hline
 -  & 3  & -  & -  &-   & 0.16 & -   & -       & 378   & 1.23 &-0.66 & 21.11$\pm$0.07  & -4.00 &  5    & y & 30.20 & - & y & - & y & ONC \\
 -  & 4  & -  & - &-  & 0.30 &  -    & -       & 394   & 1.20 &-0.72 & 20.00$\pm$1.65  & -4.00 &  6    & y & 31.32 & - & y & - & y & ONC \\
 -  & 5  & -  & - &-   &0.24  &   -  & -       & 443   & 2.42 & 0.35 & 21.76$\pm$0.39  & -0.51 &  1    & n & 28.25 & - & y & y & y & ONC \\
 -  & 8  & -  & -  &-   & 0.11 &   -  & -       & 524   & 3.31 & 0.52 & 22.37$\pm$0.31  & -4.00 &  2    & y & 28.51 & - & y & y & y & ONC \\
 -  & 13 & -  & - &-   & 0.32 &   -  & -       & 607   & 4.37 & 0.22 &       -         & -0.05 &  1    & n &   -   & - &  - & - & - & OMC$^{h}$ \\ 
 -  & 20 & -  & -  &-  & 0.29  &    -  & -       & 658   & 2.37 & 0.23 & 22.23$\pm$2.62  & -4.00 &  9    & y & 30.41 & y & y & y & y & ONC \\
 -  & 23 & -  & -  & -  & 0.32 &   -  & -       & 671   & 1.50 &-0.52 & 21.83$\pm$0.14  & -2.20 &  3    & y & 29.21 & y &  y & - & y & ONC \\
 -  & 28 & -  & - & -  & 0.34 &   -  & -       & 690   & 3.72 & 0.82 & 22.70$\pm$0.15  & -4.00 &  2    & y & 29.31 & y & - &  y & - & ONC \\
 -  & 29 & -  & - & -  & 0.24 &   -  & -       & 689   & 1.32 &-0.61 & 20.85$\pm$0.15  & -4.00 & 22    & y & 31.82 & - & y & - & y & ONC \\
 -  & 51 & -  & -  & -  & 0.30  &   -  & -       & 780   & 3.60 & 0.92 & 22.83$\pm$0.03  & -4.00 &  6    & y & 30.95 & - & - & - & y & ONC$^i$ \\
 -  & 66 & -   & - & -  &0.32  &   -  & -        & 856   & 1.47 &-0.47 & 21.65$\pm$0.02  & -4.00 &  9    & y & 30.34 & y & y & y & y & ONC \\
 -  & 70 & -   & -& -  & 0.32  &   -  &  -       & 885   & 1.44 &-0.50 & 21.61$\pm$0.02  & -4.00 &  7    & y & 30.33 & - & y & - & y & ONC \\
\hline \hline
\end{tabular}
\begin{list}{}{}
\begin{tiny}
\item[$^a$]{Source ID from \citet{zapata04a}.}
\item[$^b$]{Optical association from the work of H97
(\citealt{hillenbrand97}) and K05 (\citealt{kastner05}).}
\item[$^c$]{Infrared association from the work of HC00
(\citealt{hillenbrand00}).}
\item[$^d$]{Source membership: OMC=Orion Molecular Cloud, ONC=Orion Nebula Cluster, EG=extragalactic.}
\item[$^e$]{From visual inspection of the X-ray light curves
published by \citet{getman05a}. The parentheses denote that the
detection is uncertain.}
\item[$^f$]{Although X-ray emission from wind shocks would be expected from the massive star associated with source I, the
non-detection by Chandra is likely due to the presence of a nearly edge-on disk (\citealt{matthews10}) that absorbs the X-ray emission.}
\item[$^g$]{As discussed in Section \ref{variability-months}, the source C is filtered out by the more extended VLA configuration observations. To compute the average flux, the variability parameter and the number of detections at 33.6 GHz, we have considered the images after smoothing to a C-configuration resolution of 0.8$\arcsec$.}
\item[$^h$]{The massive BN object has an X-ray counterpart, COUP
599b, much fainter than a low-mass companion located 0.9$\arcsec$
from BN, COUP 599a (\citealt{grosso05}).}
\item[$^h$]{The radio source 13 identified by \citet{zapata04a} does not have optical or IR counterpart, but it has an X-ray counterpart (COUP 607) with low number of counts, which prevent the derivation of $N_{\rm H}$. \citet{rivilla13b} showed that this sources is likely driving a molecular outflow in the OMC1-S region, and hence we classified it as a star embedded in the OMC.}
\item[$^i$]{COUP 780 does not have optical counterpart and exhibits high extinction ($\log N_{\rm H}$=22.83 cm$^{-2}$). Given its
location, we propose that it is a particularly embedded ONC star that belongs to the Trapezium cluster surrounding the massive star
$\theta^{1}$ {\it  Ori C} (\citealt{rivilla13a}), rather than an OMC member.}
\end{tiny}
\end{list}
\end{table*}

\begin{table}
\caption{Radio variability on hour timescales between the two observations on 2010 Nov 23.}
\label{table-hours}
\tabcolsep 5.pt
\centering
\begin{tabular}{c c c}
\hline\hline
Source & $F_{\rm 523.89}/F_{\rm 523.76}$ & $\beta$ \\
\hline\hline
BN & 1.05 &  0.03 \\
I & 1.4 &   0.22 \\
n & 2.9 &  0.69 \\
F & 3.0 &   0.70\\
7 & 3.0 &   0.70 \\
A & 0.7 &  0.30 \\
C & 0.95 &   0.04\\
15 & 0.95 &   0.04\\
6 & 1.17 &  0.11\\
12 & 2.4 &   0.59\\
\hline\hline
\end{tabular}
\end{table}

\section{Comparison with lower radio frequency and connection to X-ray emission}
\label{full-sample}

With the aim of better understanding the nature of the radio emission from young stars of the ONC/OMC region, in this section we compare the properties of the 33.6 GHz emission from our monitoring with those of the 8.3 GHz emission from \citet{zapata04a}. These authors analyzed VLA observations in 4 different epochs within a larger FoV and cadence $\sim$1 year.
We have cross-correlated our sample of 18 radio sources emitting at 33.6 GHz
with their catalog of radio sources.

Furthermore, to explore the link between the radio and X-ray emission in a statistically significant sample, we study the full sample of sources emitting at both wavelengths in Orion. We have cross-correlated the full sample radio sources
with the catalog of X-ray stars provided by the very deep Chandra Orion Ultra Deep Project (COUP, \citealt{getman05a}). 

\begin{figure}
\centering
\includegraphics[angle=90,width=9.5cm]{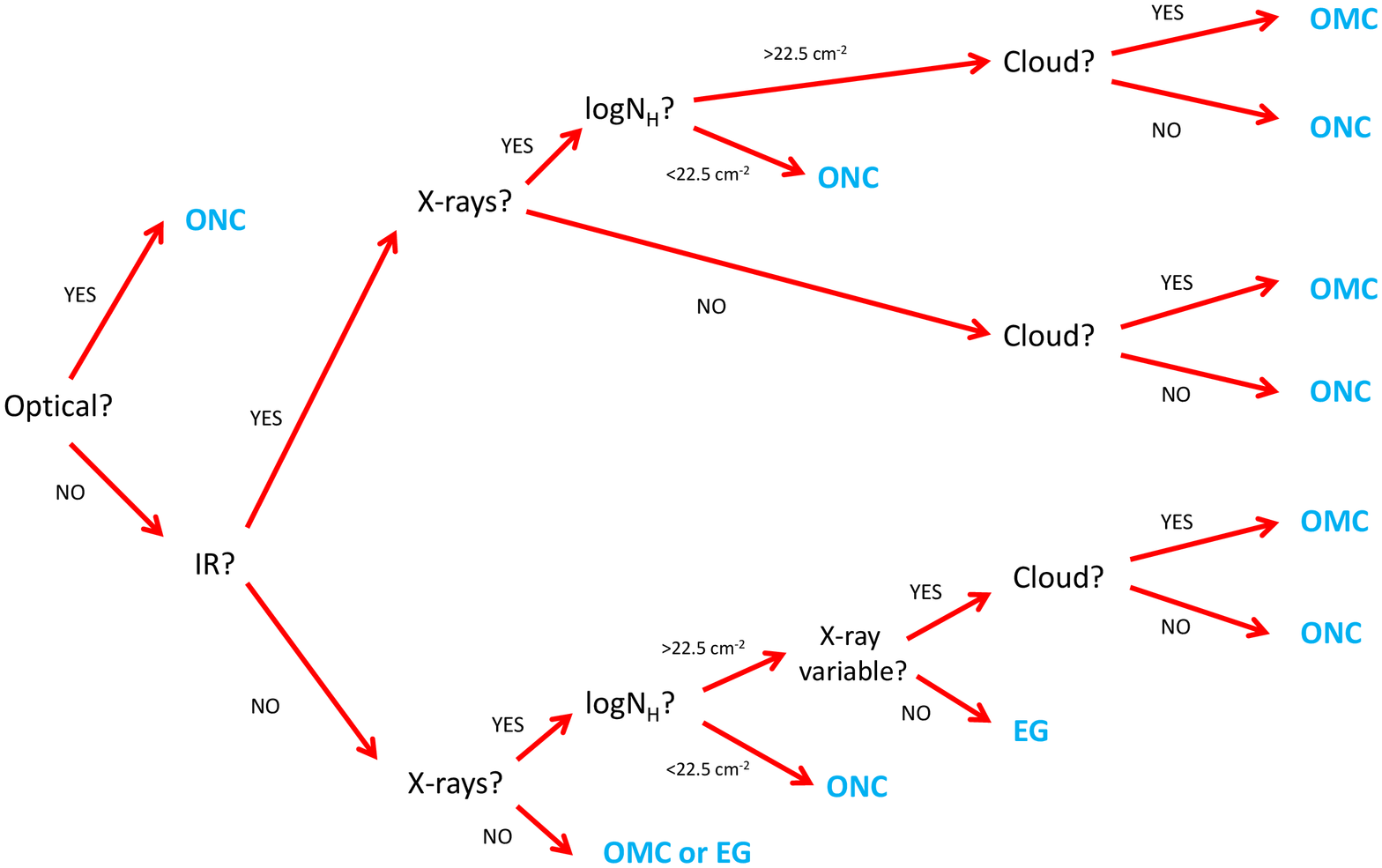}
\caption{Scheme of the method used to classify the sources as ONC or OMC members, or EG sources. We use cross-correlation with optical, IR and X-ray stellar catalogs, and comparison with the spatial distribution of molecular material (see text).}
\label{fig-scheme-membership}
\end{figure}

\begin{figure}
\centering
\includegraphics[angle=0,width=8.5cm]{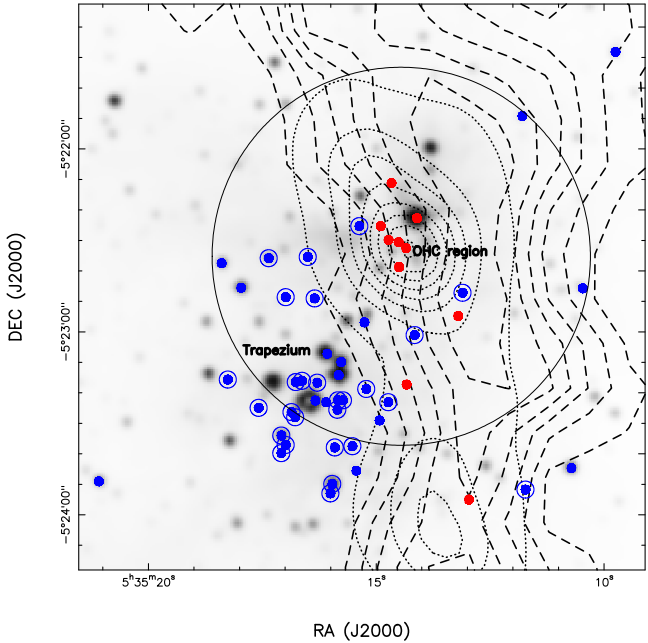}
\caption{Spatial distribution of the full sample of sources used in our analysis. Red and blue dots indicate members of the ONC and OMC, respectively. The ONC sources related with proplyds are denoted by large blue circles. The radio
source classified as EG falls outside the region shown. The greyscale-scale shows the IR K-band image
from the 2 Micron All Sky Survey (2MASS). The dashed contours trace gas from the OMC (CN N=1$-$0 emission from \citealt{rodriguez-franco98}), from 12 K km s$^{.1}$ in steps of 4 K km s$^{.1}$. The dotted contours correspond to emission at 850 $\mu$m from \citet{difrancesco08}. The large circle indicates the primary beam of the 33.6 GHz observations.}
\label{fig-spatial-distribution}
\end{figure}

The ONC/OMC region harbors different populations of radio stars: i) members of the optically visible foreground ONC illuminated by the massive stars of the Trapezium (some of them associated with proplyds); ii) stars still embedded in the background OMC in a earlier evolutionary stages; and iii) extragalactic (EG) sources. Since the emission mechanisms of each group could be different, we distinguish in our study these different populations.

\subsection{Source membership}
\label{membership}

To obtain a rough estimate of the number of expected EG sources we follow \citet{fomalont02}, who estimated the expected number of EG contaminants at 8 GHz.   
In a typical FoV at 8 GHz, the expected EG contamination is $\sim$1 source (\citealt{zapata04a}).
In the smaller FoV of our 33.6 GHz monitoring, we would expect an even lower EG contamination of $\sim$0.01 sources at 8 GHz for the most sensitive of our observations and a detection
criterion of 5$\sigma$. Since these extragalactic sources generally show non-thermal emission of the type $F \propto \nu^{\alpha}$ with $\alpha < 0$ between 1$-$100 GHz (\citealt{condon92}), the contamination at 33.6 GHz is expected to be even lower.  We therefore conclude that the impact of EG contamination in our monitoring is very low. In the case of the \citet{zapata04a} observations at 8.3 GHz, they inferred $\sim$1 EG source.

We classify the radio sources into 4 groups: ONC members without evidence of proplyds (``naked'' ONC), ONC members associated with proplyds, embedded OMC members and EG contaminants.
We cross-check the radio sources with available stellar catalogs in the optical (\citealt{odell94,hillenbrand97,kastner05}), infrared (\citealt{hillenbrand00}) and X-rays (COUP, \citealt{getman05a}). Additionally, we compare the position of the sources with the spatial distribution of the OMC, traced by the CN N=1$-$0 emission from \citet{rodriguez-franco98}. A scheme explaining the method used to classify the sources is presented in Fig. \ref{fig-scheme-membership}. It is based on 6 steps:

\begin{figure*}
\centering
\includegraphics[angle=0,width=5.55cm]{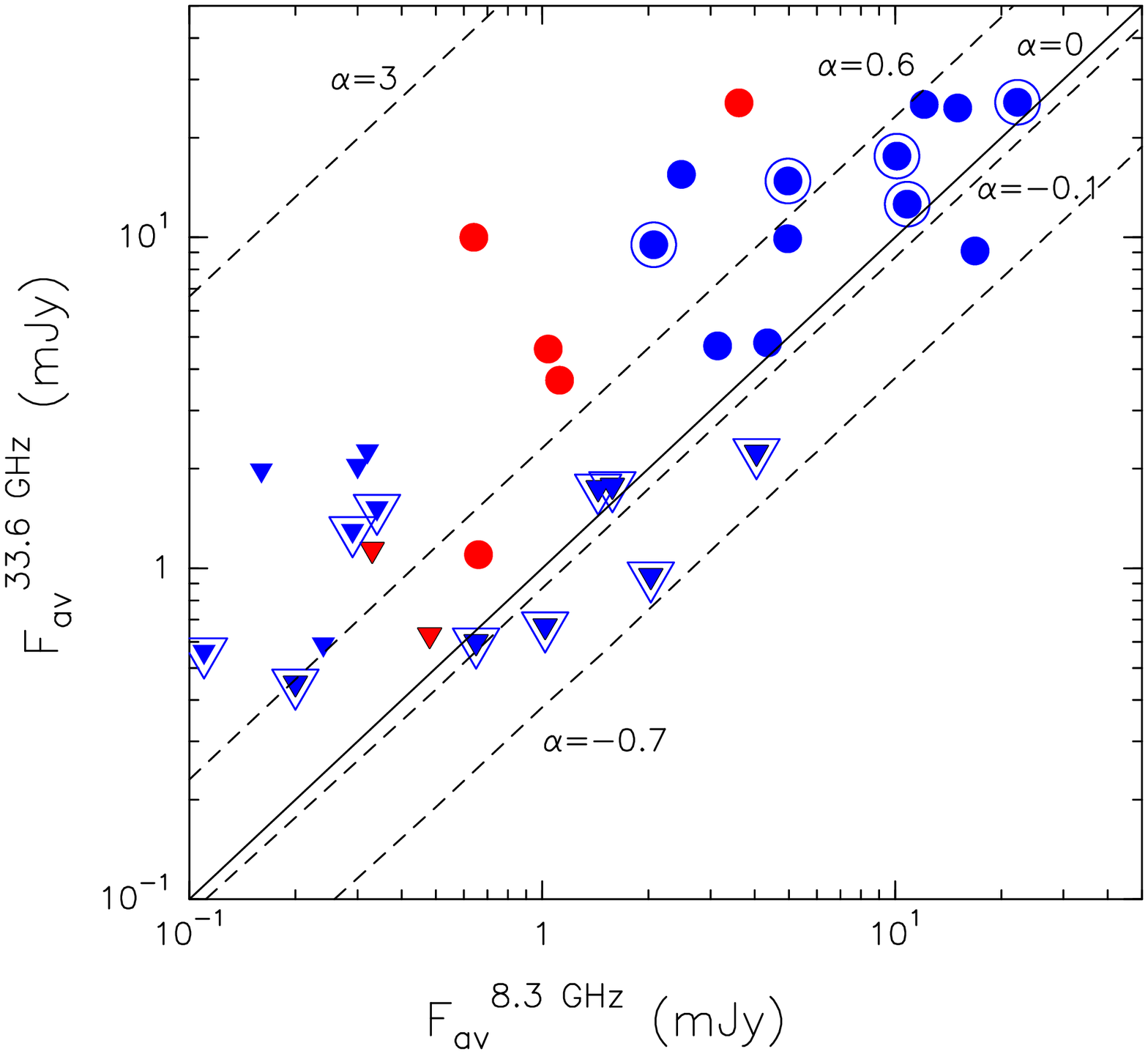}
\hspace{0.4cm}
\includegraphics[angle=0,width=5.55cm]{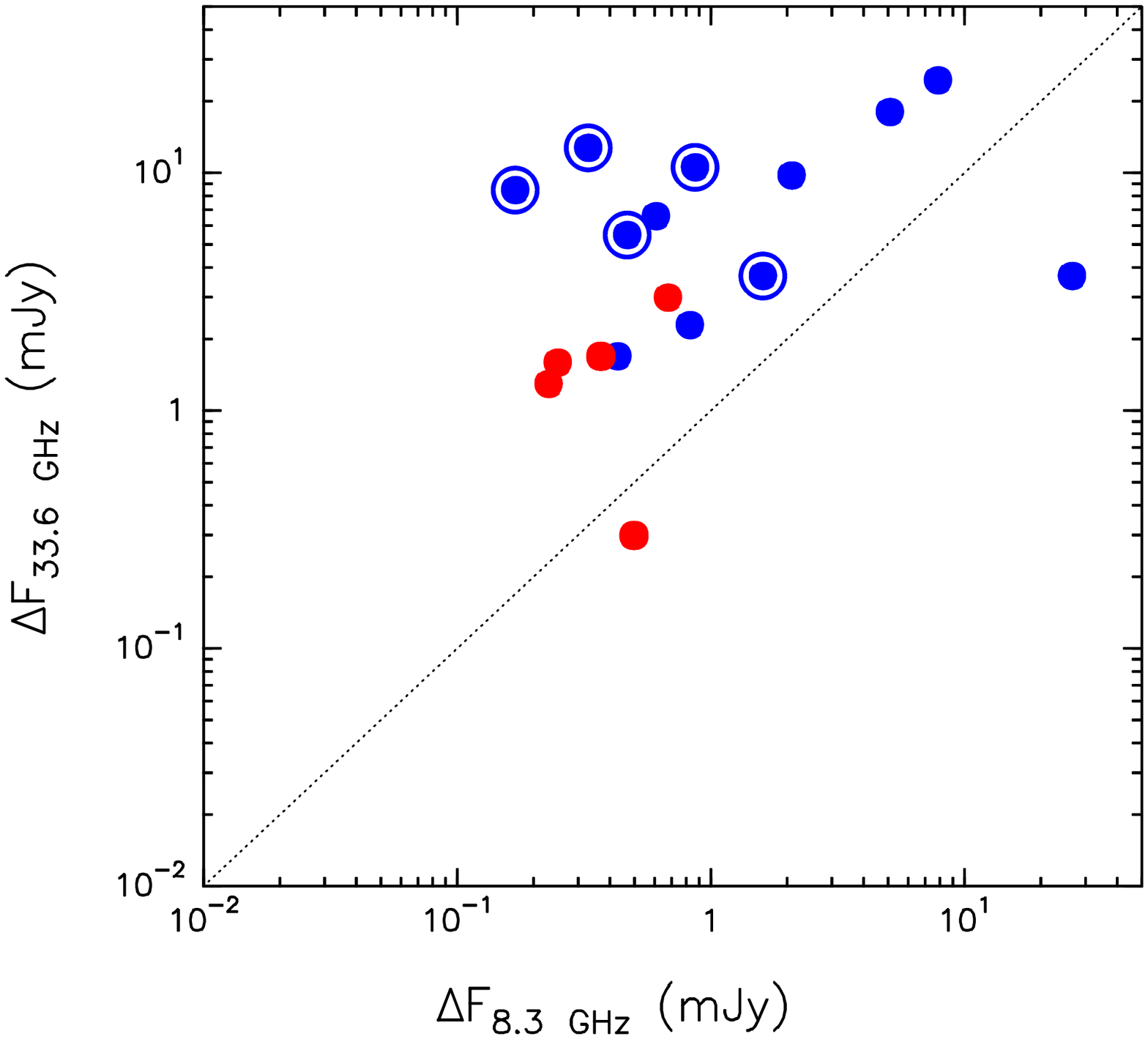}
\hspace{0.4cm}
\includegraphics[angle=0,width=5.55cm]{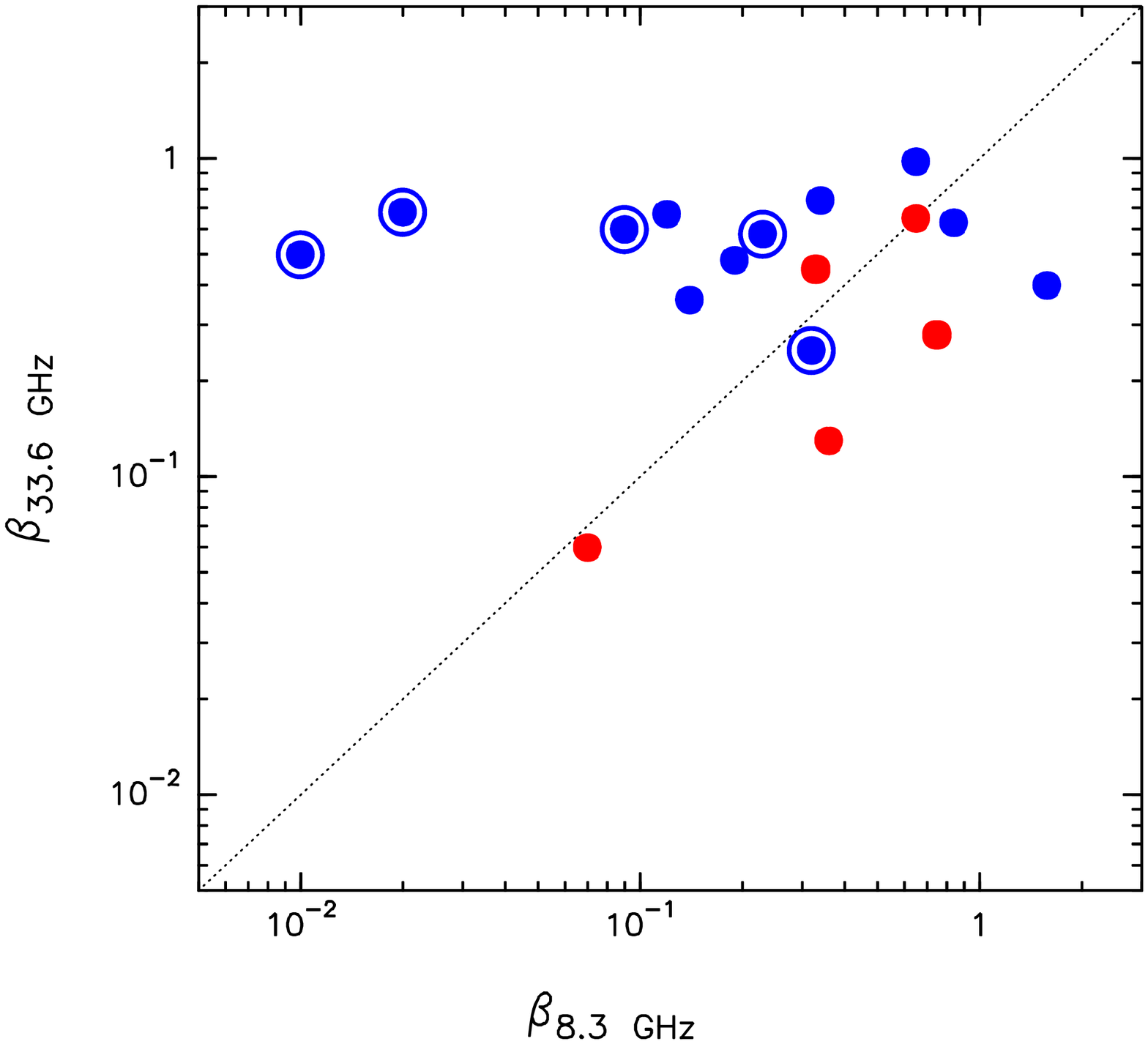}
\caption{Comparison of radio properties at 33.6 and 8.3 GHz. Red and blue dots indicate members of the ONC and OMC, respectively. The ONC sources related with proplyds are denoted by large blue circles. {\it Left panel:} Mean flux density at 33.6 GHz versus the mean flux density
at 8.3 GHz. For those sources that fall within the FoV of our 33.6
GHz observations but are only detected at 8.3 GHz, we have included
3$\sigma$ upper limits (blue solid triangles; and large empty triangles if they are related with proplyds).  The dashed lines show the relation expected for different types of emission: optically-thin
thermal dust (spectral index $\alpha \sim 3$); optically-thick,
ionized, stellar wind ($\alpha=0.6$); optically-thin free-free
($\alpha = -0.1$); and non-thermal synchrotron ($\alpha=-0.7$).  We
also indicate with a solid line the case of a flat spectrum
($\alpha=0$).  
{\em Middle panel:} Comparison between $\Delta F$ at 8.3 GHz and 33.6 GHz. The symbols are the same as in the left panel.
{\em Right panel:} Comparison between the variability parameters $\beta$ at 8.3 GHz
and 33.6 GHz. The symbols are the same as in the left panel.
}
\label{fig-radio-radio}
\end{figure*}

\begin{enumerate}

\item{The presence of optical counterpart indicates that the source is likely an ONC member. Since it is possible that some of the sources identified in this way are foreground field stars not related with the young cluster, we have additionally cross-correlated the radio sample with the list of 16 field stars from \citet{getman05b}, without any coincidence.}

\item{The presence of IR counterpart indicates that the source is likely an ONC or OMC star, because EG sources are expected to be weaker IR sources.}

\item{The value of hydrogen column density $N_{\rm H}$ derived from X-rays is in general a good indicator to discriminate between ONC and OMC members. A source with log$N_{\rm H}<$22.5 cm$^{-2}$ is considered as ONC member (\citealt{rivilla13a}).}

\item{The presence of X-ray variability is a good indicator to determine if a highly extincted source without IR counterpart is an embedded OMC member or an EG source, because young stars are expected to exhibit much higher X-ray variability. Following \citet{getman05a} we consider 3 signposts of X-ray radio variability: i) the significance of a Kolmogorov-Smirnov test ($P_{\rm KS}$), which establishes whether variations are above those expected from Poisson noise associated with a constant source; ii) the number of segments of the Bayesian block parametric model ($BBNum$) of source variability developed by \citet{scargle98}; iii) visual inspection of the X-ray light curves. We consider that a source is X-ray variable when $P_{\rm KS}<-2.0$ and/or $BBNum \geq 2$ and/or it exhibits X-ray flares in the light curves. We have obtained the values of $P_{\rm KS}$ and $BBNum$, and visually examined the X-ray light curves from \citet{getman05a} (see Table \ref{table-big}).}

\item{Finally, we compare the position of the sources with respect to the location of the OMC, traced by the CN N=1$-$0 emission from \citet{rodriguez-franco98} (see contours in Fig. \ref{fig-spatial-distribution}).}

\end{enumerate}

Additionally, to identify the ONC stars related to proplyds, we have cross-correlated the radio sample with the catalog of proplyds from \citet{kastner05}, using a counterpart radius of 0.5$\arcsec$. 
The resulting membership classification is shown in the last column of Table \ref{table-big}.
Fig.\ \ref{fig-spatial-distribution} shows the spatial distribution of the different groups of sources.

\begin{figure*}
\centering
\includegraphics[angle=0,width=8.9cm]{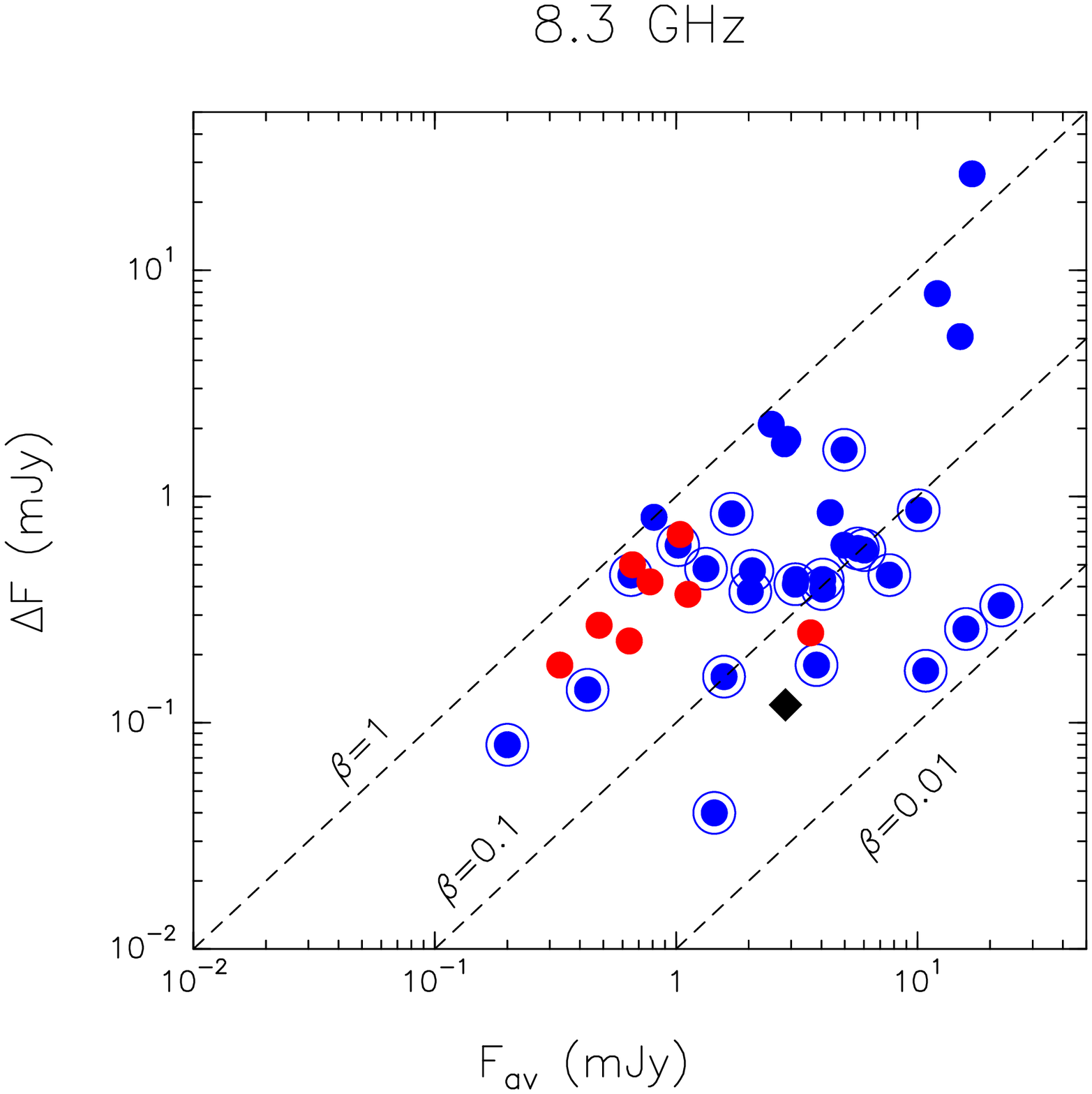}
\hspace{0.5cm}
\includegraphics[angle=0,width=8cm]{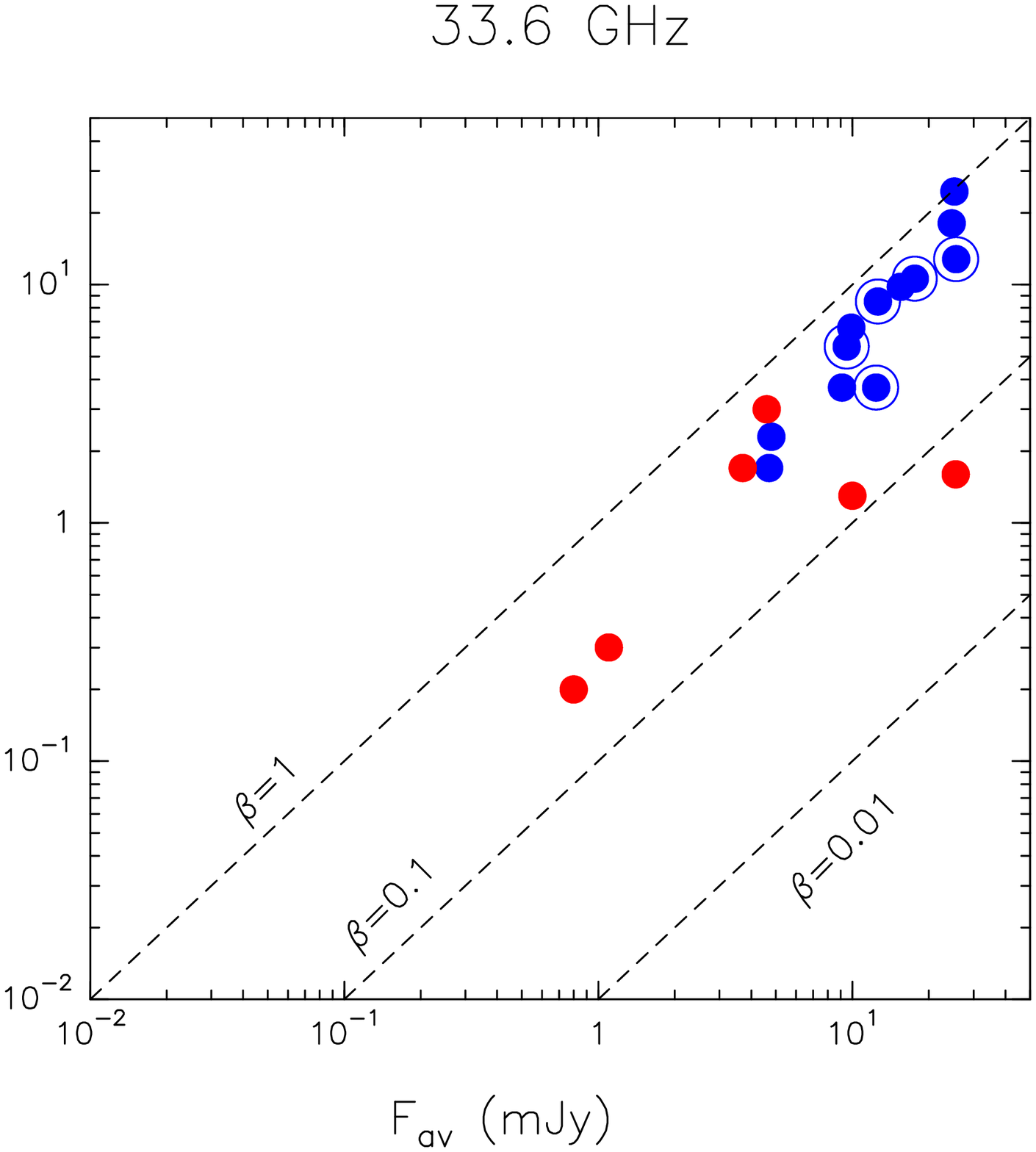}
\caption{Radio properties at 8.3 GHz (left) and 33.6 GHz (right) of the radio/X-ray sample. The black diamond corresponds to the EG source. The other colors and symbols are the same as in Fig. \ref{fig-spatial-distribution}. The dashed lines indicate values of $\beta$ of 1, 0.1 and 0.01.}
\label{fig-ratio-deltaF-Fav}
\end{figure*}

\subsection{Sample of sources emitting at 8.3 and 33.6 GHz}
\label{comparison-zapata}

The cross-correlation between our radio sample at 33.6 GHz and the radio sample at 8.3 GHz, with a counterpart radius of 0.5$\arcsec$,  shows that all the 33.6 GHz sources were also detected at 8.3 GHz, with the only exception of the new radio flaring source OHC-E. On the other hand, our monitoring did not detect emission from 25 sources from \citet{zapata04a} located within our FoV.
The radio properties at both radio frequencies are summarized in Table \ref{table-big}.
In the left panel of Fig. \ref{fig-radio-radio} we compare the average flux densities at both frequencies. We have distinguished ``naked" ONC stars, ONC stars related with proplyds and OMC stars. For those sources without 33.6 GHz detection, we have considered 3$\sigma$ upper limits measured in our image with best sensitivity.

In general, the radio sources detected by our monitoring exhibit higher 8.3 GHz flux densities than those undetected. 
This suggests that the non-detection at high frequency is likely due to lack of sensitivity.

With only one exception, the radio sources have higher average flux densities at 33.6 GHz. In the case of sources BN, I and C, we interpret that this is due to ``quiescent" thermal component that increases with frequency. In the other sources, this does not directly imply evidence of thermal origin, because the observations are not simultaneous and their emission is highly variable (Section \ref{results}).
Since this variability is likely connected to non-thermal processes, the value of $F_{\rm av}$ is very sensitive to the presence of flares, and not representative of their thermal emission.

In the middle and right panels of Fig. \ref{fig-radio-radio} we compare the values of $\Delta F$ and $\beta$ at 8.3 and 33.6 GHz. Most of the sources appear more variable at higher frequencies (15/17 have higher $\Delta F$; and 11/17 have higher $\beta$). This may indicate that the radio variability increase with frequency, although new observations of a more statistically representative sample of sources are needed to confirm this behavior.
 
We note that even the sources related with proplyds, which are expected to emit non-variable free-free or dust emission from circumstellar material, are clearly variable. This points toward the presence of a non-thermal component arising from the central PMS stars. 

\subsection{Full sample of sources emitting radio and X-rays}
\label{comparison-COUP}

We have cross-correlated of the sample radio sources (detected at 8.3 GHz and/or at 33.6 GHz\footnote{We have also included the flaring radio source ORBS detected by \citet{forbrich08} at 22 GHz.}) with the COUP catalog, searched for X-ray counterparts within 0.5$\arcsec$ of the radio sources. We have obtained a final sample of 51 sources emitting at radio and X-rays. The properties of the X-ray counterparts are shown in Table \ref{table-big}.

\subsubsection{Radio properties}

In Fig. \ref{fig-ratio-deltaF-Fav} we show the radio properties of the subsamples of X-ray sources with emission at 8.3 and 33.6 GHz.  
The ONC sources related with proplyds exhibit low absolute variability at 8.3 GHz ($\Delta F \lesssim $ 1 mJy) compared to ``naked" ONC stars, which appear more variable ($\Delta F \gtrsim$ 1 mJy). This suggests that the 8.3 GHz emission in the proplyds is dominated by thermal nearly constant emission from circumstellar material, while ``naked" ONC stars are might be dominated by variable non-thermal processes related with magnetic activity of the PMS star. This could be gyrosynchrotron emission produced by the acceleration of electrons in magnetic field reconnection events in the corona of the star (\citealt{andre96}).

However, we note that most of the proplyds with lower $F_{\rm av}$ have values of $\beta>$0.1 (left panel of Fig. \ref{fig-ratio-deltaF-Fav}), because their $\Delta F$ is a significant fraction of their $F_{\rm av}$. Therefore, it seems that the 8.3 GHz emission from proplyds, although dominated by thermal emission, could include also a non-thermal component arising from the central PMS stars. 

The OMC members show low values of absolute variability at  8.3 GHz ($\Delta F < $ 1 mJy). However, since most of them are weak sources, this low absolute variation translates into significant relative variability, with values of $\beta>$0.1, also pointing to non-thermal origin. 
Only BN is nearly constant, with low values of both $\Delta F$ and $\beta$, as occurs at 33.6 GHz. This confirms that the radio emission is likely dominated by thermal emission from the ionized gas in the massive envelope around the star.

The right panel of Fig. \ref{fig-ratio-deltaF-Fav} shows that all sources appear in general more variable at 33.6 GHz (both in $\Delta F$ and $\beta$). This may indicate that the emission at higher frequencies is more dominated by non-thermal highly variable processes.

Therefore, we conclude that the radio emission can be attributed to two different mechanisms: i) highly-variable (flaring) non-thermal radio gyrosynchrotron emission produced by accelerated electrons in the magnetospheres of low-mass PMS stellar members of both the ONC and OMC; and ii) non-variable thermal emission from the ionized gas and heated dust of the ONC proplyds illuminated by the Trapezium Cluster, or from the ionized gas in the envelopes surrounding massive stars, as in the case of the BN object.

\begin{figure*}
\centering
\includegraphics[angle=0,width=8.8cm]{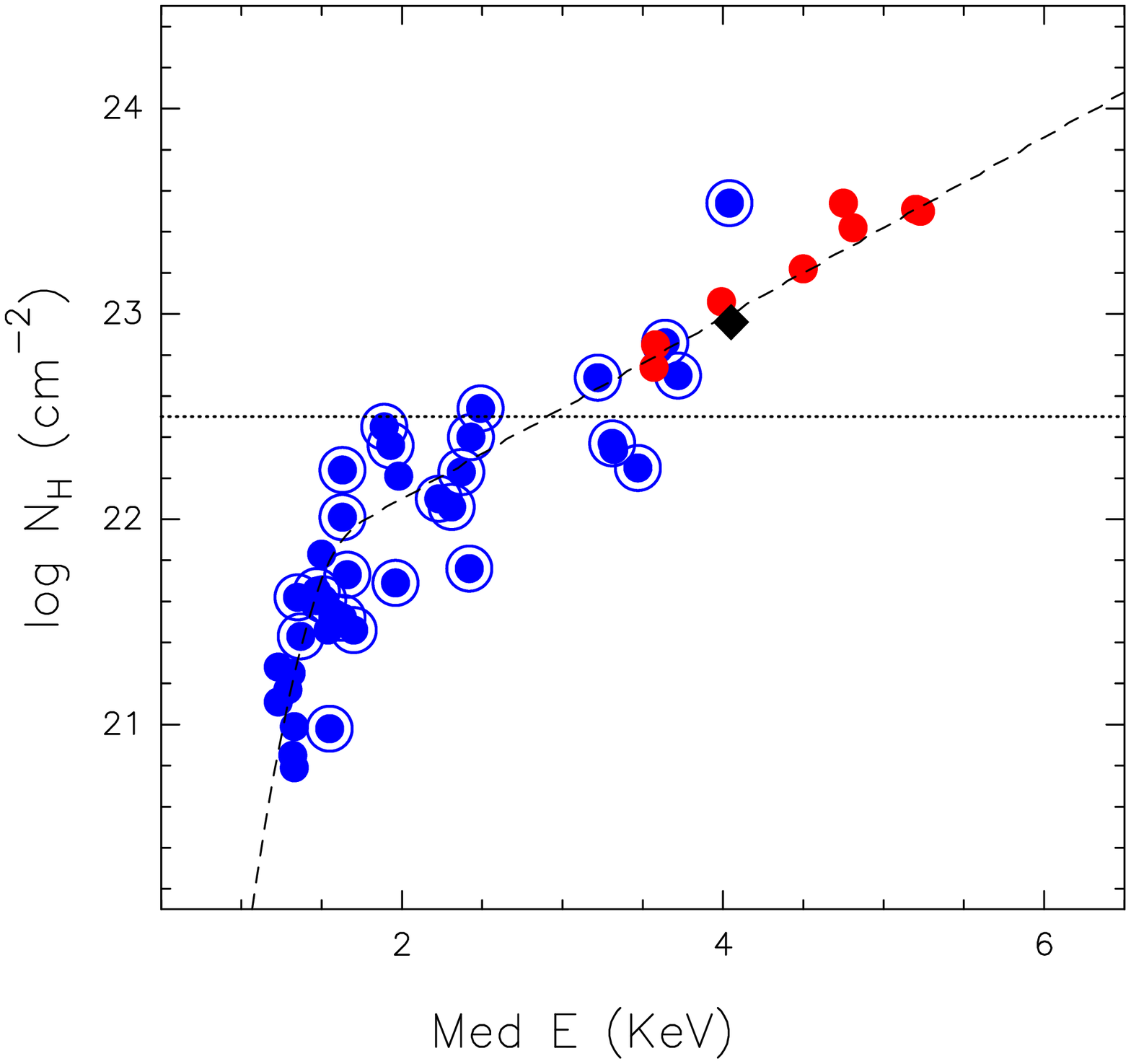}
\hspace{0.5cm}
\includegraphics[angle=0,width=8cm]{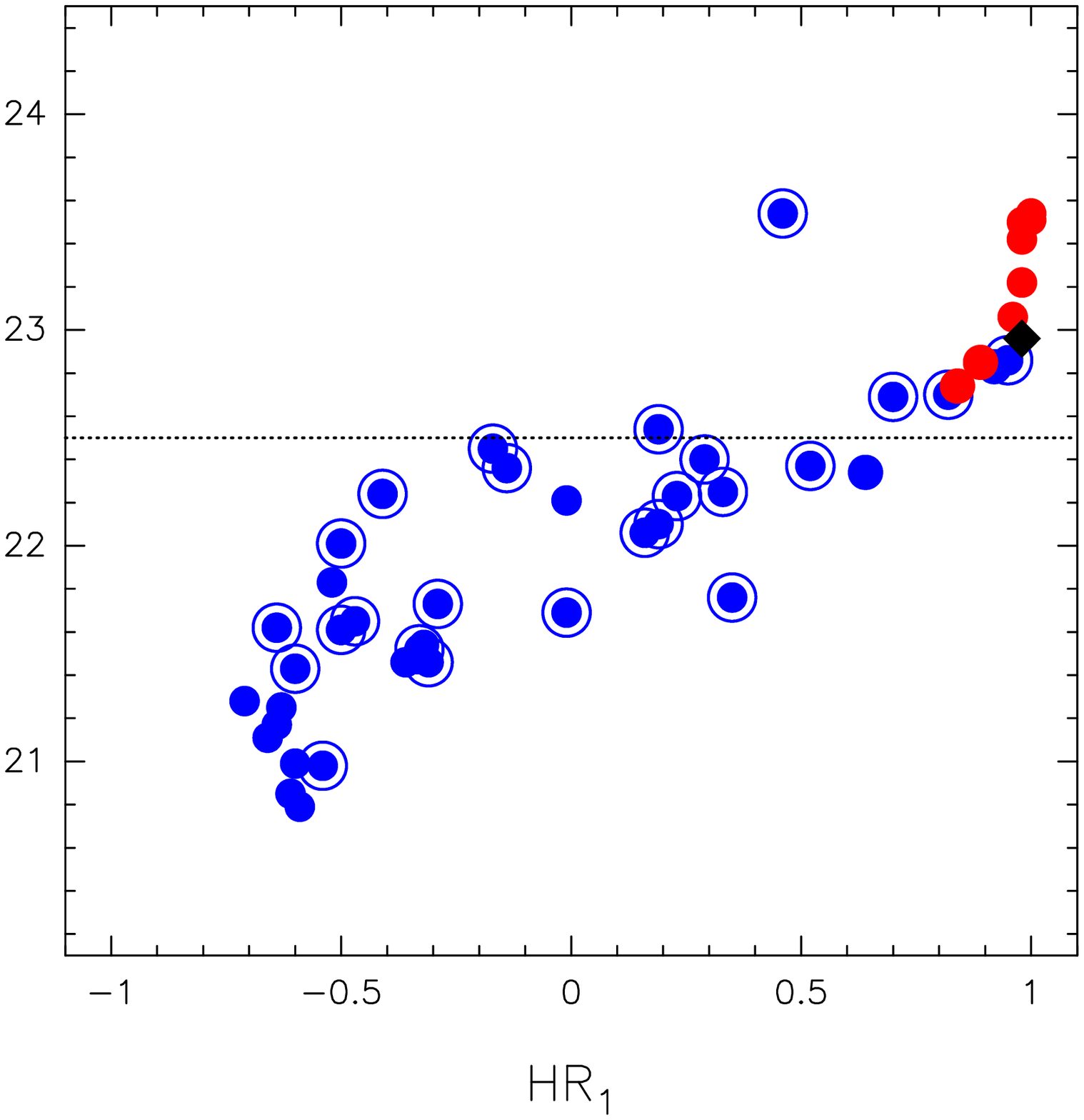}
\caption{Relation between the hydrogen column density $N_{\rm H}$ and the median energy of X-ray photons ($MedE$, left panel) and the hardness ratio ($HR_{\rm 1}$, right panel) of the members of the radio/X-ray sample. The dashed line in left panel is the empirical fit found by \citet{feigelson05}. The colors and symbols are the same as in Fig. \ref{fig-ratio-deltaF-Fav}.}
\label{fig-MedE-HR}
\end{figure*}

\subsubsection{X-ray properties}

\vspace{0.2cm}

{$\bullet$ \em Hydrogen column density $N_{\rm H}$, Hardness ratio ($HR_{\rm 1}$) and Median energy ($MedE$):} in Fig.  \ref{fig-MedE-HR} the values of median energy ($MedE$) of X-ray photons and the hardness ratio\footnote{Hardness ratio $HR_{1} = (h-s)/(h+s)$, where $h$ and $s$ refer to the counts detected in the hard (2.0$-$8.0 keV) and soft (0.5$-$2.0 keV) bands, respectively. Values closer to 1.0 indicate a hard X-ray source, and $-$1.0 a soft X-ray source.} ($HR_{\rm 1}$) of the full radio/X-ray sample as a function of $N_{\rm H}$ are shown. It is clear that the stars embedded in the OMC have higher values ($MedE>$3.5 keV and $HR_{1}>$0.84) than ONC stars.  \citet{feigelson05} found a relation between the median energy of the X-ray photons arising from the stars and the hydrogen column density $N_{\rm H}$ (left panel of Fig.\ \ref{fig-MedE-HR}). This trend is due to an absorption effect: in the embedded sources, only the harder photons are able to escape through the molecular gas and then be detectable, while softer ones (with lower energy) are absorbed. As a consequence, the embedded stars appear as harder sources. 
Among the ONC members, those related with proplyds exhibit higher values of $N_{\rm H}$ (and hence of $MedE$ and $HR_{\rm 1}$). This is likely due to the presence of circumstellar material that produces higher extinction than in ``naked" ONC stars.

\begin{figure*}
\centering
\includegraphics[angle=0,width=8cm]{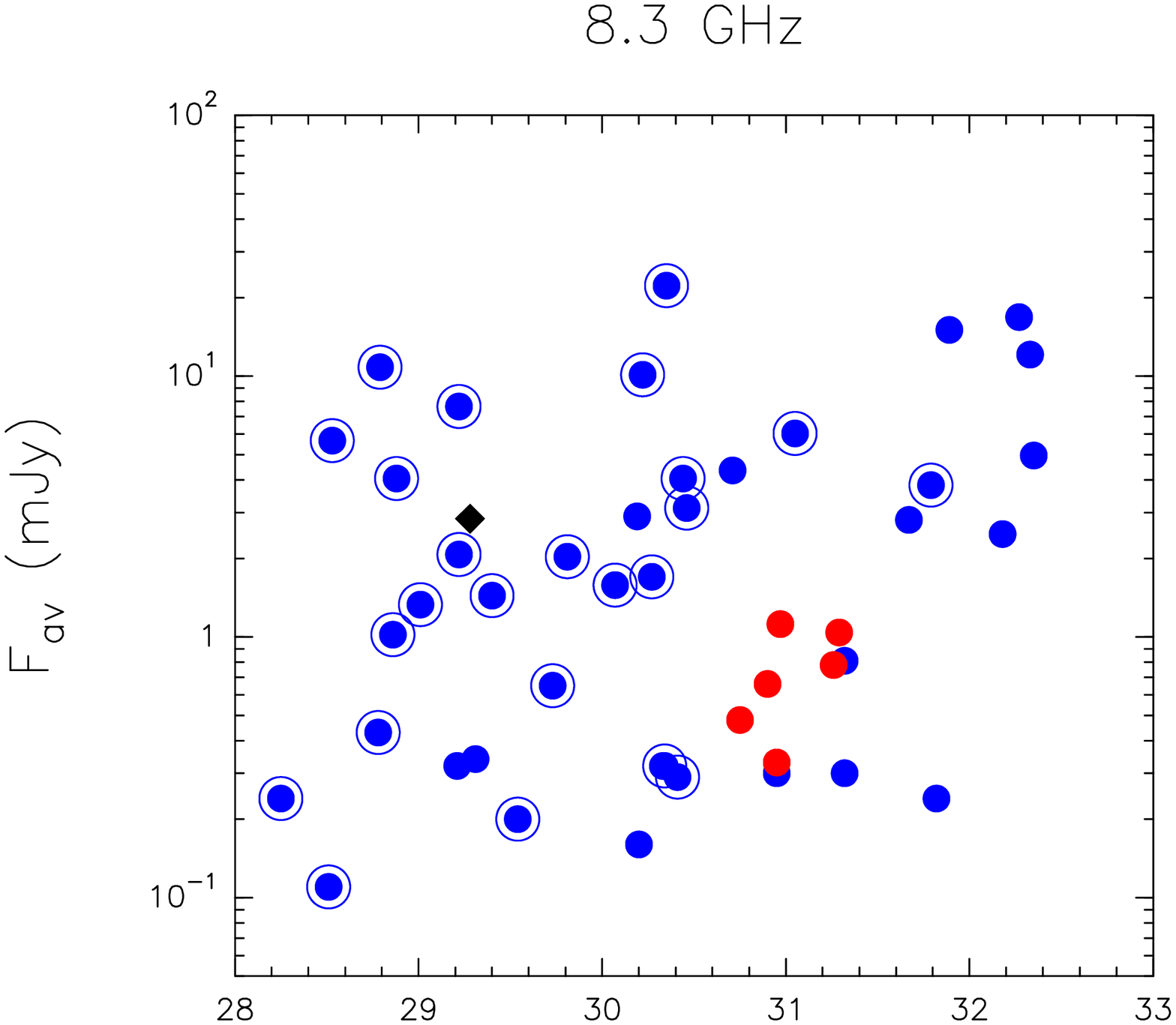}
\hspace{0.5cm}
\includegraphics[angle=0,width=7.0cm]{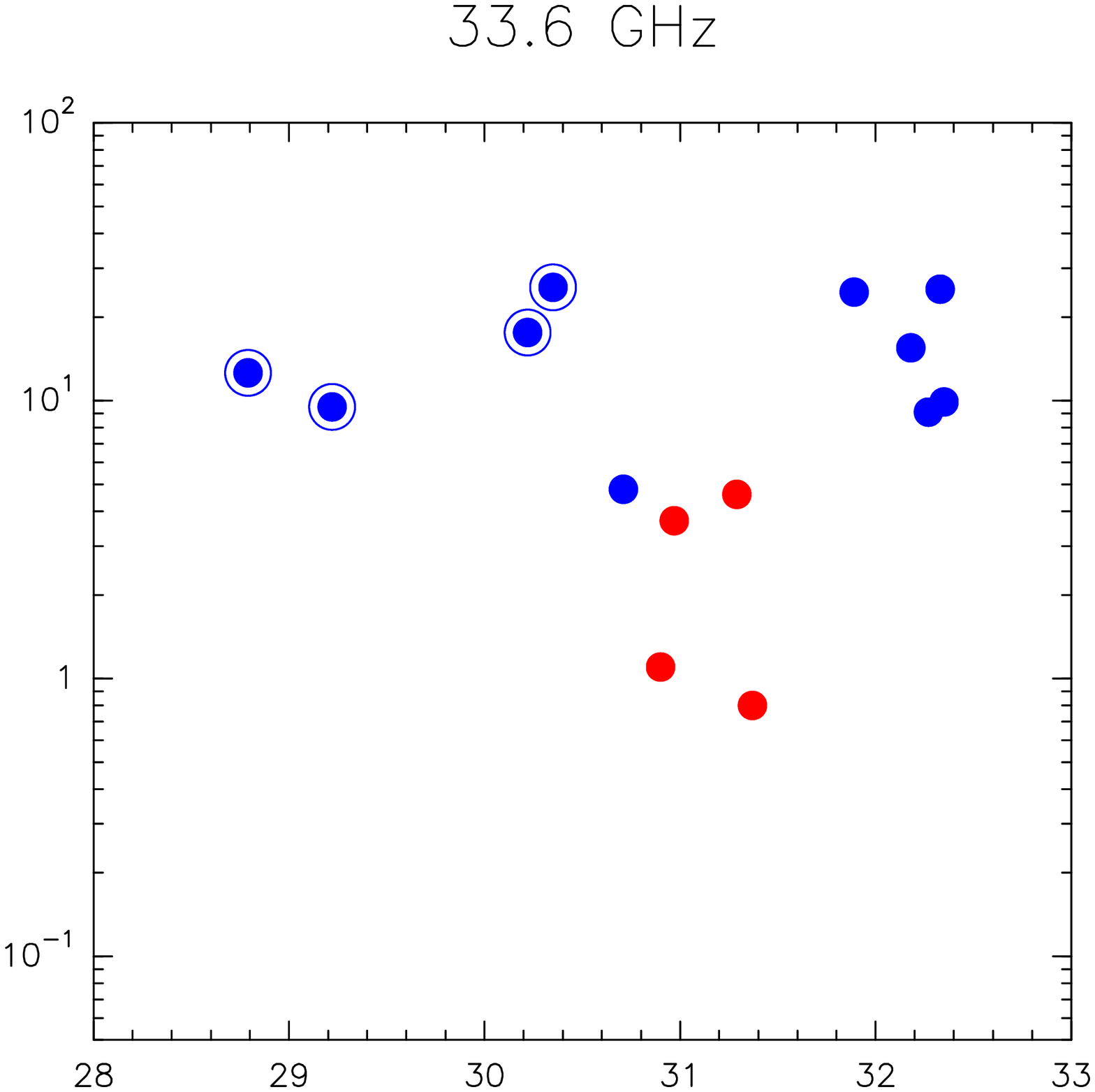}

\vspace{0.5cm}
\includegraphics[angle=0,width=8.0cm]{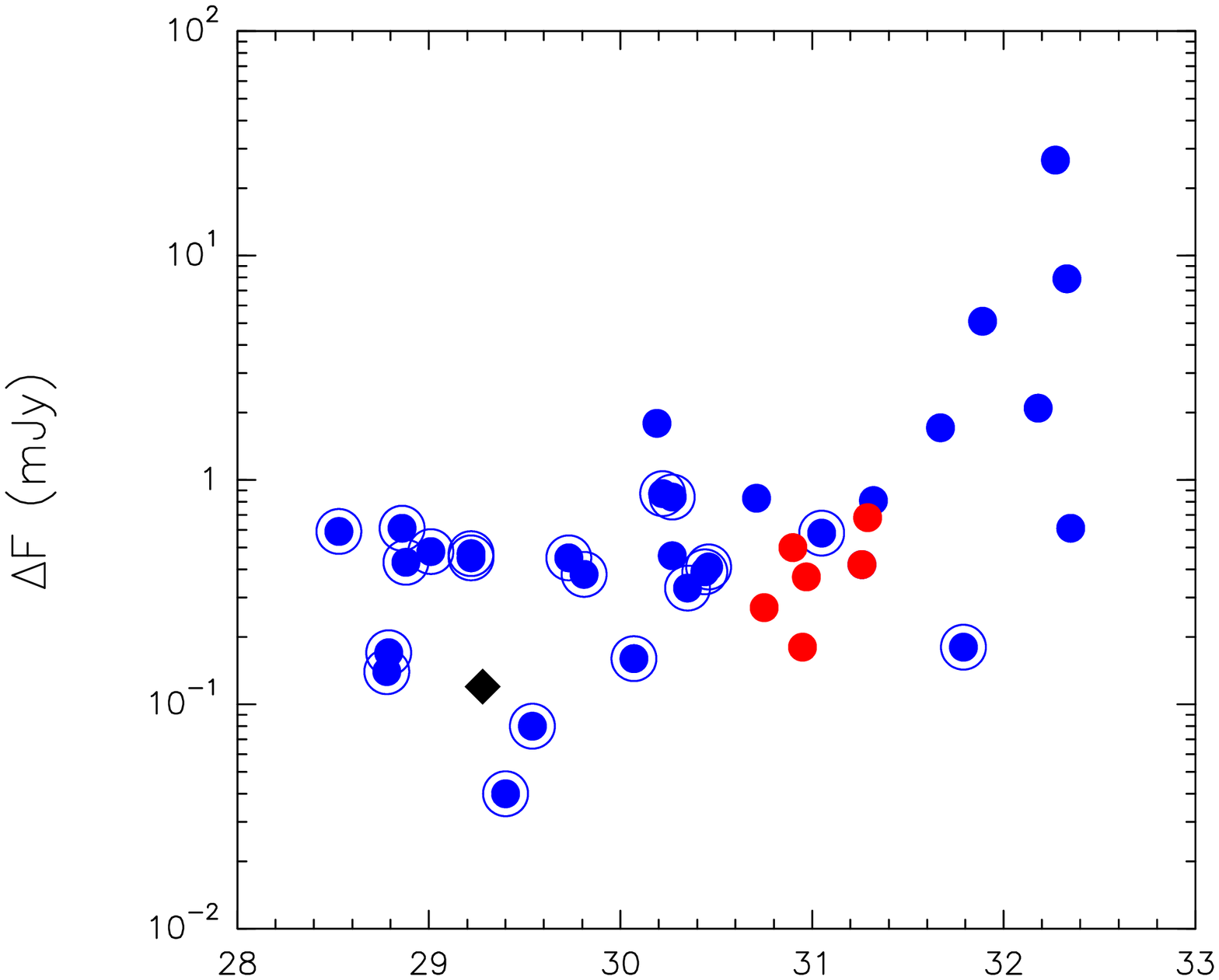}
\hspace{0.5cm}
\includegraphics[angle=0,width=7cm]{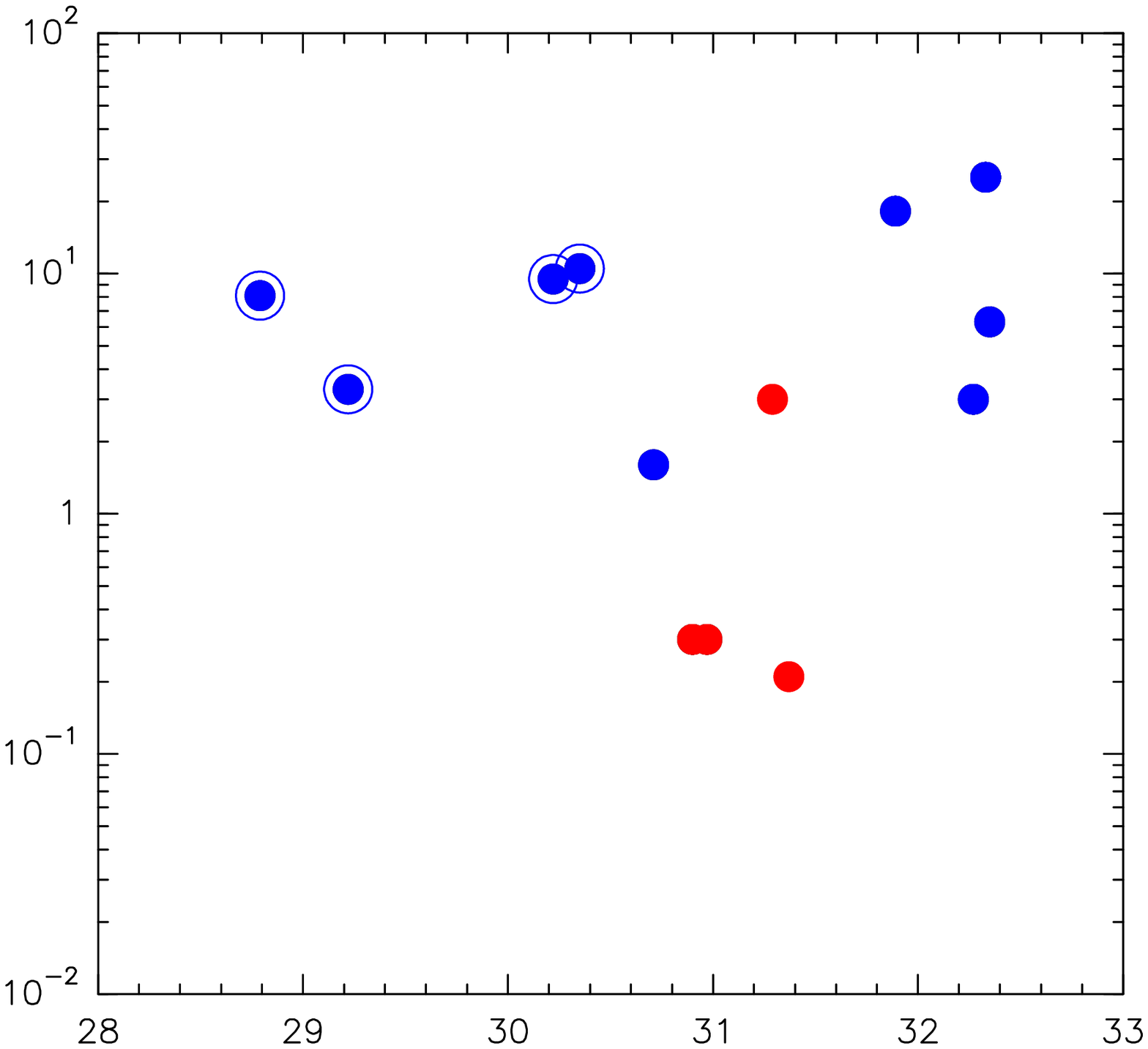}

\vspace{0.5cm}
\hspace{0.1cm}
\includegraphics[angle=0,width=7.75cm]{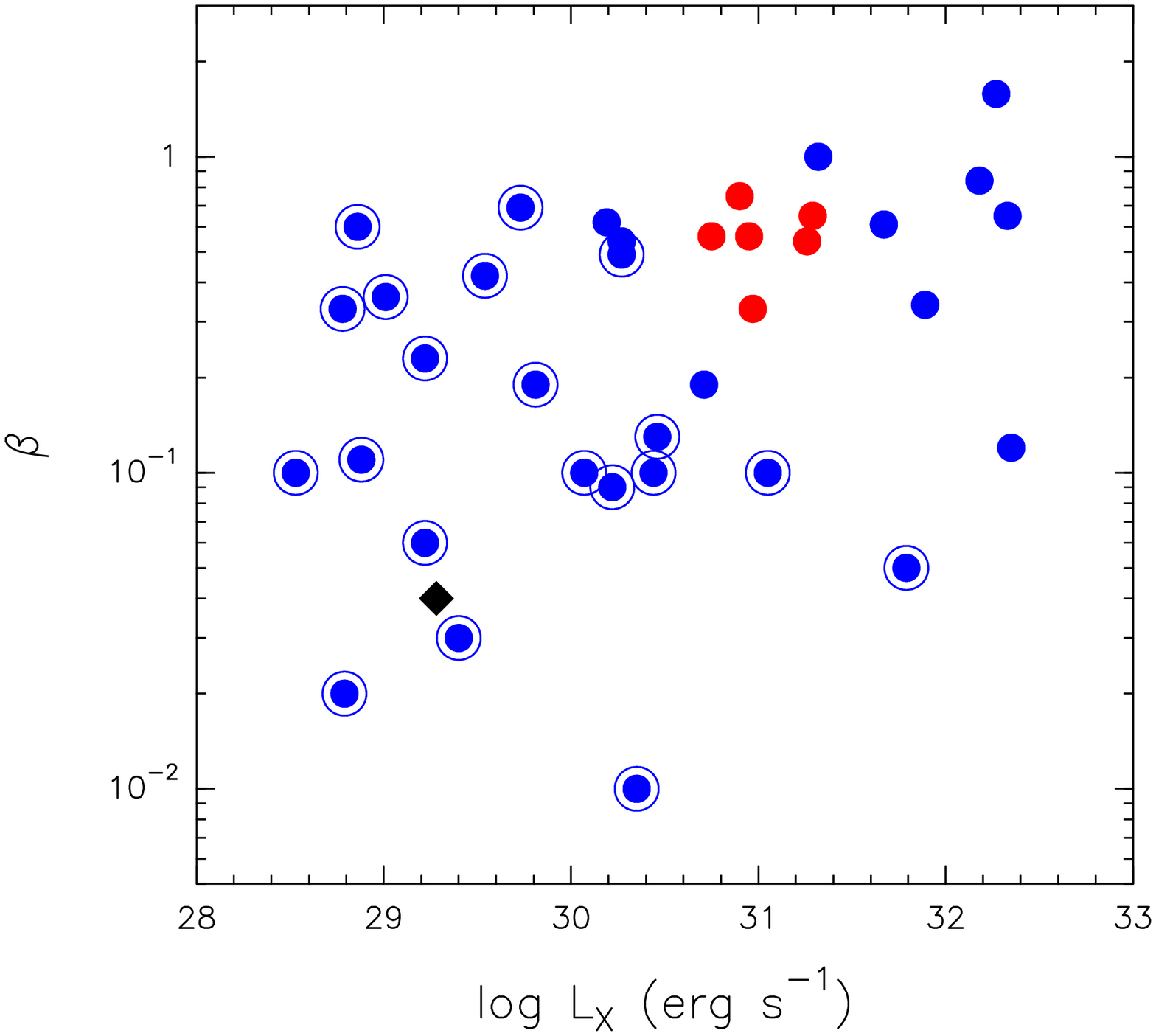}
\hspace{0.5cm}
\includegraphics[angle=0,width=7cm]{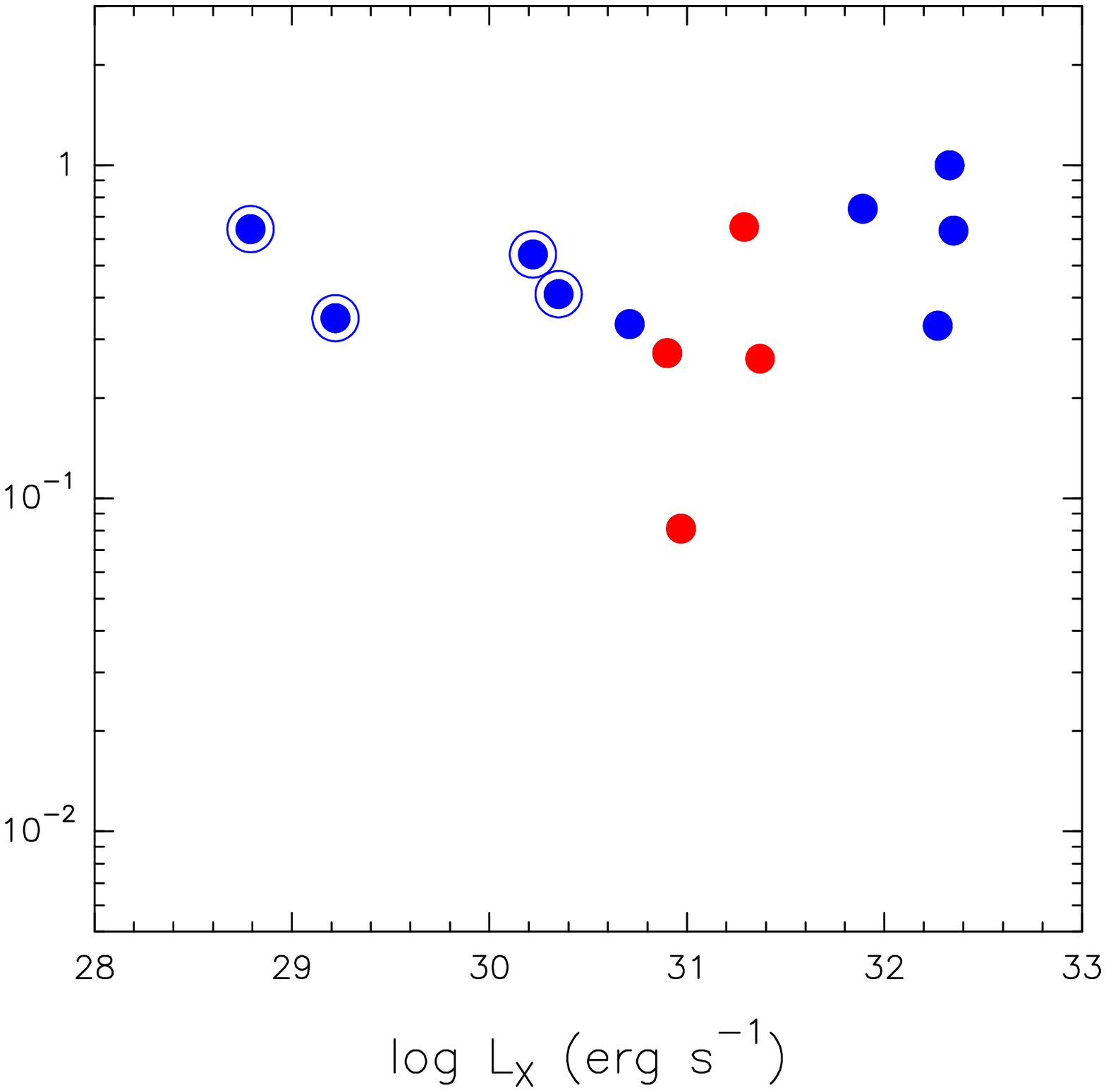}

\caption{Radio properties ($F_{\rm av}$, $\Delta F$, $\beta$) at 8.3 and 33.6 GHz versus the value of the X-ray luminosity derived from the X-ray COUP counterparts.  The colors and symbols are the same as in Fig. \ref{fig-ratio-deltaF-Fav}.}
\label{fig-radio-Lx}
\end{figure*}

\vspace{0.5cm}

{$\bullet$ \em X-ray luminosities}: In Fig. \ref{fig-radio-Lx} we show the radio properties of the radio/X-ray sample as a function of the extinction-corrected X-ray luminosity $L_{\rm X}$. We find that $L_{\rm X}^{ONC-proplyds} < L_{\rm X}^{OMC} < L_{\rm X}^{ONC-naked}$. Although this result could be physically real, we note that it should be taken with caution, because the derivation of the extinction-corrected luminosity is more uncertain for highly extincted sources. 

We also investigate whether there is a relation between the radio properties and the X-ray luminosity in the radio/X-ray sample. \citet{forbrich13} remarked that there is no clear correlation between the radio flux densities measured by \citet{zapata04a} at 8.3 GHz and the COUP X-ray luminosities (upper left panel in Fig. \ref{fig-radio-Lx}).  Similarly, we do not see a trend in our subsample at 33.6 GHz (upper right panel in Fig. \ref{fig-radio-Lx}). However, if we do not consider the proplyds a tentative trend appears: the radio flux density generally increases with the X-ray luminosity. This suggests a relation between the X-ray and radio emissions for the ``naked" ONC and OMC sources. In the case of proplyds, a significant fraction of radio emission is expected to arise from gas ionized by external illumination, which is not linked to X-ray activity. Then, a relation between the radio flux and X-ray luminosity is not expected in proplyds, as observed in Fig. \ref{fig-radio-Lx}.

\begin{figure}
\centering
\includegraphics[angle=0,width=8cm]{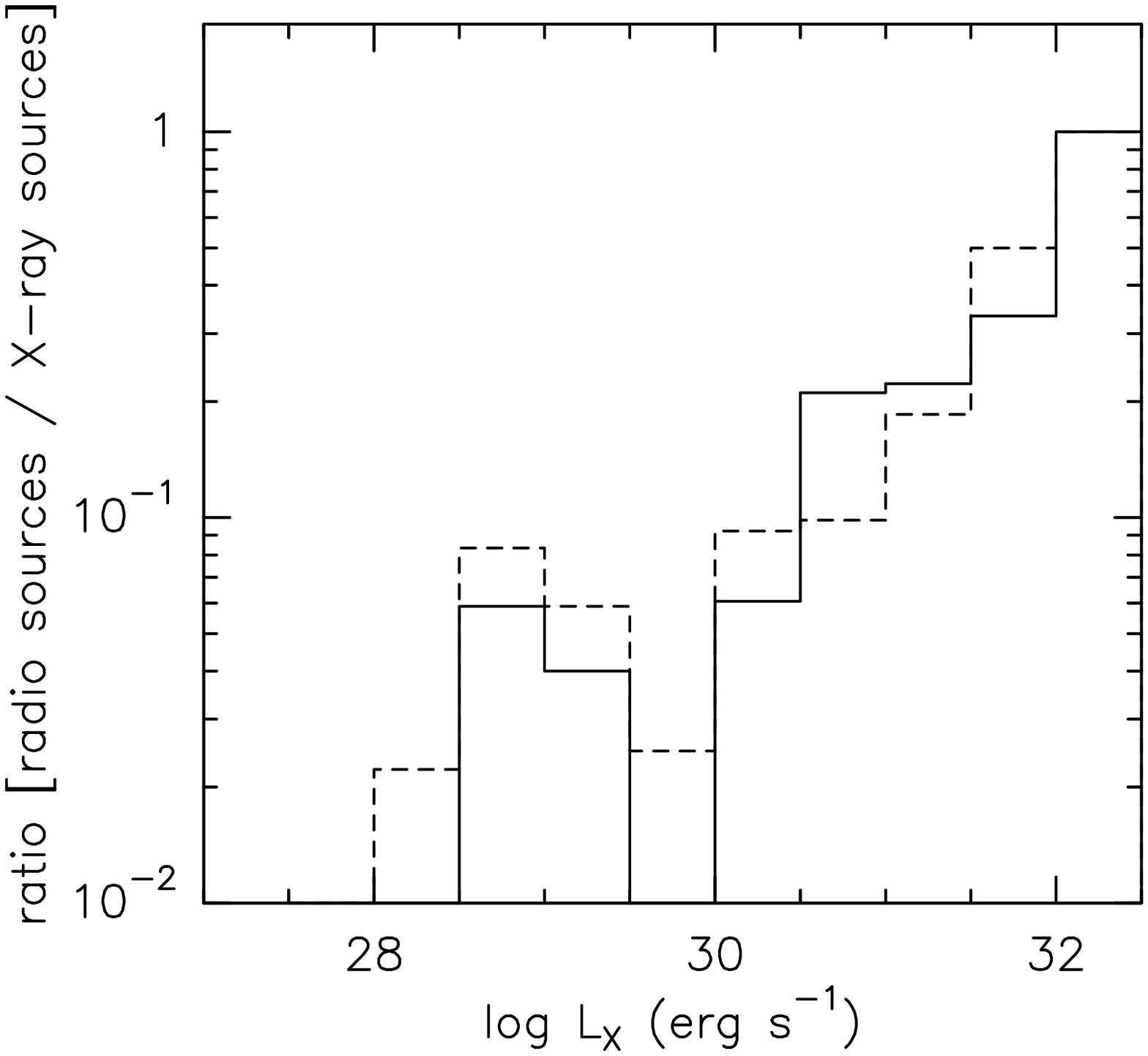}
\caption{Fraction of X-ray COUP sources detected by our radio
monitoring at 33.6 GHz (solid line) and \citet{zapata04a} monitoring
at 8.3 GHz (dashed line) as a function of the absorption-corrected
total X-ray luminosity $L_{\rm X}$. In each case, we have considered
the COUP sources that fall within the FoV and with enough counts
so that the X-ray luminosity corrected for extinction could be
reported: 159 for our monitoring and 595 for the 8.3 GHz monitoring
(see also \citealt{forbrich13}).}
\label{fig-ratio-radio-X}
\end{figure}

We have also studied the fraction of X-ray sources detected at radio wavelengths as a function of the X-ray luminosity (Fig. \ref{fig-ratio-radio-X}) for the 8.3 and 33.6 GHz monitorings. In each case, we have considered the COUP sources that fall within the FoV of each observation with enough counts so that the X-ray luminosity corrected for extinction could be reported. We find 159 X-ray sources within our FoV and 595 sources within the 8.3 GHz FoV. Both samples show a very similar behavior. The weaker X-ray sources with log$L_{\rm X}< 28.5$ erg s$^{-1}$ (200 sources, i.e., 13$\%$ of the full COUP sample) were not detected by our radio monitoring, while \citet{zapata04a} only detected 1 source. Fig. \ref{fig-ratio-radio-X} shows that the fraction of X-ray sources detected in the radio increases with X-ray luminosity.  This indicates that the radio observations have statistically detected the most luminous X-ray sources. This suggests that: i) the underlying mechanisms responsible for the X-ray and (at least some fraction of) the radio emission are somehow related; and ii) radio monitorings have been limited in the past due to sensitivity, confirming the conclusion of \citet{forbrich13}.  

The middle and lower panels of Fig. \ref{fig-radio-Lx} also show tentative trends between the radio variability ($\Delta F$ and $\beta$) and the X-ray luminosity for the sources not related with proplyds. In general, the radio variability increases towards higher $L_{\rm X}$.

However, we note that the number of sources of the ``naked" ONC and OMC subsamples is too low to draw robust general conclusions. Deeper radio observations detecting a much larger number of sources are needed to confirm these tentative trends.

\begin{figure*}
\centering
\includegraphics[angle=0,width=8cm]{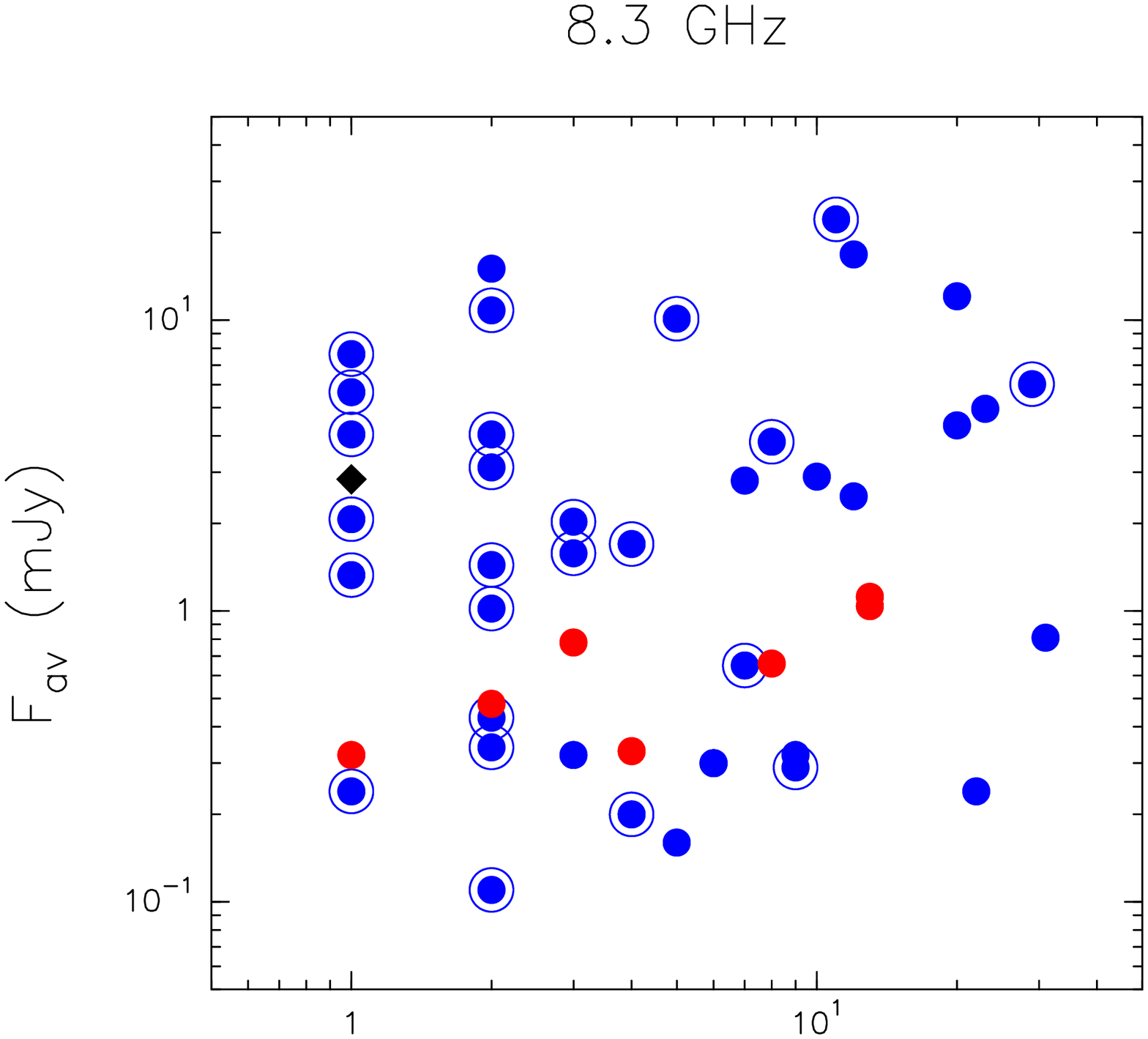}
\hspace{0.5cm}
\includegraphics[angle=0,width=7cm]{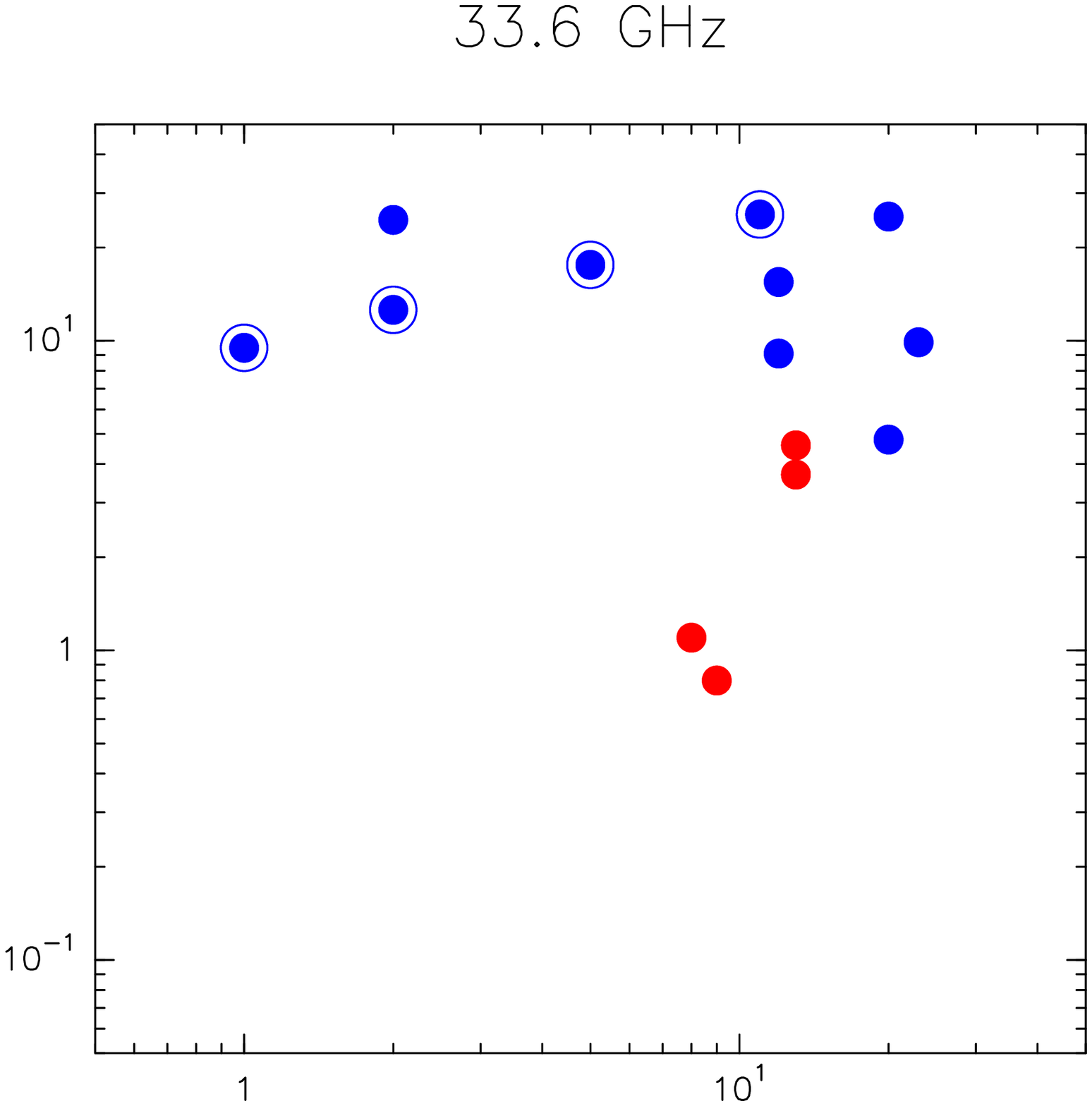}
\vspace{0.5cm}

\includegraphics[angle=0,width=8cm]{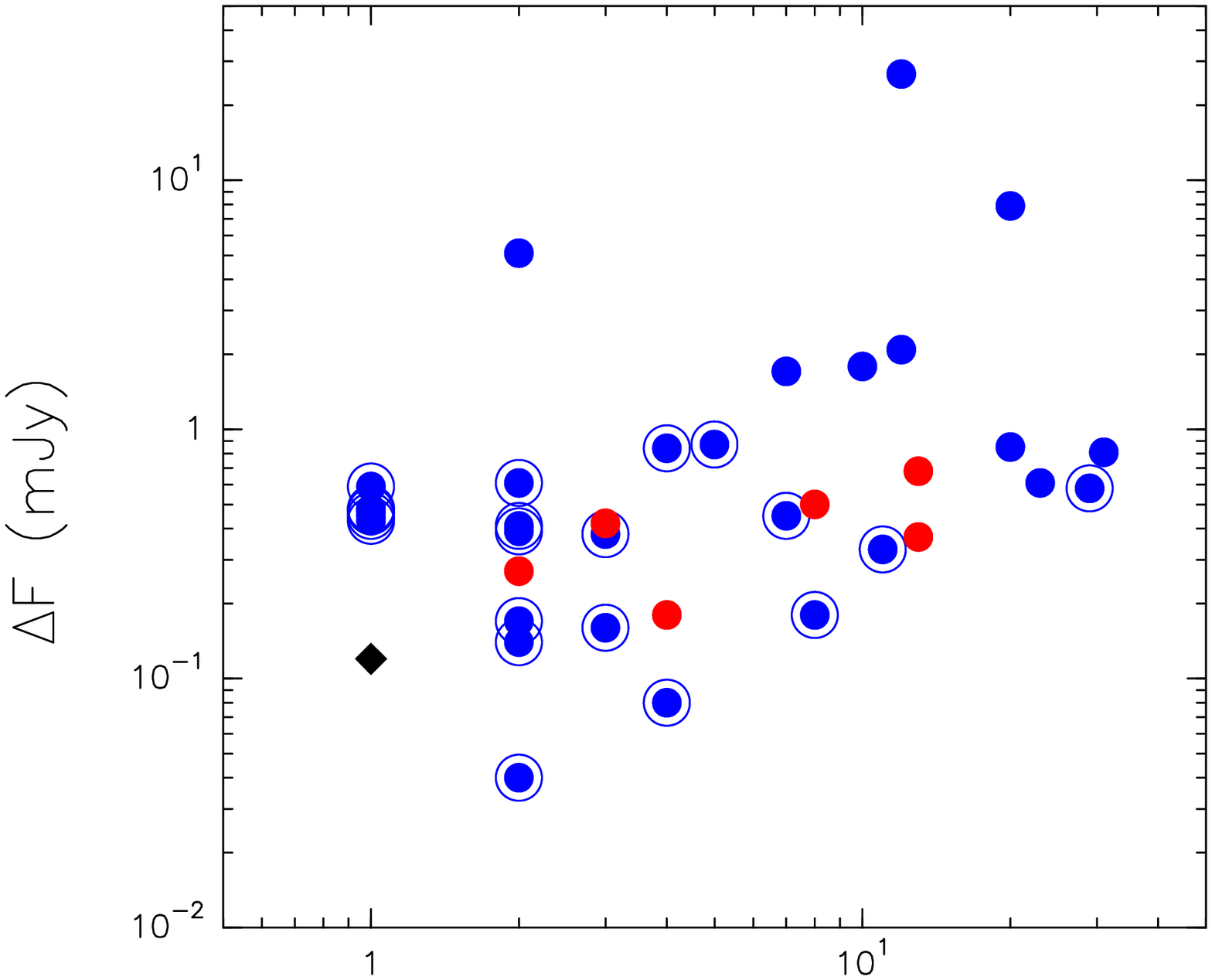}
\hspace{0.5cm}
\includegraphics[angle=0,width=7cm]{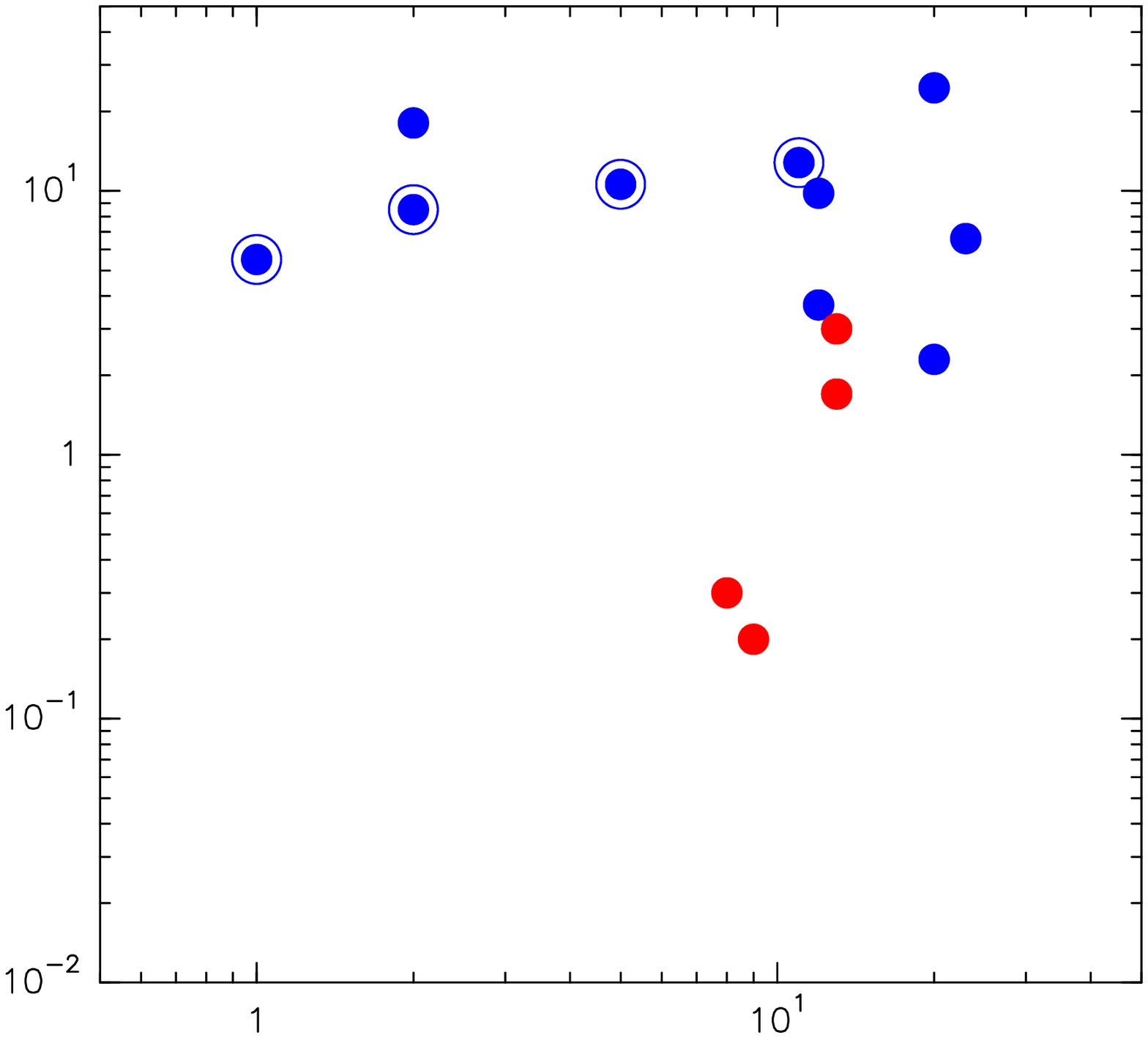}
\vspace{0.5cm}

\includegraphics[angle=0,width=7.75cm]{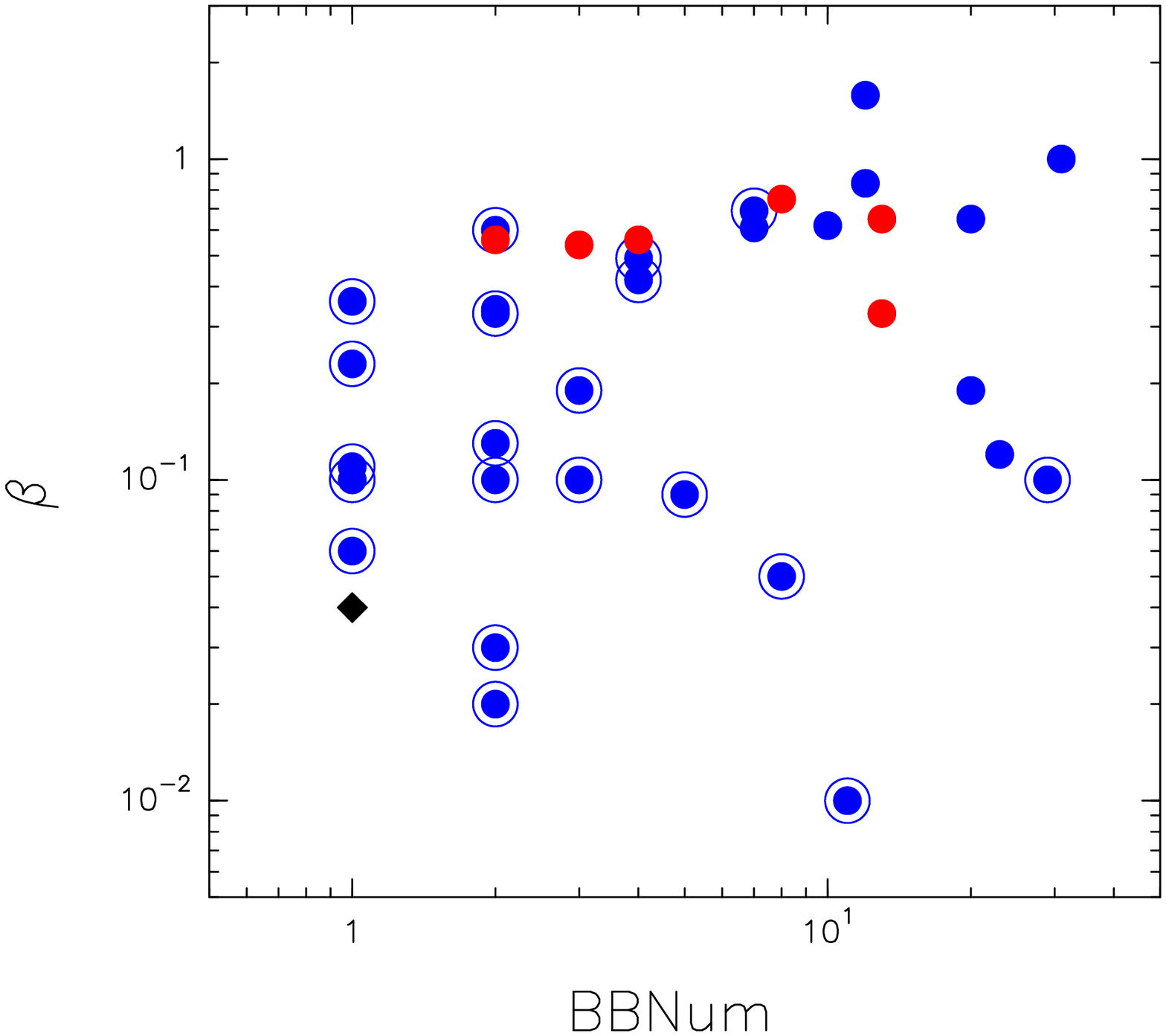}
\hspace{0.5cm}
\includegraphics[angle=0,width=7cm]{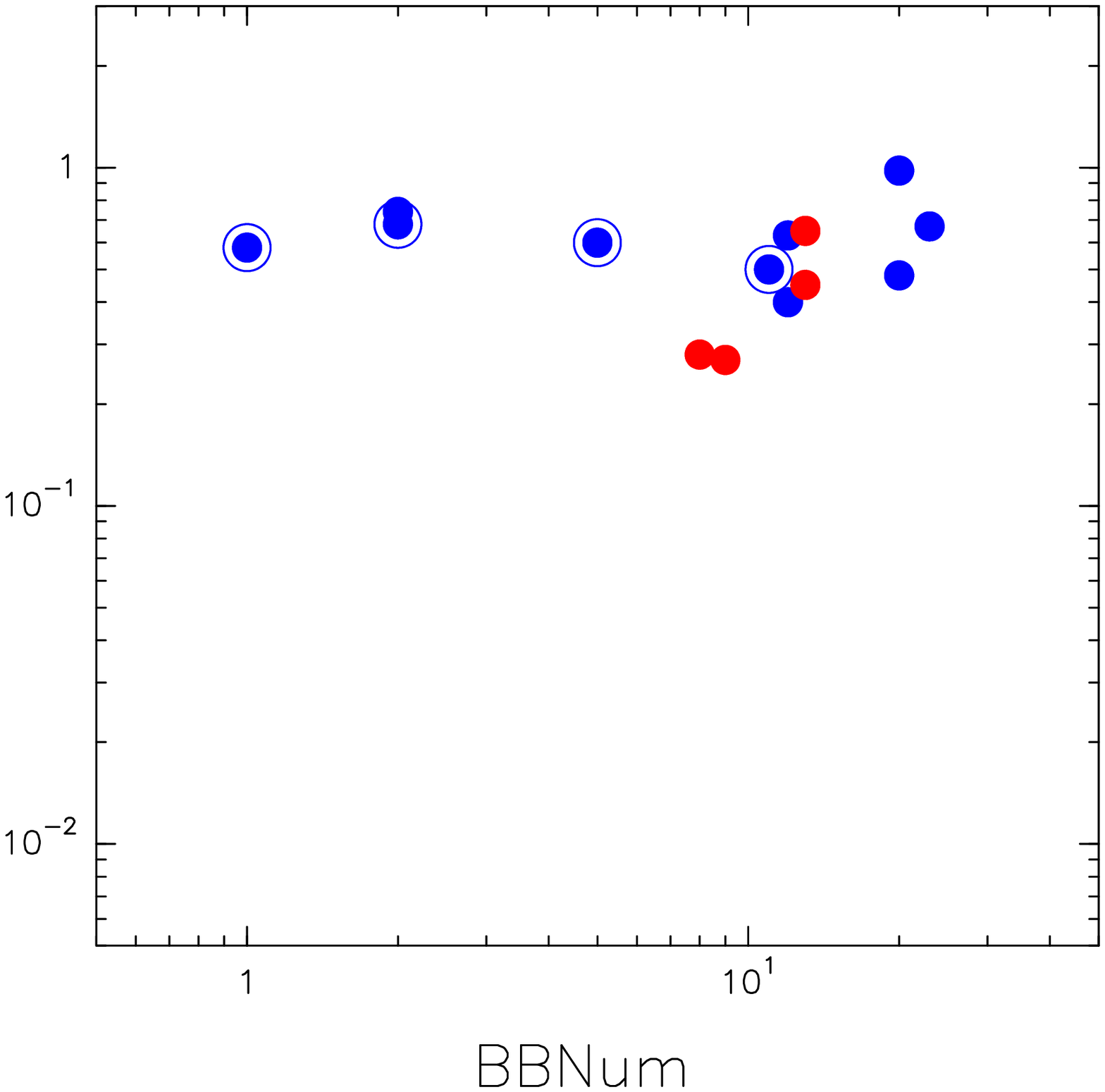}

\caption{Radio properties ($F_{\rm av}$, $\Delta F$, $\beta$) at 8.3 and 33.6 GHz versus the value of the $BBNum$ derived from the X-ray COUP counterparts, which is a proxy for the X-ray variability. The colors and symbols are the same as in Fig. \ref{fig-ratio-deltaF-Fav}.}
\label{fig-radio-BBNUM}
\end{figure*}

\vspace{0.5cm}

{$\bullet$ \em X-ray variability}: We now study whether there is a relation between the radio and the X-ray variability. We follow the 3 conditions presented in Sect. \ref{membership} to consider a source X-ray variable: $P_{\rm KS}<-2.0$ and/or $BBNum \geq 2$ and/or presence of X-ray flares in the light curves (see Table  \ref{table-big}). Only one radio source (E) of the 33.6 GHz/X-ray subsample does not exhibit X-ray variability. Namely, 93$\%$ of the stars of this subsample  are X-ray variable. Regarding the full radio/X-ray sample, 85$\%$ of the stars are X-ray variable. These fractions are higher than the fraction of X-ray variable sources of the full COUP sample, which is $\sim$60$\%$ (\citealt{getman05a}). Therefore, we have found that the radio sources are mostly associated with X-ray variable stars. This supports that (at least some of the) radio emission might be related to the same magnetic events in the coronae of PMS low-mass stars that also produce the X-ray emission.

In Fig. \ref{fig-radio-BBNUM} we show the radio properties of the radio/X-ray sample as a function of the level of X-ray variability, quantified by $BBNum$. As in Fig. \ref{fig-radio-Lx}, tentative trends appear if we do not consider the proplyds. However, given that the X-ray and radio observations were carried out at different epochs and different sensitivities, no firm conclusions can be drawn. Unlike the very deep COUP observation, which continuously observed the region during $\sim$10 days, the radio monitoring runs are much shorter, yielding a lower probability to detect large flux density variations such as flares. To better understand if radio and X-ray variability are directly related, simultaneous observations will be needed.

\section{The new radio source embedded in the Orion Hot Core: OHC-E}
\label{discussionE}

Our observations have detected twice a new radio source, OHC-E (2009
March 19 and 2011 July 09, see Fig.\ \ref{fig-OHC-E}).  Neither
\citet{felli93} nor \citet{zapata04a}, who performed observations
covering 7 months and 4 years at 5 and 15 GHz and 8.3 GHz, respectively,
nor \citet{goddi11}, who observed the same region at 45 GHz two
months before our detection (2009 January 12), detected emission
towards this source.

The source is not resolved by the beam of our observations in any
of our two detections ($\sim 0.2\arcsec$ and $\sim 0.075\arcsec$).
We can set upper limits to the deconvolved size of the emission of
$<0.1\arcsec$ ($<40$ AU) for the
Q-band emission and of $<0.04\arcsec$ ($< 10$ AU) for the Ka-band emission. Unfortunately, it is not
possible to determine the spectral index of the emission, because
observations at several radio wavelengths were not carried out
simultaneously. We derived lower limits to the brightness temperature
of $\sim 460$ K and $\sim 615$ K from the source flux density at
45.6 GHz and 33.6 GHz, respectively, and the upper limits to the
source sizes. Although these temperatures are consistent with both
thermal and non-thermal emission, our monitoring shows that the
emission is highly variable, suggesting a non-thermal origin.

\begin{figure}
\centering
\includegraphics[angle=0,width=8.0cm]{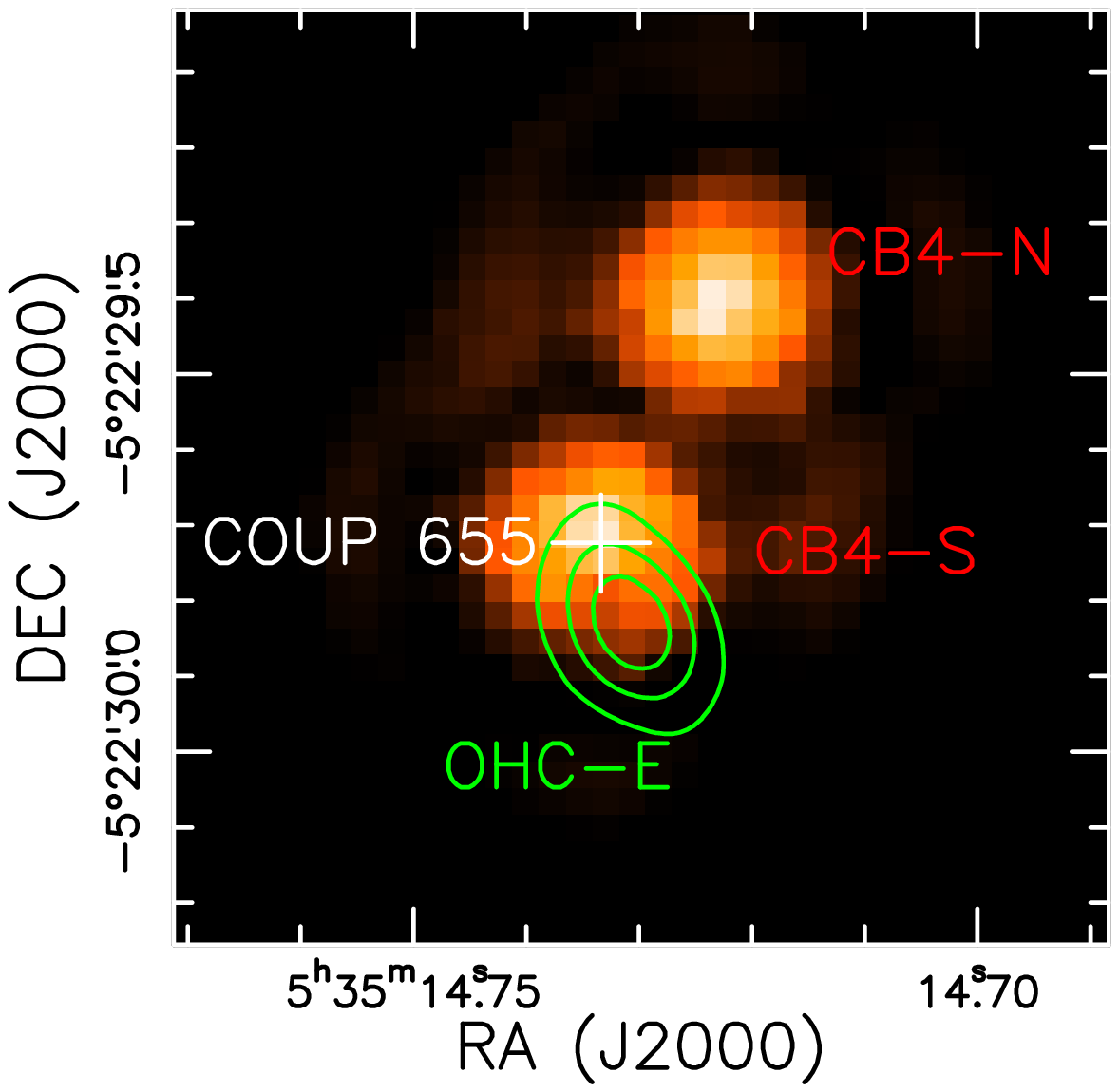}
\caption{2.15 $\mu$m HST-NICMOS image (color scale) of the region where OHC-E was detected, from the Hubble Legacy Archive (HLA). An IR binary is detected toward this region, CB4 (see also \citealt{stolovy98,simpson06}). The green contours correspond to the radio emission at 45.6 GHz  detected on our 2009 March 19 image (5$\sigma$, 10$\sigma$ and 15$\sigma$). The positions of the X-ray star COUP 655 is indicated with the white plus sign.}
\label{fig4}
\end{figure}

The position of the radio flare coincides with an embedded X-ray
low-mass pre-main sequence (PMS) star COUP 655 (Fig. \ref{fig4}).
The source OHC-E is also located very close to the southeast member of a binary stellar
system (CB4, Fig. \ref{fig4}) observed with high angular
resolution by the Near Infrared Camera and Multi-Object Spectrometer (NICMOS) onboard the Hubble Space telescope (HST) (see also \citealt{stolovy98}; \citealt{simpson06}).
Therefore it seems that both the radio and the X-ray emission are related to this star. We can give a rough estimate for the mass of this star using
the canonical relation for X-ray young stars, log[$L_{\rm X}$/$L_{\rm
bol}$]=$-$3.0 (\citealt{pallavicini81}). Assuming an stellar age typical for a massive star forming region of 5$\times$10$^{5}$ yr, OHC-E would have a mass of $\sim$1 M$_{\odot}$ (using the \citealt{siess00} stellar models), supporting that it is a low-mass star. Hence, the variability observed both in radio and X-rays wavelengths can be related with magnetic activity in the corona of this PMS low-mass star.

\section{Source 12: the binary $\theta^{1}$ {\it Ori A}}
\label{source-12}

$\theta^{1}$ $Ori$ $A$ is a well known binary system, with a B0.5
primary and a low-mass companion (e.g., \citealt{close13}).
As observed in the binary WR140 (\citealt{williams90}), the binarity
can produce a smooth periodic variation of the radio emission caused
by the combination of two geometrical effects: i) variation of the
free-free opacity due to the ionized envelope of the primary star
as the companion orbits; ii) variation of the stellar activity (and
hence the non-thermal emission) inversely proportional to the
separation between components.

\begin{figure}
\centering
\includegraphics[angle=0,width=8cm]{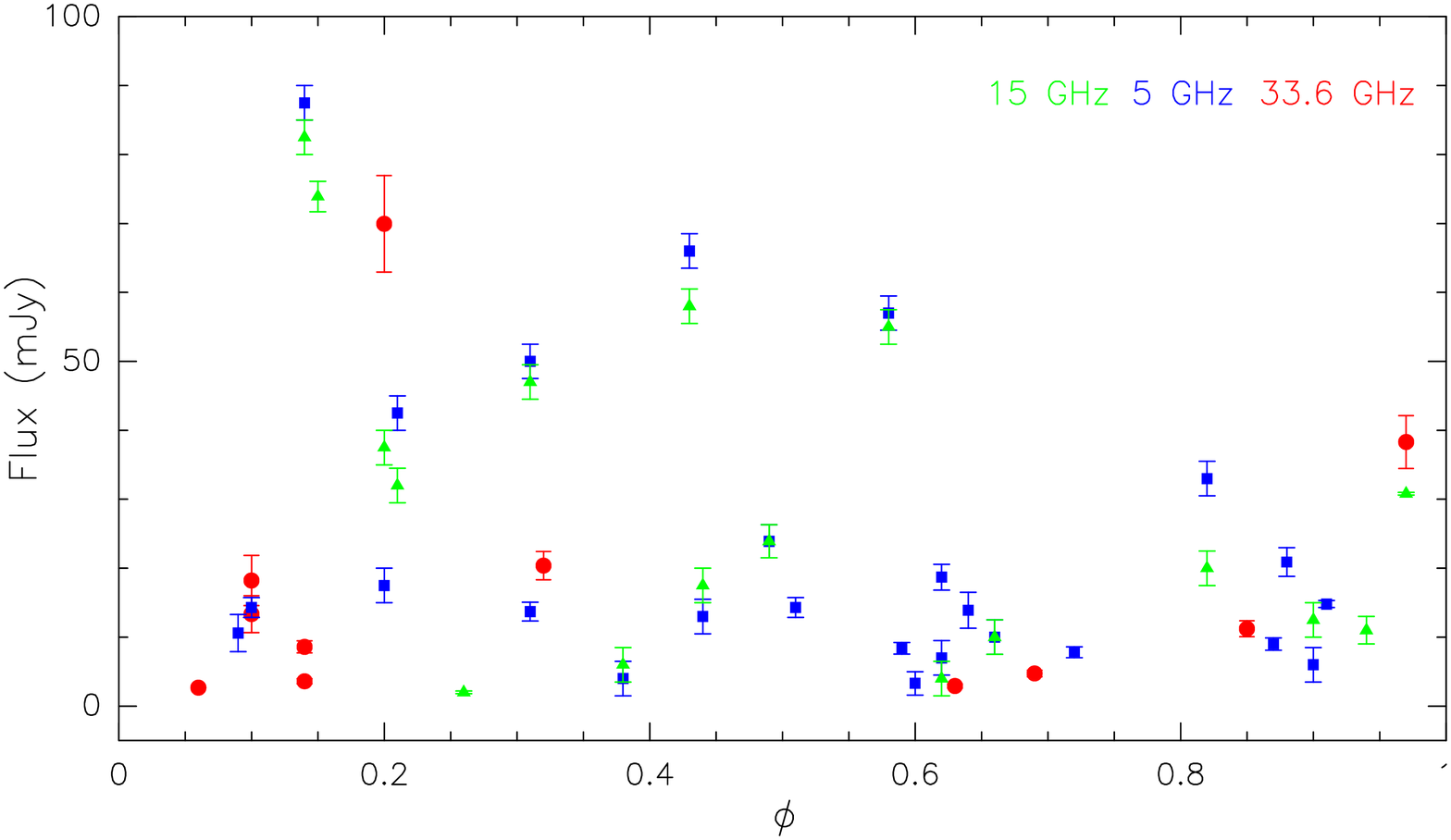}
\caption{Flux densities of source 12 as a function of the orbital
period at 33.6 GHz (red dots, this paper), 5 and 15 GHz (blue squares
and green triangles, respectively, from \citealt{felli91,felli93}).}
\label{fig-source12}
\end{figure}

\citet{felli91,felli93} monitored the radio emission from this
binary at 5 and 15 GHz.
We plot in Fig.\ \ref{fig-source12} their flux
densities and those of our monitoring as a function of the orbital phase $\phi$. We
have used the orbital parameters $P$=65.4325 and
$T_{\rm 0}$=JD 2446811.95 (\citealt{bossi89}).  The flux density at 5 and 15 GHz peaks near periastron ($\phi\sim$0.15), 
with lower levels at $\phi\sim$0.1 and $\phi=$0.6 to
0.9. Our higher frequency flux densities are
consistent with those at lower frequencies, with a peak after
periastron at $\phi\sim$0.2.  As \citet{felli93} noted and our data
at higher frequency confirm, the orbital modulation model may make
the main peak be detected always at the same position $\phi\sim$0.15$-$0.2.
However, the presence of large scatter in the radio emission cannot
be explained with this model.  \citet{felli93} set an upper limit
for the variability timescale of 10 to 20 days. Our monitoring has
revealed variability at shorter timescales of hours (Section
\ref{short-variability}). This is in agreement with non-thermal
emission due to stellar activity, perhaps in addition to orbital
modulation.  Therefore, we conclude that although certain orbital
modulation may be present, producing the observed flux density peak,
there is also a non-thermal emission component arising likely from
the low-mass companion that varies independently of the orbital
phase.

\section{Rate of radio flaring activity}
\label{flaring-rate}

There are only a few radio flares detected from young stellar objects. This could indicate that these phenomena are rare events, or alternatively, that the sensitivity of the observations carried out so far has limited their detection. Our monitoring allows us to evaluate which is the most likely scenario. We have detected flares in sources OHC-E, F, n, 7 and 12. Other sources (e.g. D, E, G, 6, 11, 12 and 25) showing high levels of variability through different epochs separated by longer timescales, might be also flaring sources, but with our data we are not able to confirm short-term variability.

\citet{bower03} parametrized the rate of radio flares (flares day$^{-1}$) in the entire ONC/OMC region based on their data and the previous observations of variable sources by \citet{felli93}. Since the number of flares depends on the field of view (FoV) of a particular observation, and of the stellar density of the observed region, we reformulate the expression for the rate of radio flares, $N_{\rm RF}$, to:

\begin{equation}
\label{flarerate}
\centering
N_{\rm RF}=\gamma \left(\frac{F_{\rm 3\sigma}}{100\,mJy}\right)^{\alpha} \left(\frac{A_{\rm FoV}}{A_{\rm tot}}\right) \left(\frac{\Sigma_{\rm FoV}}{\overline{ \Sigma}}\right),
\end{equation}

\noindent where $F_{3\sigma}$ is the threshold detection limit
(considered as 3 times the RMS of the observation); $\alpha$ is
the spectral index of the emission; $A_{\rm FoV}$ and $A_{\rm tot}$
are the areas of the observed region and the full ONC/OMC region,
respectively; and $\Sigma_{\rm FoV}$ and $\overline{\Sigma}$ are
the surface stellar density in the observed region and the mean
stellar density of the entire ONC/OMC, respectively. The parameter
$\gamma$ is a constant that \citet{bower03} estimated to be between
0.01 and 0.1.

Since the flare rate is a function of the stellar density, it is
convenient to observe the more crowded region of a cluster to enhance
the chances of detecting radio flares. The region with highest
stellar density in Orion is the OHC (\citealt{rivilla13a}). Considering
a $\sim$25$\arcsec\times$25$\arcsec$ region centered in the stellar
density peak, using the typical power-law index of non-thermal
emission of $\alpha=-1$, the mean RMS noise of our monitoring (0.3
mJy), the approximated size of the ONC/OMC cluster of 15 pc$\times$15
pc, and the census of stars in the region provided by COUP, we
obtained that $N_{\rm RF}\sim$0.017-0.17 flare day$^{-1}$.

We have detected 2 clear flares (OHC-E and source n\footnote{The
detections of sources H and D in the OHC region show also tentative
evidences of flaring emission towards these stars (see Section
\ref{comparison-zapata})}.) in this area in 14 observations, so a
rough estimate of the flaring rate is $\sim$0.14 flare observation$^{-1}$.
Assuming that the typical duration of the radio flares is several
hours to days (\citealt{andre96,bower03,forbrich08}), this would
be approximately equivalent to 0.14 flare day$^{-1}$, which is
similar to the upper value from Eq.\ \ref{flarerate}.  Therefore,
our results suggest that $\gamma$ is closer to 0.1 rather than to
0.01, and consequently that the number of detectable flares is
significant. Therefore, our multi-epoch monitoring confirms that
the presence of radio flares is not a rare phenomenon in crowded
young stellar clusters.

Obviously, the detection of radio flares in the OHC is favored by
its high density of embedded low-mass stars. But, according to Eq.\
\ref{flarerate}, even in less dense regions the number of detectable
flares would be significant, especially if the sensitivity is
enhanced.  The improved capabilities of the VLA and ALMA may be
expected to reveal many more radio flares arising from young low-mass
star clusters. A single polarization ALMA observation at 90 GHz
(band 3) with a full bandwidth of 7.5 GHz and 50 antennas, can reach
a 8 $\mu$Jy sensitivity limit in only 2.3 hours of on-source observing
time\footnote{According to the ALMA sensitivity calculator available in
the ALMA Observing Tool.}
Considering a FoV of $\sim$25$\arcsec\times$25$\arcsec$, this ALMA
observation may find $\sim$ 6 radio flares day$^{-1}$, which
represents $\sim 25$\% of the X-ray sources detected by Chandra in
the region. This would confirm that radio flares are common events,
similarly to the X-ray flares detected by Chandra.

Future observations are clearly needed to derive a better estimate
of this radio flaring rate, through the detection of many more radio
flares. The flaring information for tens or even hundreds of PMS
stars will provide a complete statistical description of radio
short-term variability. Beyond the purely scientific interest, this
would have important technical implications for interferometric
imaging (\citealt{bower03}). The classical interferometric imaging
techniques assume a constant sky in image reconstruction. However,
this assumption would be violated by the presence of many variable
sources in the field. This would lead to a reduced dynamic range
of the image (\citealt{stewart11}). Also, it would become difficult
to concatenate multiple observations of the same region to obtain
deeper images. Therefore, a more accurate determination of the radio
flare rate would help to understand to what extent this may affect
deeper VLA and ALMA observations of young stellar clusters.

\section{Summary and Conclusions}
\label{conclusions}

In this work we have presented a multi-epoch radio monitoring of
the ONC/OMC region carried out with the VLA at high centimeter frequencies (33
and 45 GHz).  We have detected 19 radio sources, mainly concentrated
in the Orion Hot Core and Trapezium regions.  Two of them are related
with massive stars: sources BN and I. The flux densities of the BN source and
source C (related with a proplyd) are compatible with constant thermal emission. The source I, besides a constant thermal component, shows
tentative evidence of radio variability at both short- and long-term
timescales.  The remaining 16 sources show long-term (month-timescale)
variability, but it is not yet clear whether these comprise multiple
short-term events not covered by our monitoring
cadence.  Indeed, we have confirmed radio flares (i.e., short-term
radio variability on timescales of hours to days) in 5 sources: F,
7, n, 12 and the new source OHC-E, previously undetected at radio
wavelengths.

We have complemented our radio sample with other radio detections at 8.3 GHz
from the literature, and cross-correlated it with the X-ray COUP catalog to obtain the full sample of sources emitting radio and X-ray in the ONC/OMC region. The radio emission from young stars can be explained by a combination of 2 different mechanisms: i) non-variable thermal emission produced by ionized gas and/or heated dust from the ONC proplyds and the massive objects BN and I; and ii) variable (flaring) non-thermal gyrosynchrotron emission produced by accelerated electrons in the stellar corona of PMS low-mass members of the ONC and OMC.
We have found several hints relating this variable radio emission with the X-ray activity.

Our study of the radio variability of $\theta^{1} {\it Ori A}$ concludes that there is evidence of a non-thermal emission component arising likely from the low-mass companion. Moreover, certain orbital modulation may be present in this binary, producing the observed flux density peak.

We have derived a rough estimate of the radio flaring rate in the densest cluster in the region, which is embedded in the Orion Hot Core. We have obtained
$\sim$0.14 flares day$^{-1}$. This value is consistent with a empirical estimate assuming the sensitivity and FoV of our observations and the stellar density of the region. This confirms that radio flares are not rare phenomena during the earliest stages of star formation as previously thought, but relatively common events similarly to the well-known X-rays flares.

Our results have shown that the radio monitorings to date has been strongly
limited by sensitivity, detecting mainly those sources with higher
X-ray luminosity. 
This implies that the new capabilities of the VLA
and ALMA offer a unique opportunity to detect a much larger population
of radio sources in young stellar clusters.  New observations with
improved sensitivity and better angular resolution will provide
crucial information about the origin and nature of the radio emission,
and they will reveal how radio and X-ray phenomena are connected.

Furthermore, the presence of multiple variable radio sources
would have important implications for interferometric
imaging, since the classical techniques assume a constant sky.  
A more accurate determination of the radio flare rate would help to understand how this variability can affect the upcoming VLA and ALMA observations in young stellar clusters.

\section*{Acknowledgments}
\begin{small}
This work has been partially funded by Spanish projects
AYA2010-21697-C05-01 and FIS2012-39162-C06-01, and Astro-Madrid
(CAM S2009/ESP-1496), and CSIC grant JAE-Predoc2008. I.J-S.
acknowledges the funding received from the People Programme (Marie
Curie Actions) of the European Union's Seventh Framework Programme
(FP7/2007-2013) under REA grant agreement number PIIF-GA-2011-301538. J.S-F. acknowledges the funding received from the project AYA2011-30147-C03-03.
The National Radio Astronomy Observatory is a facility of the
National Science Foundation operated under cooperative agreement
by Associated Universities, Inc.
\end{small}

\bibliographystyle{mn2e}
\bibliography{bib_orion}

\end{document}